\documentclass[twocolumn]{aastex631}
\pdfoutput=1

\usepackage{newtxtext,newtxmath}
\usepackage{graphicx}	
\usepackage{amsmath}	
\usepackage{multirow}   
\usepackage{amssymb}	
\usepackage{xspace}
\usepackage{comment}    

\newcommand{\ergs}{\textrm{erg\,s$^{-1}$}\xspace}

\newcommand{\perdegsq}{\textrm{deg$^{-2}$}}
\newcommand{\kms}{\textrm{km s$^{-1}$}}
\newcommand{\ergscmsq}{\textrm{erg s$^{-1}$ cm$^{-2}$}}
\newcommand{\GGaia}{$G_{\textrm{Gaia}}$}
\newcommand{\Grr}{$G_{\textrm{Gaia}}-r$}
\newcommand{\sigmax}{\ensuremath{\sigma_{\textrm{max}}}}

\newcommand{\lya}{\textrm{Ly}\ensuremath{\alpha}}

\newcommand{\ciii}{\textrm{C}~\textsc{iii}]}
\newcommand{\civ}{\textrm{C}~\textsc{iv}}
\newcommand{\ha}{\textrm{H}\ensuremath{\alpha}}
\newcommand{\hb}{\textrm{H}\ensuremath{\beta}}
\newcommand{\hgam}{\textrm{H}\ensuremath{\gamma}}
\newcommand{\oii}{[\textrm{O}\,\textsc{ii}]}
\newcommand{\oiilam}{[\textrm{O}~\textsc{ii}]\ensuremath{\lambda\lambda}3726,3729}
\newcommand{\oiii}{[\textrm{O}~\textsc{iii}]}
\newcommand{\nii}{[\textrm{N}~\textsc{ii}]}
\newcommand{\mgii}{\textrm{Mg}~\textsc{ii}}
\newcommand{\oiiilam}{[\textrm{O}~\textsc{iii}]\ensuremath{\lambda}5007}
\newcommand{\niilam}{[\textrm{N}~\textsc{ii}]\ensuremath{\lambda}6584}
\newcommand{\nevlam}{[\textrm{Ne}~\textsc{iv}]\ensuremath{\lambda}3425}

\graphicspath{{./}}

\begin{document}

\title{Identifying Quasars from the DESI Bright Galaxy Survey}

\shorttitle{DESI BGS AGN Target Selection}
\shortauthors{S.~Juneau et al.}

\correspondingauthor{St\'ephanie Juneau}
\email{stephanie.juneau@noirlab.edu}



\author[0000-0002-0000-2394]{S.~Juneau}
\affiliation{NSF NOIRLab, 950 N. Cherry Ave., Tucson, AZ 85719, USA}

\author{R.~Canning}
\affiliation{Institute of Cosmology and Gravitation, University of Portsmouth, Dennis Sciama Building, Portsmouth, PO1 3FX, UK}

\author[0000-0002-5896-6313]{D.~M.~Alexander}
\affiliation{Centre for Extragalactic Astronomy, Department of Physics, Durham University, South Road, Durham, DH1 3LE, UK}
\affiliation{Institute for Computational Cosmology, Department of Physics, Durham University, South Road, Durham DH1 3LE, UK}

\author[0000-0002-4940-3009]{R.~Pucha}
\affiliation{Department of Physics and Astronomy, The University of Utah, 115 South 1400 East, Salt Lake City, UT 84112, USA}
\affiliation{Steward Observatory, University of Arizona, 933 N, Cherry Ave, Tucson, AZ 85721, USA}

\author[0000-0003-1251-532X]{V.~A.~Fawcett}
\affiliation{School of Mathematics, Statistics and Physics, Newcastle University, Newcastle, UK}

\author{A.~D.~Myers}
\affiliation{Department of Physics \& Astronomy, University  of Wyoming, 1000 E. University, Dept.~3905, Laramie, WY 82071, USA}

\author[0000-0002-2733-4559]{J.~Moustakas}
\affiliation{Department of Physics and Astronomy, Siena College, 515 Loudon Road, Loudonville, NY 12211, USA}

\author{O.~Ruiz-Macias}
\affiliation{Institute for Computational Cosmology, Department of Physics, Durham University, South Road, Durham DH1 3LE, UK}

\author[0000-0002-5954-7903]{S.~Cole}
\affiliation{Institute for Computational Cosmology, Department of Physics, Durham University, South Road, Durham DH1 3LE, UK}

\author[0000-0003-0230-6436]{Z.~Pan}
\affiliation{Kavli Institute for Astronomy and Astrophysics at Peking University, PKU, 5 Yiheyuan Road, Haidian District, Beijing 100871, P.R. China}

\author{J.~Aguilar}
\affiliation{Lawrence Berkeley National Laboratory, 1 Cyclotron Road, Berkeley, CA 94720, USA}

\author[0000-0001-6098-7247]{S.~Ahlen}
\affiliation{Physics Dept., Boston University, 590 Commonwealth Avenue, Boston, MA 02215, USA}

\author[0000-0002-3757-6359]{S.~Alam}
\affiliation{Tata Institute of Fundamental Research, Homi Bhabha Road, Mumbai 400005, India}

\author[0000-0003-4162-6619]{S.~Bailey}
\affiliation{Lawrence Berkeley National Laboratory, 1 Cyclotron Road, Berkeley, CA 94720, USA}

\author{D.~Brooks}
\affiliation{Department of Physics \& Astronomy, University College London, Gower Street, London, WC1E 6BT, UK}

\author[0000-0001-8996-4874]{E.~Chaussidon}
\affiliation{Lawrence Berkeley National Laboratory, 1 Cyclotron Road, Berkeley, CA 94720, USA}

\author{C.~Circosta}
\affiliation{Department of Physics \& Astronomy, University College London, Gower Street, London, WC1E 6BT, UK}

\author{T.~Claybaugh}
\affiliation{Lawrence Berkeley National Laboratory, 1 Cyclotron Road, Berkeley, CA 94720, USA}

\author{K.~Dawson}
\affiliation{Department of Physics and Astronomy, The University of Utah, 115 South 1400 East, Salt Lake City, UT 84112, USA}

\author[0000-0002-1769-1640]{A.~de la Macorra}
\affiliation{Instituto de F\'{\i}sica, Universidad Nacional Aut\'{o}noma de M\'{e}xico,  Cd. de M\'{e}xico  C.P. 04510,  M\'{e}xico}

\author[0000-0002-4928-4003]{Arjun~Dey}
\affiliation{NSF NOIRLab, 950 N. Cherry Ave., Tucson, AZ 85719, USA}

\author{P.~Doel}
\affiliation{Department of Physics \& Astronomy, University College London, Gower Street, London, WC1E 6BT, UK}

\author[0000-0003-2371-3356]{K.~Fanning}
\affiliation{Kavli Institute for Particle Astrophysics and Cosmology, Stanford University, Menlo Park, CA 94305, USA}
\affiliation{SLAC National Accelerator Laboratory, Menlo Park, CA 94305, USA}

\author[0000-0002-2890-3725]{J.~E.~Forero-Romero}
\affiliation{Departamento de F\'isica, Universidad de los Andes, Cra. 1 No. 18A-10, Edificio Ip, CP 111711, Bogot\'a, Colombia}
\affiliation{Observatorio Astron\'omico, Universidad de los Andes, Cra. 1 No. 18A-10, Edificio H, CP 111711 Bogot\'a, Colombia}

\author{E.~Gaztañaga}
\affiliation{Institut d'Estudis Espacials de Catalunya (IEEC), 08034 Barcelona, Spain}
\affiliation{Institute of Cosmology and Gravitation, University of Portsmouth, Dennis Sciama Building, Portsmouth, PO1 3FX, UK}
\affiliation{Institute of Space Sciences, ICE-CSIC, Campus UAB, Carrer de Can Magrans s/n, 08913 Bellaterra, Barcelona, Spain}

\author[0000-0003-3142-233X]{S.~Gontcho A Gontcho}
\affiliation{Lawrence Berkeley National Laboratory, 1 Cyclotron Road, Berkeley, CA 94720, USA}

\author{G.~Gutierrez}
\affiliation{Fermi National Accelerator Laboratory, PO Box 500, Batavia, IL 60510, USA}

\author[0000-0003-1197-0902]{C.~Hahn}
\affiliation{Department of Astrophysical Sciences, Princeton University, Princeton NJ 08544, USA}

\author{K.~Honscheid}
\affiliation{Center for Cosmology and AstroParticle Physics, The Ohio State University, 191 West Woodruff Avenue, Columbus, OH 43210, USA}
\affiliation{Department of Physics, The Ohio State University, 191 West Woodruff Avenue, Columbus, OH 43210, USA}
\affiliation{The Ohio State University, Columbus, 43210 OH, USA}

\author{R.~Kehoe}
\affiliation{Department of Physics, Southern Methodist University, 3215 Daniel Avenue, Dallas, TX 75275, USA}

\author[0000-0003-3510-7134]{T.~Kisner}
\affiliation{Lawrence Berkeley National Laboratory, 1 Cyclotron Road, Berkeley, CA 94720, USA}

\author[0000-0001-6356-7424]{A.~Kremin}
\affiliation{Lawrence Berkeley National Laboratory, 1 Cyclotron Road, Berkeley, CA 94720, USA}

\author{A.~Lambert}
\affiliation{Lawrence Berkeley National Laboratory, 1 Cyclotron Road, Berkeley, CA 94720, USA}

\author[0000-0003-1838-8528]{M.~Landriau}
\affiliation{Lawrence Berkeley National Laboratory, 1 Cyclotron Road, Berkeley, CA 94720, USA}

\author[0000-0001-7178-8868]{L.~Le~Guillou}
\affiliation{Sorbonne Universit\'{e}, CNRS/IN2P3, Laboratoire de Physique Nucl\'{e}aire et de Hautes Energies (LPNHE), FR-75005 Paris, France}

\author[0000-0003-4962-8934]{M.~Manera}
\affiliation{Departament de F\'{i}sica, Serra H\'{u}nter, Universitat Aut\`{o}noma de Barcelona, 08193 Bellaterra (Barcelona), Spain}
\affiliation{Institut de F\'{i}sica d’Altes Energies (IFAE), The Barcelona Institute of Science and Technology, Campus UAB, 08193 Bellaterra Barcelona, Spain}

\author[0000-0002-4279-4182]{P.~Martini}
\affiliation{Center for Cosmology and AstroParticle Physics, The Ohio State University, 191 West Woodruff Avenue, Columbus, OH 43210, USA}
\affiliation{Department of Astronomy, The Ohio State University, 4055 McPherson Laboratory, 140 W 18th Avenue, Columbus, OH 43210, USA}
\affiliation{The Ohio State University, Columbus, 43210 OH, USA}

\author[0000-0002-1125-7384]{A.~Meisner}
\affiliation{NSF NOIRLab, 950 N. Cherry Ave., Tucson, AZ 85719, USA}

\author{R.~Miquel}
\affiliation{Instituci\'{o} Catalana de Recerca i Estudis Avan\c{c}ats, Passeig de Llu\'{\i}s Companys, 23, 08010 Barcelona, Spain}
\affiliation{Institut de F\'{i}sica d’Altes Energies (IFAE), The Barcelona Institute of Science and Technology, Campus UAB, 08193 Bellaterra Barcelona, Spain}

\author{A.~Muñoz-Gutiérrez}
\affiliation{Instituto de F\'{\i}sica, Universidad Nacional Aut\'{o}noma de M\'{e}xico,  Cd. de M\'{e}xico  C.P. 04510,  M\'{e}xico}

\author[0000-0001-6590-8122]{J.~Nie}
\affiliation{National Astronomical Observatories, Chinese Academy of Sciences, A20 Datun Rd., Chaoyang District, Beijing, 100012, P.R. China}

\author[0000-0003-3188-784X]{N.~Palanque-Delabrouille}
\affiliation{IRFU, CEA, Universit\'{e} Paris-Saclay, F-91191 Gif-sur-Yvette, France}
\affiliation{Lawrence Berkeley National Laboratory, 1 Cyclotron Road, Berkeley, CA 94720, USA}

\author[0000-0002-0644-5727]{W.~J.~Percival}
\affiliation{Department of Physics and Astronomy, University of Waterloo, 200 University Ave W, Waterloo, ON N2L 3G1, Canada}
\affiliation{Perimeter Institute for Theoretical Physics, 31 Caroline St. North, Waterloo, ON N2L 2Y5, Canada}
\affiliation{Waterloo Centre for Astrophysics, University of Waterloo, 200 University Ave W, Waterloo, ON N2L 3G1, Canada}

\author{C.~Poppett}
\affiliation{Lawrence Berkeley National Laboratory, 1 Cyclotron Road, Berkeley, CA 94720, USA}
\affiliation{Space Sciences Laboratory, University of California, Berkeley, 7 Gauss Way, Berkeley, CA  94720, USA}
\affiliation{University of California, Berkeley, 110 Sproul Hall \#5800 Berkeley, CA 94720, USA}

\author[0000-0001-7145-8674]{F.~Prada}
\affiliation{Instituto de Astrof\'{i}sica de Andaluc\'{i}a (CSIC), Glorieta de la Astronom\'{i}a, s/n, E-18008 Granada, Spain}

\author[0000-0002-3500-6635]{C.~Ravoux}
\affiliation{Aix Marseille Univ, CNRS/IN2P3, CPPM, Marseille, France}
\affiliation{IRFU, CEA, Universit\'{e} Paris-Saclay, F-91191 Gif-sur-Yvette, France}
\affiliation{Universit\'{e} Clermont-Auvergne, CNRS, LPCA, 63000 Clermont-Ferrand, France}

\author[0000-0001-5589-7116]{M.~Rezaie}
\affiliation{Department of Physics, Kansas State University, 116 Cardwell Hall, Manhattan, KS 66506, USA}

\author{G.~Rossi}
\affiliation{Department of Physics and Astronomy, Sejong University, Seoul, 143-747, Korea}

\author[0000-0002-9646-8198]{E.~Sanchez}
\affiliation{CIEMAT, Avenida Complutense 40, E-28040 Madrid, Spain}

\author[0000-0002-3569-7421]{E.~F.~Schlafly}
\affiliation{Space Telescope Science Institute, 3700 San Martin Drive, Baltimore, MD 21218, USA}

\author{D.~Schlegel}
\affiliation{Lawrence Berkeley National Laboratory, 1 Cyclotron Road, Berkeley, CA 94720, USA}

\author{M.~Schubnell}
\affiliation{Department of Physics, University of Michigan, Ann Arbor, MI 48109, USA}
\affiliation{University of Michigan, Ann Arbor, MI 48109, USA}

\author[0000-0002-6588-3508]{H.~Seo}
\affiliation{Department of Physics \& Astronomy, Ohio University, Athens, OH 45701, USA}

\author[0000-0002-3461-0320]{J.~Silber}
\affiliation{Lawrence Berkeley National Laboratory, 1 Cyclotron Road, Berkeley, CA 94720, USA}

\author[0000-0002-2949-2155]{M.~Siudek}
\affiliation{Institute of Space Sciences, ICE-CSIC, Campus UAB, Carrer de Can Magrans s/n, 08913 Bellaterra, Barcelona, Spain}

\author{D.~Sprayberry}
\affiliation{NSF NOIRLab, 950 N. Cherry Ave., Tucson, AZ 85719, USA}

\author[0000-0003-1704-0781]{G.~Tarl\'{e}}
\affiliation{University of Michigan, Ann Arbor, MI 48109, USA}

\author[0000-0002-4135-0977]{Z.~Zhou}
\affiliation{National Astronomical Observatories, Chinese Academy of Sciences, A20 Datun Rd., Chaoyang District, Beijing, 100012, P.R. China}

\author[0000-0002-6684-3997]{H.~Zou}
\affiliation{National Astronomical Observatories, Chinese Academy of Sciences, A20 Datun Rd., Chaoyang District, Beijing, 100012, P.R. China}



\begin{abstract}
The Dark Energy Spectroscopic Instrument (DESI) cosmology survey includes a Bright Galaxy Survey (BGS) which will yield spectra for over ten million bright galaxies ($r<20.2$ AB mag). 
The resulting sample will be valuable for both cosmological and astrophysical studies. However, the star/galaxy separation criterion implemented in the nominal BGS target selection algorithm excludes quasar host galaxies in addition to bona fide stars. While this excluded population is comparatively rare ($\sim 3-4$ per square degrees), it may hold interesting clues regarding galaxy and quasar physics. Therefore, we present a target selection strategy that was implemented to recover these missing active galactic nuclei (AGN) from the BGS sample. The design of the selection criteria was both motivated and confirmed using spectroscopy. The resulting BGS-AGN sample is uniformly distributed over the entire DESI footprint. According to DESI survey validation data, the sample comprises 93\% quasi-stellar objects (QSOs), 3\% narrow-line AGN or blazars with a galaxy contamination rate of 2\% and a stellar contamination rate of 2\%. 
Peaking around redshift $z=0.5$, the BGS-AGN sample is intermediary between quasars from the rest of the BGS and those from the DESI QSO sample in terms of redshifts and AGN luminosities. The stacked spectrum is nearly identical to that of the DESI QSO targets, confirming that the sample is dominated by quasars. We highlight interesting small populations reaching $z>2$ which are either faint quasars with nearby projected companions or very bright quasars with strong absorption features including the \lya\ forest, metal absorbers and/or broad absorption lines. 
\end{abstract}



\section{Introduction}
\label{sec:intro}

Obtaining a complete census of active galactic nuclei (AGN) is of fundamental importance for our understanding of black hole growth as well as for a complete picture of galaxy formation and evolution \citep{Madau+2014,Somerville+2015}. The presence of an AGN can have significant impacts on its host galaxy. For example, AGN feedback is thought to regulate star formation by heating up and/or expelling gas from the host galaxy \citep{Fabian2012,Harrison+2018}. Furthermore, AGN play a crucial role in the cosmological evolution of black holes as the likely main drivers of the growth of supermassive black holes across cosmic time \citep{Marconi+2004,Alexander+2012}. As a result, understanding cosmic history requires to study a representative galaxy population including active phases of black hole accretion ranging from moderate AGN luminosities to the most luminous instances known as quasars or quasi-stellar objects (QSOs) as there might be distinct modes of black hole activity \citep[e.g.,][]{Heckman+2014}.

Observationally, one method to identify AGN is searching for spectral signatures of AGN-ionized gas such as broad emission lines ($>1000$~\kms) arising from high-velocity gas near the black hole from the so-called broad line region (BLR), and highly-ionized but lower velocity gas from the narrow-line region (NLR) with typical line widths in the range $200-400$~\kms \citep{Osterbrock+2006}. However, the broad lines may be missing if the central region is obscured by intervening dusty material along the line of sight, whereas the NLR can be extended on scales of hundreds of parsecs to $\sim10$ kiloparsecs \citep[e.g.,][]{Bennert+2002,Hainline+2014}. 
AGN and quasars with both broad and narrow lines in their spectra are classified as Type~1 whereas those with only narrow lines are classified as Type~2 \citep{Khachikian+1974}. In addition to broad lines, the spectra of Type~1 quasars typically feature a prominent blue power-law component attributed to the accretion disk \citep{Elvis+1994}.

The wide area cosmological survey conducted with the Dark Energy Spectroscopic Instrument \citep{Overview+2022} will obtain spectra for approximately 40 million galaxies and quasars \citep{DESI+2016a}. While DESI is primarily a Stage IV cosmology experiment \citep{Levi+2013}, the wealth of spectroscopic data is bound to generate a large number of astrophysical discoveries. 
The selection of DESI targets is based on photometric properties and was carefully optimized for each tracer. The dark-time tracers consist of luminous red galaxies \citep[LRGs]{Zhou+2023}, emission-line galaxies \citep[ELGs]{Raichoor+2023}, and quasars \citep[QSOs]{Chaussidon+2023}.  

At the low redshift end, the Bright Galaxy Survey (BGS) will achieve a high spectroscopic sampling for galaxies spanning $0<z<0.6$ over a $\sim$14,000 square degree footprint \citep{Hahn+2023}. The BGS consists of a magnitude-limited component (BGS Bright, $r<19.5$), augmented by a component that reaches fainter magnitudes (BGS Faint, $z\lesssim22.2$) 
but that further includes color-dependent fiber magnitude cuts to ensure a high redshift success rate ($>95\%$). Together, the BGS Bright and BGS Faint samples (BGS-BF) will produce redshifts for $>10$ million galaxies during the course of the DESI survey.

During early Survey Validation \citep[SV;][]{DESI_SV+2023}, we discovered that the star-galaxy separation step of the BGS target selection removes quasars which present photometric signatures of a dominant point source and, as a result, appear stellar-like rather than galaxy-like. While some of those quasars are eventually selected as DESI QSOs \citep{Chaussidon+2023}, the latter favor $z>1$ targets. Thus, there remained an intriguing sub-population of bright AGN that risk being unaccounted by failing the BGS selection criteria due to their star/galaxy cut but also failing the QSO target selection criteria due to being at lower redshifts and/or possibly having a non-PSF morphology.
We thus developed a new sample selection algorithm to recover these missed luminous AGN and create the BGS-AGN sample, which is complementary to BGS-BF by design. 

In this paper, we describe how we solved this problem by defining a set of color and magnitude cuts trained on a spectroscopically confirmed sample from the Sloan Digital Sky Survey (SDSS), and applied to the DESI One-percent survey portion of SV. 
The photometric and spectroscopic observations are introduced in Section~\ref{sec:data}, followed by the primary BGS target selection and dedicated BGS-AGN target selection in Sections~\ref{sec:probl} and \ref{sec:selection}, respectively. Results include the target selection validation from the One-percent survey and high signal-to-noise stacked spectra (Section~\ref{sec:validation}). Lastly, we present physical properties of the BGS-AGN sample in Section~\ref{sec:properties}, which demonstrate that BGS-AGN targets essentially bridge the gap between the rest of the BGS galaxy sample and the QSO sample. Lastly, we summarize our main findings in Section~\ref{sec:conclusions}. We use AB magnitudes throughout this manuscript.

\section{Observations}
\label{sec:data}

\subsection{Tractor photometric catalogs}

The DESI survey target selection is based on photometric catalogs from the ninth data release (DR9) of the Legacy Surveys\footnote{\url{https://www.legacysurvey.org/dr9/description/}} \citep[LS; see overview by][]{Dey+2019}. The total LS footprint surpasses 14,000 square degrees and is divided into a northern portion covered by the Beijing-Arizona Sky Survey \citep[BASS;][]{Zou+2017} in the $g$ and $r$ bands and by the Mayall $z$-band Legacy Survey (MzLS) in the $z$ band. The southern portion is covered by the Dark Energy Camera Legacy Survey (DECaLS) in all $g$, $r$, $z$ bands. The photometry is extracted by running the Tractor pipeline \citep{Lang+2016} on a detection image made by coadding those three broad bands, and then fixing the shape and size to measure fluxes and their uncertainties in each filter. For a nominal galaxy exponential profile with a half-light radius of $0.45\arcsec$, DECaLS reaches median $5\sigma$ detection limits of $g=23.72$, $r=23.27$, and $z=22.22$ AB magnitudes. Similarly, the BASS/MzLS surveys reach $g=23.48$, $r=22.87$, and $z=22.29$ AB magnitudes.

The ground-based optical and near-infrared photometry was augmented by forced photometry in the mid-infrared spanning $3-22$~microns. The latter is based on images from the Wide-field Infrared Survey Explorer (WISE) satellite survey \citep{Wright+2010} and its NEOWISE extension \citep{Mainzer+2014} that were reprocessed by \citet{Meisner+2021} to create the unWISE maps in the $W1$ and $W2$ channels \citep[also see][]{Meisner+2018,Schlafly+2019}. The redder $W3$ and $W4$ photometry is extracted from the original WISE maps (AllWISE) because the NEOWISE extension and unWISE maps only exist for the first two channels. Lastly, the LS are matched to Gaia DR2 photometry in the broad $G$ band \citep[hereafter \GGaia;][]{Gaia+2018}.

In particular, Gaia-detected sources are fixed to a point spread function (PSF) morphological type when running the Tractor pipeline if:
\begin{equation}
    (G_{\rm Gaia} \leq 18\ \&\ \textsc{AEN} < 10^{0.5}) | (G_{\rm Gaia} \leq13)
\end{equation}
where AEN is the ASTROMETRIC\_EXCESS\_NOISE from Gaia DR2. While this condition is built-in when generating the photometric catalog, it is not explicitly used when developing the BGS target selection \citep{Hahn+2023} nor the BGS-AGN criteria described below in Section~\ref{sec:selection} so we discuss its impact on the BGS-AGN sample separately (Section~\ref{sec:inherit}).

\subsection{DESI Targeting}

\begin{deluxetable*}{lrllrll} 
\tablecaption{DESI samples and targeting summary.}\label{tab:targets}
\tablehead{
\colhead{Sample name} & \multicolumn{2}{c}{\tt SV3\_DESI\_TARGET} & & \multicolumn{2}{c}{\tt SV3\_BGS\_TARGET} & \colhead{Reference} \\
\cline{2-3} \cline{5-6}
\colhead{} \vspace{-0.2 cm} & \colhead{Bit} & \colhead{Bit name} & & \colhead{Bit} & \colhead{Bit name} & \colhead{} 
}
\startdata
BGS-BF  & 60 & {\tt BGS\_ANY} & & 0 & {\tt BGS\_FAINT}  & \citet{Hahn+2023} \\
        &    &                & & 1 & {\tt BGS\_BRIGHT} &   \\
\hline
BGS-AGN & 60 & {\tt BGS\_ANY} & & 2 & {\tt BGS\_WISE}   & This work \\
\hline
QSO     & 2  & {\tt QSO}      & & | &   |               & \citet{Chaussidon+2023} \\
\enddata
\end{deluxetable*}

DESI targets comprise both dark time and bright time tracers as well as secondary targets. 
A detailed description of DESI targeting is provided by \citet{Myers+2023}. In this work, we focus on BGS and QSO targets. The DESI {\tt QSO} target class is designed to favor $z>1$ QSOs and is intended to be observed during dark time \citep{Chaussidon+2023}. By contrast, the BGS is intended to be observed during bright time \citep{Hahn+2023}.
However, some targets are assigned to both the bright and dark time programs for data calibration purposes. In addition, a given object may be targeted by more than one target class if it meets the respective selection criteria of two or more classes. We describe how the target information is captured and quantify the overlap between for objects that fulfill more than one targeting classes.

The DESI targeting information is encoded using bitmasks.\footnote{\url{https://desidatamodel.readthedocs.io/en/latest/bitmasks.html\#target-masks}} The primary targeting column is called {\tt DESI\_TARGET} for the main survey, and {\tt SV3\_DESI\_TARGET} for the SV3 survey (similarly for SV1, and SV2). The BGS is further divided into Bright, Faint and AGN components and this targeting information is encoded as {\tt BGS\_TARGET} for the main survey and {\tt SV3\_BGS\_TARGET} for the SV3 survey. and we compile the information most directly relevant to this work in Table~\ref{tab:targets}.

By definition, there is no overlap between BGS-BF and BGS-AGN. However, there is a small overlap of these classes with the QSO targets. Looking at the number of unique targets with a good fiber status\footnote{\texttt{COADD\_FIBERSTATUS=0}} in SV3, we obtain N(BGS-BF)=259,181 with only 92 also targeted as QSO (0.035\%). For BGS-AGN, there are 519 unique targets targets with 72 also targeted as QSO (14\%). We compare the physical properties of the BGS-BF, BGS-AGN, and the QSO samples in Section~\ref{sec:properties}. 

\subsection{Spectroscopic catalogs}

We use spectroscopic catalogs from SDSS DR16 \citep{Ahumada+2020} and from DESI SV \citep{DESI_SV+2023,DESI_EDR+2023} to motivate and validate the target selection criteria. We further use spectra from the DESI SV campaign for visual inspection, to select representative examples and to create stacked spectra.

\subsubsection{SDSS DR16}\label{sec:sdsssample}

In the case of SDSS DR16, we cross-matched the main redshift table (SpecObj) to the LS DR9 photometry using a $1\arcsec$ matching radius, resulting in a sample of 4,531,048 objects. We use the SDSS pipeline classification provided (CLASS) to split the SDSS spectroscopic sample into STAR (795,683), GALAXY (2,612,295) and QSO (861,906). In the case of stars and galaxies, we further require ZWARNING=0 to ensure a reliable spectrum. This SDSS-LS test sample has the advantage to include a large number of objects but is limited to brighter magnitudes compared to the DESI survey depth. By cross-matching to the LS photometry, we can apply the same criteria as the DESI targeting to select subsamples of interest to predict the DESI targeting outcome before validating with the actual DESI SV observations.

\subsubsection{DESI SV catalogs}\label{sec:svsample}

The BGS-AGN target selection was included starting with the One-Percent survey portion of SV, which was the third phase \citep[SV3; described by][in their Section 2.3]{Myers+2023}. 
We use the SV3 redshift catalog as the validation dataset (Section~\ref{sec:validation}). The redshift catalogs produced by the Redrock pipeline (Fuji version; Bailey et al., in preparation) include best-fit redshift, error estimates, quality flags and the spectral classification based on the best-fitting template (SPECTYPE=STAR, GALAXY, QSO). 
Restricting to BGS-BF targets, the healpix redshift catalog from the SV3 survey includes a total of 274,890 entries with a valid fiber status\footnote{We apply COADD\_FIBERSTATUS=0.} for 259,181 unique BGS-BF targets. This total includes 154,974 unique BGS Bright targets and 104,207 BGS Faint targets.

We augment the default pipeline redshift catalogs with catalogs from the QSO afterburner {\tt QuasarNet} \citep{Busca+2018} and the broad Mg II finder pipelines \citep[see Section 6.2 of][for details]{Chaussidon+2023}. We compare the spectral classification from both classifiers and the redshift value from {\tt QuasarNet}.\footnote{The MgII classifier uses the Redrock redshift as an input and does not compute an independent redshift estimate.} \citet{Alexander+2023} found that using a modified pipeline with these additional classifiers identified 94\% of visually inspected quasars relative to 86\% with the standard Redrock pipeline alone. We therefore examine the results from both approaches in this work as well (Sections~\ref{sec:rr}, \ref{sec:afterburn}).

In addition to redshift and spectral type information, we gather emission line fluxes resulting from spectral fitting with version 2.5.0 of the \texttt{FastSpecFit}\footnote{\url{https://fastspecfit.readthedocs.io/en/latest/}} package (J.~Moustakas et al., in prep.; \citealt{Moustakas2023}). \texttt{FastSpecFit} models the continuum using a stellar population component and a smooth continuum component where the latter accounts for potential non-stellar contributions (such as an AGN accretion disk power-law) and for possible effects from flux calibration or other instrumental errors. The \texttt{FastSpecFit} package additionally fits emission lines using Gaussian profiles with either narrow lines or a combination of narrow and broad lines for the allowed transitions such as the Balmer lines of Hydrogen and Helium. In this work, we use rest-frame optical lines such as \hb, \oiiilam\ (hereafter \oiii) to assess the ionization properties via the \oiii/\hb\ ratio, and estimate the AGN luminosity via the \oiii\ luminosity. We correct the latter for dust attenuation via the Balmer decrement \ha/\hb\ ratio assuming a reference intrinsic ratio of 3.0 as expected for AGN ionized gas \citep{Osterbrock+2006}. We further extend toward the rest-frame UV by searching for broad \mgii$\lambda\lambda$2796,2803 doublet emission, which is fit separately from the optical lines by \texttt{FastSpecFit}. We use version 3.2 of the EDR value-added catalog\footnote{\url{https://fastspecfit.readthedocs.io/en/latest/fuji.html}}, except for some edge cases that we needed to refit as described in Appendix~\ref{app:finalclass}.

\subsection{DESI SV spectra}

We use the spectra from SV3 for visual inspection of all BGS-AGN targets (Section~\ref{sec:vi}) and to produce representative examples as well as stacked spectra to demonstrate average spectral properties. The spectra were reduced by the Redrock pipeline. We used the Everest internal reduction for visual inspection, and the more recent Fuji reduction for the rest of the analysis in this paper including the final Redrock classification results, afterburner classifier results and the example spectra and stacked spectra. The Fuji version has now been publicly released as the DESI Early Data Release \citep[EDR;][]{DESI_EDR+2023}. The spectra shown in this paper can be retrieved via the main DESI data access page\footnote{\url{https://data.desi.lbl.gov/doc/releases/edr/}} or via the Spectra Access and Retrievable Catalog Lab \citep[SPARCL,][]{Juneau+2024}.\footnote{\url{https://astrosparcl.datalab.noirlab.edu/}}

To maximize the spectral quality, we employ spectra that have been coadded per location on the sky \citep[\emph{HEALPix coadds}][]{Gorski+2005} though we note that the spectra are coadded separately for the dark and bright programs. While BGS is designed to be conducted primarily during bright time, some targets are also observed during dark time for calibration and quality assurance purposes. 
When there are multiple spectra for a single object, we select the best spectrum using a SPEC\_PRIMARY flag determined by the redshift warning (ZWARN) and ranked by the value TSNR2\_QSO, which a proxy for the signal-to-noise ratio (SNR) for QSO spectra \citep{Guy+2023}. In SV3, there are 782 coadded spectra for BGS-AGN targets with no fiber warning (COADD\_FIBERSTATUS=0). We visually examined all spectra but we report statistics for 519 unique objects in the remainder of this Paper to avoid double counting.

\section{DESI BGS Target Selection and Missing AGN}
\label{sec:probl}

As detailed by \citet{Hahn+2023}, the BGS target selection process involved magnitude and fiber-magnitude limits, star-galaxy separation, spatial masking around bright stars and other data quality cuts. Generally speaking, BGS-Bright is nearly magnitude limited at $r<19.5$, while BGS-Faint extends to fainter magnitudes but has additional color cuts to maximize redshift success rate of those fainter targets. The most relevant criterion for AGN and quasar subpopulations is the star-galaxy separation step because it is challenging to remove stars without also removing point-source dominated quasars and bright AGN.

\subsection{Star-galaxy separation}

\begin{figure}
\begin{centering}
\includegraphics[width=0.47\textwidth]{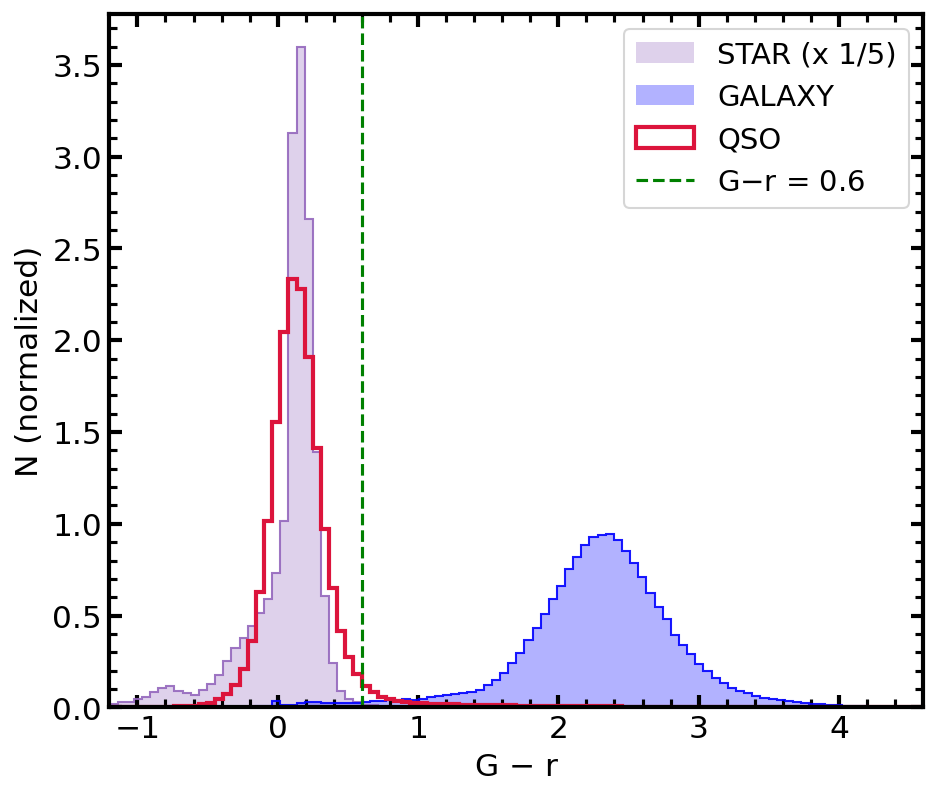}
\caption{Distribution of morphology-sensitive \Grr\ magnitude difference for SDSS galaxies with spectral classification of STAR (light purple), GALAXY (blue) or QSO (open red histogram). Given the much larger number of stars and their narrow distribution, we only show one fifth of that subsample. The vertical green dashed line is the adopted value to reject stars from the BGS target selection. Galaxies are either undetected in \GGaia\ or tend to have \Grr~$>0.6$.
\label{fig:grr_hist}}
\end{centering}
\end{figure}

The star-galaxy separation criterion was based on the comparison between the Gaia $G$ magnitude extracted assuming a narrow PSF profile (appropriate for stars) and the total Tractor $r$ magnitude accounting for the full light profile which can be extended for galaxies. Stars and objects otherwise dominated by a PSF-like morphology such as QSOs are expected to have a small difference \Grr\ $< 0.6$ while resolved galaxies will typically have an excess in the profile integrated $r$ magnitude and tend to have \Grr\ $> 0.6$ \citep{Ruiz-Macias+2020,Ruiz-Macias+2021}. This is supported by the distributions of \Grr\ values for the spectroscopically classified SDSS sample with LS DR9 photometry from Section~\ref{sec:sdsssample} as shown in Figure~\ref{fig:grr_hist}. This star-galaxy separation criterion is only applicable to objects with a Gaia detection.

We note that the majority (88.5\%) of SDSS galaxies are not detected in \GGaia\ and so are not shown in Figure~\ref{fig:grr_hist} and the criterion in not applicable in such cases. Galaxies with \Grr\ $>0.6$ account for 10\%, and only 1.5\% are below the threshold. In contrast, most (89\%) SDSS stars have a \GGaia\ detection and they display a very narrow distribution in \Grr. Among all 795,683 stars, 88\% are below the 0.6 threshold while only 0.8\% are above and the remaining 11\% lack a \GGaia\ detection. The QSO distribution is slightly broader than that of stars but the majority of QSOs with a \GGaia\ detection would be cut out by the threshold. However, a significant fraction of QSOs lack a \GGaia\ detection (43.5\%; not shown), which leaves 54.3\% of the SDSS QSO subsample below the threshold and 2.2\% above. Some of the cases undetected by Gaia will be selected as part of the BGS-BF and/or QSO targets.

\subsection{SDSS sample with DESI BGS selection criteria}

\begin{figure*}
\begin{centering}
\includegraphics[width=\textwidth]{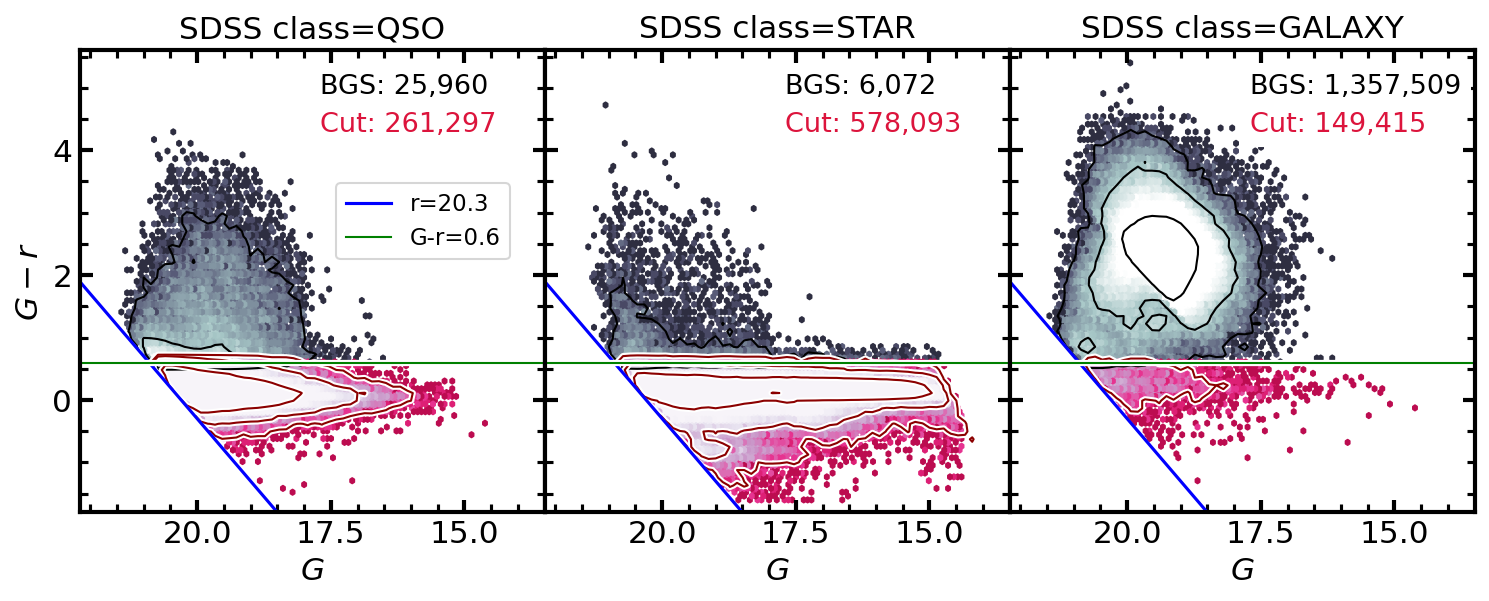}
\caption{SDSS-LS cross-matched sample on the \Grr\ vs. \GGaia\ plane divided according to the SDSS spectral class: (left) QSOs, (middle) Stars, (right) Galaxies. This figure illustrates some of the selection criteria of the BGS-BF sample (grey color map) relative to the complement sample that is cut (red color map) based on \Grr~$>0.6$ (green horizontal line). For reference, the solid blue line corresponds to $r=20.3$ chosen to be slightly fainter than the formal BGS faint limits ($r<20.175$ and $r<20.22$) to minimize overlap with the contours. The number density contours are spaced in increments of 1~dex with the outermost contour corresponding to 10 objects per bin. Numbers are labeled and show that the \Grr\ criterion removes the majority of stars (99\%), a modest percentage of galaxies (10\%), but also unfortunately removes a significant percentage of QSOs (91\%).
\label{fig:grr_g_missed}}
\end{centering}
\end{figure*}

We apply the BGS target selection criteria \citep{Hahn+2023} with the exception of the \Grr\ $>0.6$ cut to the SDSS-LS test sample with spectroscopic classification (Section~\ref{sec:sdsssample}) to investigate the impact of that criterion on the three spectral types. 
Figure~\ref{fig:grr_g_missed} shows the resulting bivariate distributions of \Grr\ as a function of \GGaia. For each spectral type (class in SDSS nomenclature), we compare objects that meet the BGS Bright or BGS Faint criteria (BGS-BF; in grey) with those that meet all the other criteria except for the star-galaxy separation and are therefore cut from BGS-BF (in red). We find that the \Grr\ $>0.6$ criterion shown with the horizontal green line removes 99\% of SDSS stars (578,093/584,165), 10\% of SDSS galaxies (149,415/1,506,924), and 91\% of SDSS QSOs (261,297/287,257). We note that criteria in $r$ magnitudes result in strong diagonal features and namely the BGS-Faint limit is around $r\sim20.22$ (the solid blue line corresponds to $r=20.3$ for visual reference given the bin size).

\subsection{Missing QSOs due to non-PSF morphology}\label{sec:psfqso}

One could expect that a significant fraction of the QSOs that are cut from the BGS sample (in red on the left-hand panel of Figure~\ref{fig:grr_g_missed}) may be recovered as DESI QSO targets. We check this by cross-referencing with potential DESI QSO targets based on the QSO targeting algorithms \citep{Chaussidon+2023, Myers+2023} applied to the LS DR9 photometry. We find that 246,885 out of the 261,297 (94.5\%) SDSS spectra with a QSO class that were cut due to the \Grr\ criterion are potential QSO targets.
Among the remaining 14,412 \emph{missing} SDSS QSOs that are not potential DESI QSO targets, a majority (56\%) have a resolved (i.e., non-PSF) morphology according to the Tractor MORPHTYPE.\footnote{The morphological type is called TYPE in the original Tractor catalogs but was renamed MORPHTYPE in the DESI catalogs to distinguish it from the spectral type SPECTYPE.} As detailed by \citet{Chaussidon+2023}, a PSF morphology is required for the nominal selection of DESI QSO targets as well as other photometric criteria tailored to preferentially select cosmological tracer QSOs at $z>1$, which are expected to appear point-like in imaging. 

However, the subpopulation of 8,115 quasars with an extended morphology could be physically interesting as possibly consisting of quasars with a resolved host galaxy component, in which case we would expect them to lie at lower redshifts and/or probe moderate luminosity AGN. We indeed find a preferentially lower redshift (median of $z=0.46$) for non-PSF morphological types compared to their counterparts with a PSF morphological type (median of $z=1.57$). This sets our expectations for the BGS AGN sample relative to the bulk of the DESI QSO target sample, which we will verify in Section~\ref{sec:zdistr}.

Given that not all the missing QSOs will be recovered by the DESI QSO target selection, we need to develop criteria to distinguish between stars and QSOs that are independent from \Grr\ given the strong overlap of this quantity between stars and QSOs. We use combinations of optical and infrared colors to achieve this goal. 
The BGS-AGN selection that we introduce in this work (Section~\ref{sec:selection}) would target 7,179 out of the 8,115 missing QSOs with a non-PSF morphology (88\%). Overall, it would target 10,671 (4.1\%) of the SDSS QSOs cut with \Grr$<0.6$. When taken together with the potential DESI QSO targets, we obtain a 97.3\% potential recovery rate.

\section{BGS AGN target selection}
\label{sec:selection}

The BGS AGN target selection relied on both precursor SDSS data and on early DESI SV1 data and was therefore implemented starting with the SV3 phase of survey validation and carried over the main DESI survey \citep{Myers+2023}. 

\subsection{Main AGN criteria}

We identify AGN hosts primarily via the presence of a hot dust component, which we infer by comparing the WISE infrared colors to optical colors, and by requiring a red $W1-W2$ color \citep[e.g.,][]{Stern+2012,Wu+2012}. 
We further select for the presence of a strong point source component via the comparison of Gaia and Tractor photometry. 
%

Relative to the BGS design paper \citep{Hahn+2023}, we adopt a similar approach in terms of applying magnitude, color and quality cuts on the photometry but we employ the opposite strategy to preferentially select objects that may be point-source dominated according to \Grr. We use predictions from the SDSS-LS DR9 sample together with SV1 spectra that were visually inspected to define the following BGS AGN criteria:

\begin{align} \label{eq:agn}
        (z - W2) - (g - r) &> -0.5   \\
        (z - W1) - (g - r) &> -0.7   \\
        (W1 - W2) &> -0.2 \\
        (G_{\rm Gaia} - r) &< 0.6
\end{align}

\begin{figure*}
\begin{centering}
\includegraphics[width=\textwidth]{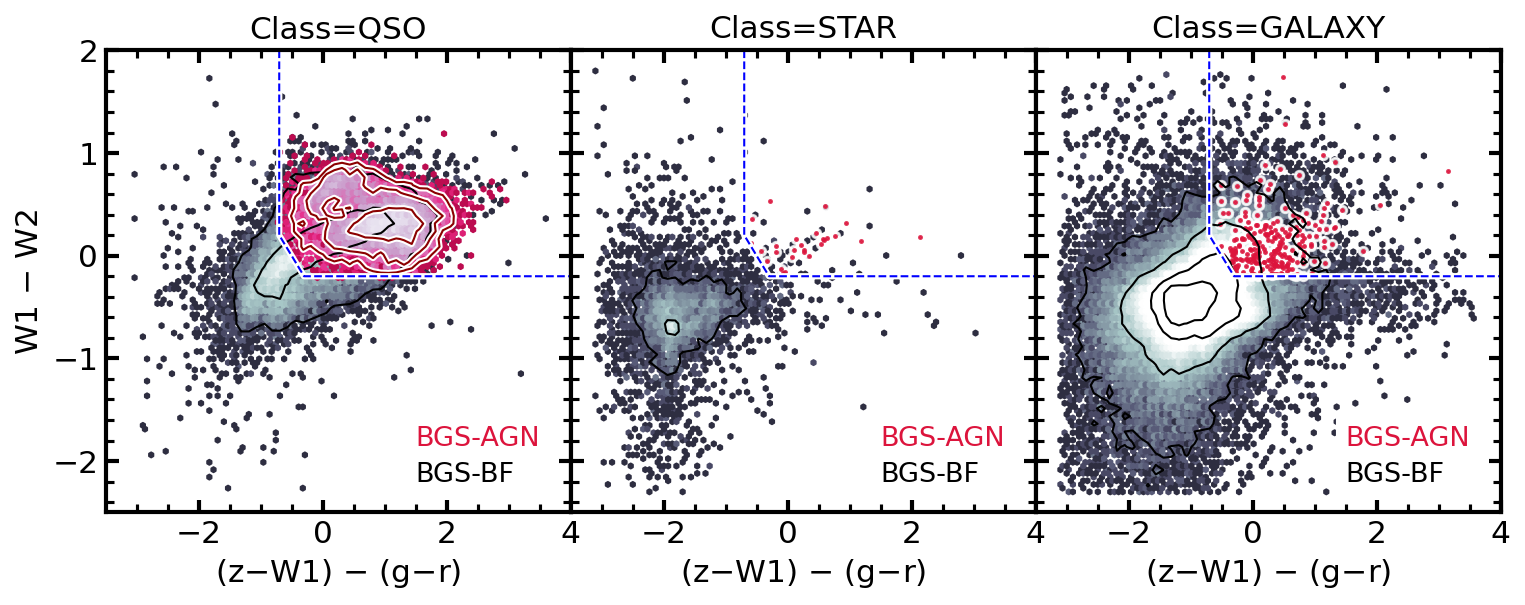}
\caption{$W1-W2$ color as a function of $(z-W1)-(g-r)$ from the LS DR9 photometry for each SDSS spectral class: QSO (left-hand panel), STAR (middle panel) and GALAXY (right-hand panel). 
The BGS-BF sample and BGS-AGN sample are shown in grey and red colors, respectively. BGS-AGN with spectral class of STAR or GALAXY are plotted with individual symbols rather than histograms and contours due to the small number of data points.
The blue dashed lines delineate the three color-based BGS-AGN selection criteria from Section~\ref{sec:selection}. The number density contours are spaced in increments of 1~dex (0.5~dex) for the BGS-BF (BGS-AGN) sample, with the outermost contour corresponding to 10 objects per bin. Notably, the BGS-AGN sample extends the QSO population toward more QSO-like colors and away from galaxy-like colors.}\label{fig:sdss_w1w2}
\end{centering}
\end{figure*}

Figure~\ref{fig:sdss_w1w2} demonstrates the first three cuts for the SDSS-LS DR9 sample after applying the selection criteria for BGS-BF (grey) and for the new BGS-AGN sample introduced here and in Section~\ref{sec:quality} (red). Namely, the $(z - W1) - (g - r)$ cut is shown directly on the horizontal axis (vertical dashed line), the $W1-W2$ cut is made along the vertical axis (horizontal dashed line) while the last cut yields the diagonal dashed line creating the top-right selection box for the BGS-AGN sample. We can see that the QSOs that happen to be selected as part of BGS-BF form a sequence in this parameter space and that BGS-AGN QSOs appear to follow and extend the same sequence away from the bulk of galaxies. For this SDSS-LS DR9 sample, applying the BGS-AGN selection yields a sample comprising 97.4\% QSOs, 0.3\% stars and 2.3\% galaxies.

\subsection{Quality criteria}\label{sec:quality}

We apply data quality criteria to ensure adequate imaging coverage, valid fluxes and the absence of masked pixels due to, e.g., proximity to a bright source in the $W2$ band.

Specifically, we require at least one observation in all of $g, r, z$ bands, with physical values of flux and inverse variance (ivar):
\begin{eqnarray}
    {\rm nobs}_i  &> 0 \quad{\rm for}~i = g, r, z \\
    {\rm flux}_i  &> 0 \quad{\rm for}~i = g, r, z \\
    {\rm ivar}_i  &> 0 \quad{\rm for}~i = g, r, z.
\end{eqnarray}

Given the importance of the WISE bands for a reliable AGN identification, we place comparatively stronger constraints on the $W1$ and $W2$ bands:
\begin{align}
    {\rm S/N}_i  &> 10 \quad{\rm for}~i = W1, W2 \\
    {\rm WISEM2} &= 0
\end{align}
where WISEM2 indicates whether the object touches a pixel in the WISEMASK\_W2 bright star mask\footnote{WISE masks are documented here: \\ \url{https://catalog.unwise.me/files/unwise\_bitmask\_writeup-03Dec2018.pdf}} as recorded in the Tractor photometric catalog (MASKBITS).\footnote{\url{https://www.legacysurvey.org/dr9/bitmasks/\#maskbits}}

Next, we apply magnitude cuts similar $-$ but not identical $-$ to the rest of the BGS target selection \citep{Hahn+2023}:
\begin{align}
    r_{\rm fibtot} &> 15 \\
    r_{\rm fiber} &< 22
\end{align}
\begin{align}
\begin{cases}
    16<r<17.5 & \text{if PSF} \\
    16<r<19.5 & \text{if not PSF}\\
    19.5\leq r<20.3 & \text{if not PSF \&}~(r_{\rm fiber} < 21.5)
    \end{cases}
\end{align}
where $r$, $r_{\rm fiber}$ and $r_{\rm fibtot}$ are respectively the $r$-band magnitude, the $r$-band fiber magnitude, and the total $r$-band fiber magnitude corrected for aperture loss.


Unlike the BGS selection, we further require an entry in the Gaia DR2 catalog (\GGaia\ $\ne 0$) and impose the following Gaia magnitude cut:
\begin{equation}
    G_{\rm Gaia}>16.
\end{equation}

Lastly, we apply two more quality flags from the photometric Tractor catalog (MASKBITS) to avoid objects near bright stars (bit ``BRIGHT'') or in areas identified as part of a globular cluster (bit ``CLUSTER'').

\subsection{Criterion inherited from the photometry}
\label{sec:inherit}

\begin{figure}
\begin{centering}
\includegraphics[width=0.45\textwidth]{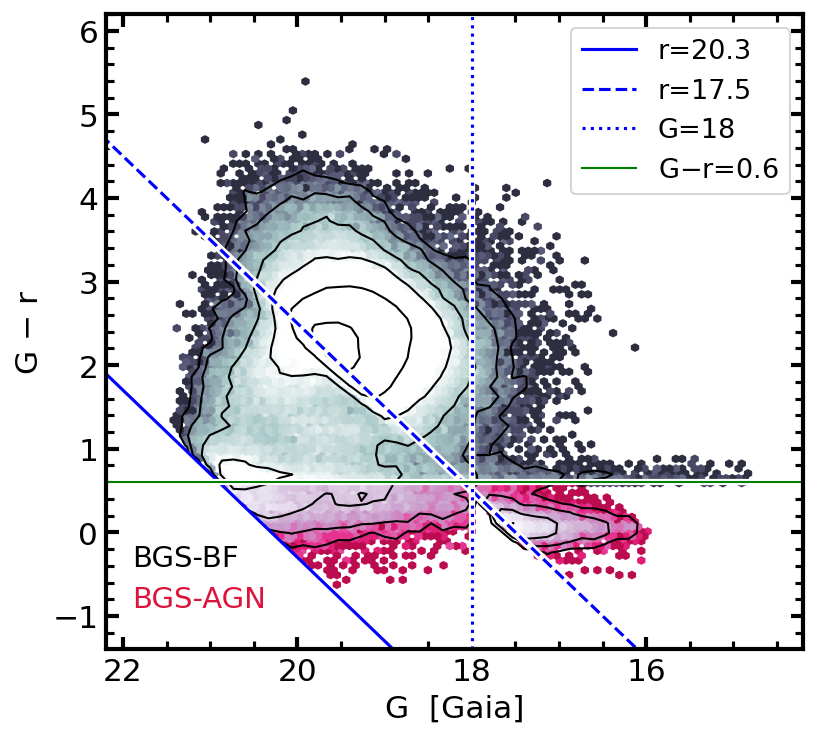}
\caption{SDSS cross-matched sample on the plane defined by \Grr\ as a function of \GGaia, illustrating some of the selection criteria of the BGS-BF sample (grey color map) and BGS-AGN sample (red color map). These include a cut based on \Grr$=0.6$ (green horizontal line), and cuts based on $r$ band magnitudes (dashed and solid blue lines values of $r=17.5$ and $20.3$, respectively). Furthermore, the LS DR9 Tractor photometry catalog includes a threshold at \GGaia=18 (dotted lines) in the sense that objects brighter than this value are assigned a PSF morphological type. The number density contours are spaced in increments of 0.5~dex with the outermost contour corresponding to 10 objects per bin. 
\label{fig:sdss_grr_g}}
\end{centering}
\end{figure}

We re-examine the bivariate distribution of \Grr\ as a function of \GGaia\ after applying all the BGS-BF and BGS-AGN criteria to the SDSS-LS sample including the quality cuts (Figure~\ref{fig:sdss_grr_g}). For objects fainter than \GGaia\ $>18$ we can see a continuation of the contours of a small population of QSOs selected by BGS-Faint (close to the limiting magnitudes around $r\sim20.22$). At the bright end ($r<17.5$, diagonal blue dashed line), the selection includes PSF morphology quasars, some of which may be bright due to their lower redshifts while others might overlap with intrinsically luminous quasars which will also be selected as QSO targets.

The small gap between \GGaia\ $<18$ and $r>17.5$ is explained by a criterion inherited from the Tractor photometry, which forces a morphological type of `PSF' for objects with a bright Gaia magnitude:  \GGaia\ $<18$ (dotted blue line) while our criteria only allowed PSF morphology for objects brighter than $z<17.5$ (dashed blue line). This photometric criterion was not initially considered during the sample selection. It impacts the subset of objects with $16<$~\GGaia~$<18$ and with $r>17.5$, which are currently excluded from the sample (small triangular region in Figure~\ref{fig:sdss_grr_g}).

\subsection{Sky coverage of BGS AGN targets}

\begin{figure*}
\begin{centering}
\includegraphics[width=0.8\textwidth]{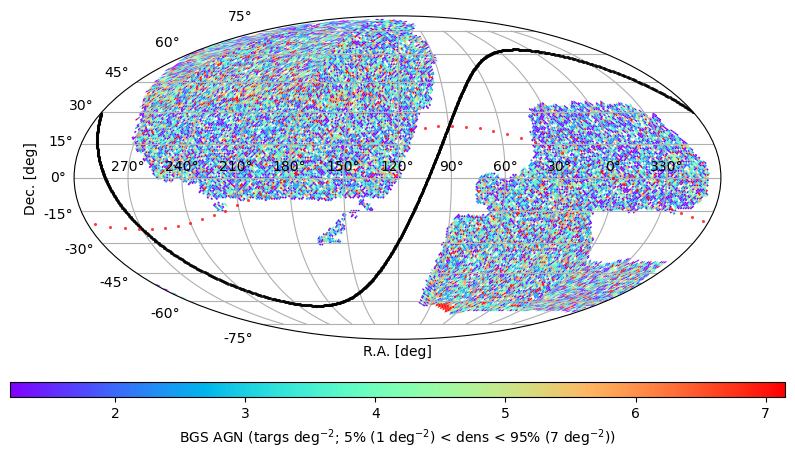}
\caption{Sky density map of BGS AGN targets. As indicated by the colorbar, the target number density is fairly uniform across the entire photometric footprint with 90\% of the values in the range $1 - 7$~targets per square degree. The solid black line (red dotted line) delineates the Galactic plane (Ecliptic plane).
\label{fig:skymaps}}
\end{centering}
\end{figure*}

After applying the full set of BGS AGN selection criteria, we compute the target number density, which is applicable for both the SV3 and the main DESI surveys over the photometric footprint. As displayed in Figure~\ref{fig:skymaps} we find a uniform sky distribution with a mean around $3 - 4$ targets per square degree and with 90\% of the values in the range $1 - 7$ targets per square degree. Importantly, there is no enhanced target density close to the Galactic plane (solid black line), which could occur if there were significant stellar contamination.

These modest values indicate that BGS AGN targets are rare relative to other target classes such as the full BGS targets ($\sim 80$~\perdegsq) and QSO targets ($\sim 320$~\perdegsq) \citep{Myers+2023}. While not numerous, they still contribute to increasing the completeness of bright galaxies observed by DESI, and can be of particular interest for galaxy and black hole studies. Namely, they are part of the DESI AGN Summary Catalog (Canning et al., in preparation).

\section{Survey Validation Results}
\label{sec:validation}

In this section, we describe how we used the DESI SV3 observations to validate the BGS AGN sample selection.  Namely, we check the spectral type (\S\ref{sec:class}) and redshift accuracy (\S\ref{sec:zaccuracy}) of the BGS AGN spectra compared to results from visual inspection. We further compare the average spectral shape to that of the average DESI QSO (\S\ref{sec:bgsstacks}) and confirm the presence of AGN ionization using an emission line diagnostic diagram (\S\ref{sec:bpt}).

\subsection{Spectral classification}\label{sec:class}

We assess the BGS AGN selection by checking the spectral classification from three different methods: (1) visual inspection; (2) default classification from the Redrock pipeline; (3) classification from the two QSO classifiers that are run in post-processing: {\tt QuasarNet} (QN) and the MgII classifier. 

\subsubsection{Visual inspection}\label{sec:vi}

\begin{figure*}
\begin{centering}
\includegraphics[width=0.95\textwidth]{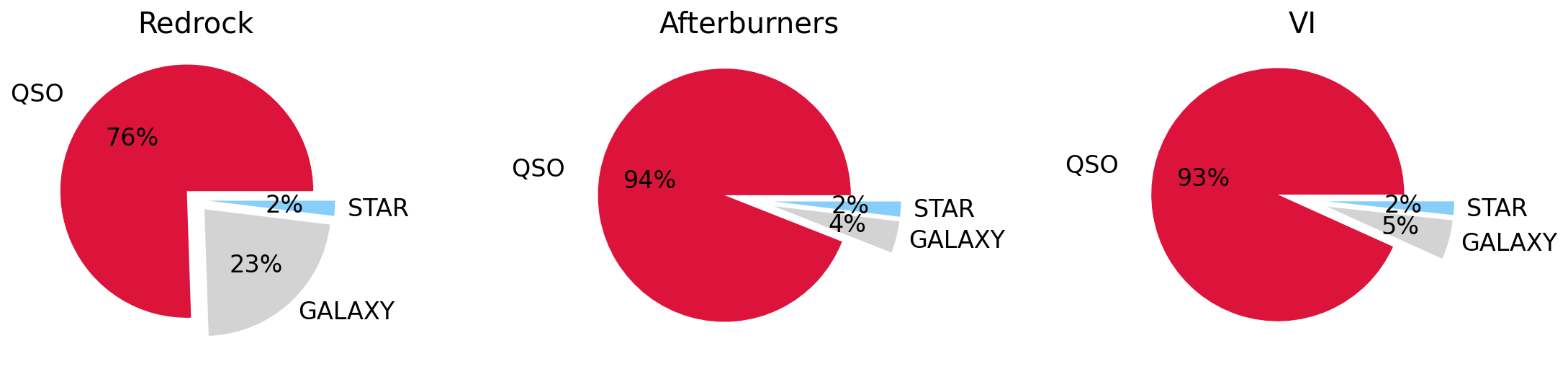}
\caption{Spectral Classification of the BGS-AGN SV3 sample according to the default Redrock pipeline (left), the QN and MgII afterburner classifiers (center), and from visual inspection (right). While Redrock misses several cases, they are recovered using the afterburners, which yield results very similar to visual inspection.
 \label{fig:pie_class}}
\end{centering}
\end{figure*}

We conducted a visual inspection campaign to examine each BGS-AGN spectrum and determine the spectral type (STAR, GALAXY or QSO), the redshift and the quality of the spectra (from 0 being unusable to 4 being the best quality). We followed a very similar method as for the DESI QSO targets \citep{Alexander+2023}. Each spectrum is independently examined by two people, and a third person reviews all the results to merge them, resolving conflicts if they arise. Because the BGS-AGN targets tend to be bright by selection, most of the spectra have a high signal-to-noise ratio and their redshift and spectral class can readily be identified visually. As a result, 516 out of 519 visual classifications were assigned a high quality ($\geq 2.5$) translating into a high confidence in the visual inspection results. The three cases with low quality were all found to be stellar contaminants (Appendix~\ref{app:contamination}).

In addition to the three main spectral types (STAR, GALAXY, QSO), visual inspectors were asked to note if the spectrum appeared to be a narrow-line AGN (Type~2 AGN) based on the presence of a clear \nevlam\ line or based on a visual assessment of the typical strong emission lines (e.g., elevated \oiii/\hb\ ratio). 
For the 519 unique BGS-AGN targets in SV3, the initial visual spectral classification yield 485 QSOs (93.4\%), 12 Type~2 AGN candidates (2.3\%), 13 galaxies (2.5\%) and 9 stars (1.7\%).

However, we consider the visual Type~2 classification tentative so we examine quantitative emission line ratio classification in Section~\ref{sec:bpt}, display individual cases in Appendix~\ref{app:finalclass}, and summarize the outcome here as well as in Table~\ref{tab:spectype}. Based on emission line measurements, we relabel one Type~2 to GALAXY due to the emission lines being very narrow ($\sigma<$100~\kms) and typical of low-mass star-forming galaxies. Conversely, one spectrum labeled as GALAXY was changed to Type~2 due to having strong \oiii/\hb\ together with a greater line width and tentative \nevlam\ and \mgii\ emission. Furthermore, one spectrum assigned as GALAXY that was flagged as ``Type~1?'' indicating tentative broad lines has now been confirmed as a QSO with a more recent DESI spectrum (Figure~\ref{fig:redqso}).

Three of the 13 spectra assigned a GALAXY spectral type are consistent with blazars with weak or no emission lines, a blue continuum, and a clear detection in the Very Large Array Sky Survey \citep[VLASS;][shown in Figure~\ref{fig:blazar}]{Lacy+2020}. Lastly, three of the Type~2 may instead be narrow-line Seyfert 1 (NLSy1), which are intermediate in terms of line widths (Figure~\ref{fig:nls1}). If we consider those as ``narrow-line'' AGN for the purpose of the final spectral classification, we obtain: 487 QSOs (93.8\%), 14 narrow-line AGN or blazar candidates (2.7\%), 9 galaxies (1.7\%) and 9 stars (1.7\%). Considering them as QSO would slightly increase our QSO success rate to 94.4\%.

\subsubsection{Redrock pipeline}\label{sec:rr}

The Redrock pipeline assigns a SPECTYPE of STAR, GALAXY or QSO based on the best-fit model template. In this work, we report the results from the Fuji spectral reduction used to produce the DESI EDR \citep{DESI_EDR+2023}. When comparing Redrock with visually classified spectra from the first phase of SV (SV1) for DESI QSO targets, \citet{Alexander+2023} found that the Redrock pipeline tends to correctly identify stars but occasionally misidentifies QSOs as galaxies such as cases with both a significant stellar continuum and AGN/QSO emission lines. This is consistent with the main findings for our sample of BGS AGN spectra. Redrock assigns a very similar percentage of stars (1.9\% versus 1.7\%) as the visually inspection results but assigns a lower QSO percentage (75.5\% compared to 93.4\%), and conversely a higher percentage of galaxies (22.5\%). While Redrock missed $\sim19$\% of the quasars, the redshift was not necessarily incorrect for such a high fraction. We evaluate the redshift success rate separately in Section~\ref{sec:zaccuracy}.

\subsubsection{Afterburner QSO classifiers}\label{sec:afterburn}

Similarly to the QSO target identification procedure outlined by \citet{Chaussidon+2023}, we use the results from two QSO classifiers that were ran in post-processing after the Redrock pipeline. The QN classifier \citep{Busca+2018,Farr+2020} is a machine-learning approach that assigns a probability of spectral emission line features being consistent with a QSO giving a confidence from 0 to 1, as well as determining a redshift for the spectrum. In this work, we apply the same criterion as for the QSO targets, which is to require that at least one of the main expected quasar emission lines considered (\lya, \civ, \ciii, \mgii, \ha, \hb) is assigned a confidence over 0.95 
(\texttt{C\_LINE\_BEST>0.95}).\footnote{An original threshold value of 0.5 was reported by \citet{Chaussidon+2023} but we subsequently revised it to 0.95 based on calculating classification precision and completeness as a function of C\_LINE\_BEST values (Canning et al., in prep.)}

The second classifier is based on the presence of a broad \mgii\ line. 
As described by \citet{Chaussidon+2023}, the MgII Classifier algorithm fits a Gaussian within a 250~\AA\ window centered at the expected position of \mgii\ given the Redrock redshift. There are three criteria to accept the presence of a broad \mgii\ line: (i) the improvement of $\chi^2$ must be better than 16, (ii) the width of the Gaussian greater than 10~\AA, and (iii) the amplitude-over-noise of the Gaussian be greater than three. If these conditions are met, the spectrum is reclassified as a QSO.

Using the combined QN and MgII classifiers yields very similar numbers and percentages as visual inspection: namely 488 QSOs (94\%), 21 galaxies (4\%) and 10 stars (1.9\%). There is no distinction for potential Type~2 AGN with neither the standard Redrock pipeline nor this modified QSO classifier pipeline so the 21 ``galaxies'' include some narrow-line AGN candidates as determined by the visual classification (Section~\ref{sec:vi}).

We show the proportion of each spectral type according to each method in Figure~\ref{fig:pie_class}. In conclusion, either visual classification or using the afterburner QSO classifiers improve the completeness of QSO classification by 19-20\% among the BGS AGN sample. We note that the spectral type of all BGS-AGN objects that were identified as QSOs by the default Redrock pipeline were confirmed during the visual inspection campaign. We discuss the small remaining contamination by stars and galaxies in Appendix~\ref{app:contamination}.

In detail, while the numbers of QSOs found by visual inspection and by the modified QSO classifier pipeline are similar, there are some small discrepancies in classification. Namely, there are six afterburner QSOs that were visually classified as either Type~2 (N=5; Figure~\ref{fig:sy2}) or as NLSy1 (TARGETID=39633413522588456; Figure~\ref{fig:nls1}). These cases were all found to have signatures of AGN and therefore this does not represent a major disagreement in the physical nature of those objects. Conversely, there were five visually identified QSOs not classified as such by the afterburners. One of the five was assigned an incorrect redshift from Redrock likely due to prominent broad absorption lines (BALs): TARGETID=39627939041510701 (Figure~\ref{fig:balQsoMakerMissed}). The remaining four spectra are shown in Figure~\ref{fig:QsoMakerMissed} and have relatively lower signal-to-noise ratios and/or redder continua relative to normal blue quasars. Overall, the reliable QSO recovery by the combined QN and MgII classifiers supports their application to the BGS AGN target class for the main DESI survey.

\subsection{Redshift accuracy}\label{sec:zaccuracy}

In addition to the spectral classification, a main objective of visual inspection is assessing the redshift. The {\tt prospect}\footnote{\url{https://desi-prospect.readthedocs.io/en/latest/index.html}} visual inspection tool presents a few redshift solutions from the Redrock pipeline as a starting point with an overlaid best-fit model and the option to interactively adjust the redshift with both a coarse and a fine slider. We consider redshifts with a visual quality flag greater or equal to 2.5 to be reliable, and use them as the reference truth to compare with the values determined from the Redrock pipeline and QN classifier. While in marginal quality cases the VI redshift could itself be wrongly assigned, we note that the majority of the BGS-AGN quasar spectra have a very high quality above 3.5 (480/485=99\%) with only 1\% having been assigned a moderately good quality flag ($2.5-3.5$).

The MgII classifier can change the spectral type but it uses the Redrock redshift value. However, QuarsarNet computes likely redshifts based on its own models and probability calculations. If the QN redshift differs from the Redrock redshift by more than 0.05 and the QSO identification has a high confidence (\texttt{C\_LINE\_BEST>0.95}), the Redrock pipeline is run again with only QSO templates and using the QN redshift as a prior to measure a new redshift for the quasar spectrum \citep{Chaussidon+2023}. For QSO targets, accurate redshifts are defined as within 3000\,\kms\ of the true redshift \citep[compared to 1000\,\kms\ for galaxies;][]{Lan+2023}. 

We define the redshift accuracy by:
\begin{equation}
dz = \frac{|z_{RR} - z_{VI}|}{(1+z_{VI})}
\end{equation}\label{eq:dz}
where $z_{VI}$ is the redshift from visual inspection and $z_{RR}$ is the redshift from Redrock. Below, we consider the Redrock redshifts before and after applying the QN classifier to compare their relative accuracy.

Finally, we quantify the percentage of quasars with good redshift accuracy as the percentage with good VI quality ($\geq2.5$) and a difference $dz\leq0.010$ (which corresponds to 3000\,\kms). Among 485 visually identified quasars, we find that 99.4\% have a good redshift accuracy after using the QN priors, compared to 95.3\% based on the original Redrock redshifts. Similarly, we find that 99.6\% of the 488 quasars identified with the combined Redrock+MgII+QN criteria have a good redshift accuracy, which is also a 4\% improvement relative to the original Redrock redshifts (95.5\%). The Redrock algorithm only identified 392 quasars ($\sim20$\% incomplete relative to the combined and VI methods) but 99.0\% of that sample have a good redshift accuracy. Based on these results, we conclude that the most complete and accurate redshift selection of quasars for the BGS AGN sample can be obtained by combining Redrock with the MgII and QN classifiers and that using QN priors can improve the fraction of targets with an accurate redshift by 4\% to reach $>99$\%.

\subsection{Stacked spectra}
\label{sec:bgsstacks}

To examine the typical spectral features of BGS AGN targets, we create stacked spectra 
following the same procedure as \citet{Alexander+2023}. Namely, the spectra are shifted to the rest-frame, aligned on a common wavelength grid, and normalized at a rest-frame wavelength of 3000\,\AA. They are then stacked by taking the median at each spectral bin. 

\begin{figure*}
\begin{centering}
\includegraphics[width=0.98\textwidth]{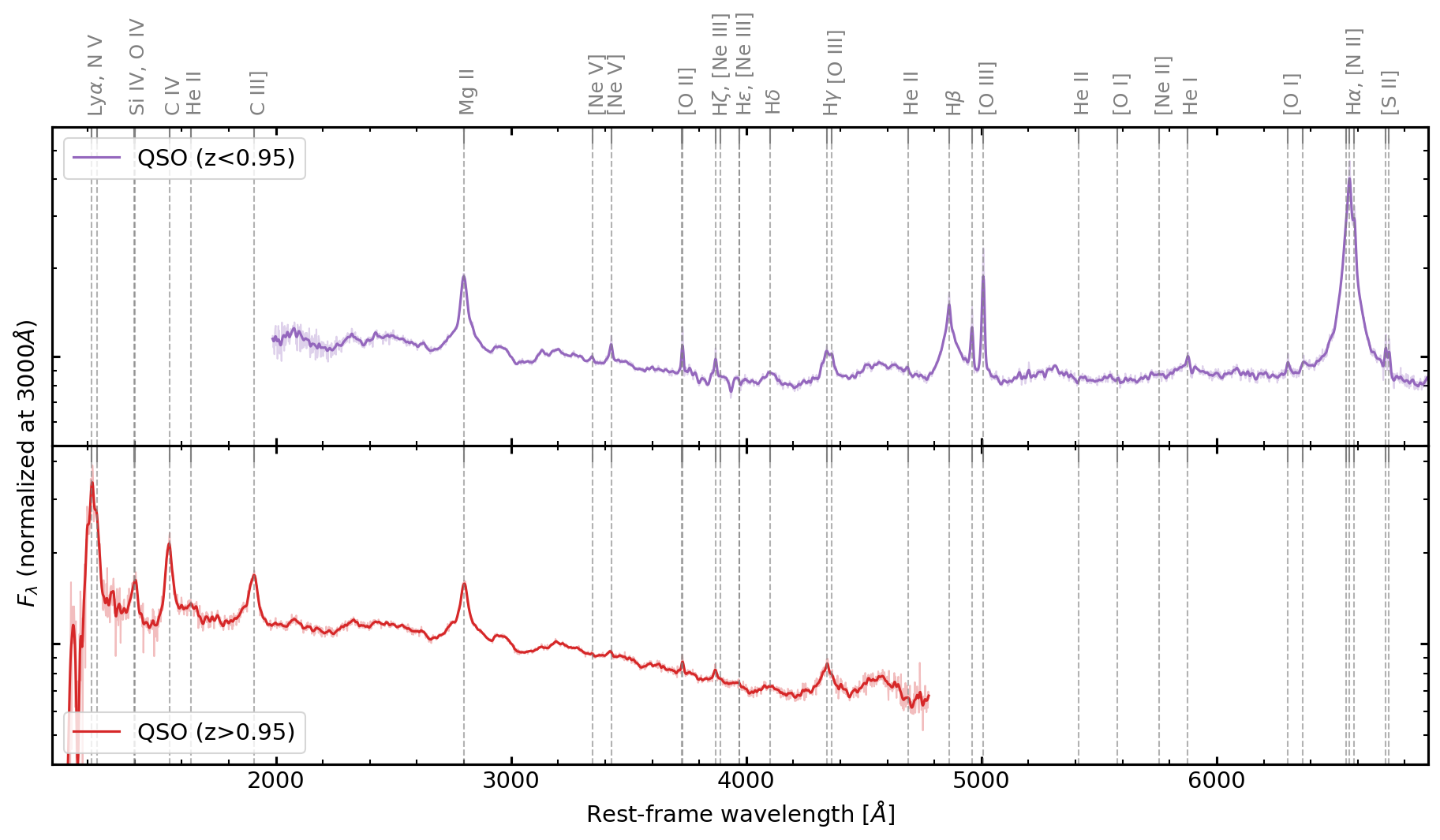}
\caption{Stacked spectra of BGS-AGN targets visually identified as QSOs split between the bulk of the sample at low redshift ($z<0.95$; $N=395$) and the remaining higher redshift QSOs ($z\geq0.95$; $N=90$). In each panel, the light line shows the stacked spectrum, and the darker line is a smoothed version with a Gaussian kernel ($\sigma$ = 2~pixels). Vertical dashed lines mark the expected location of emission lines, as labeled at the top.
 \label{fig:stacks}}
\end{centering}
\end{figure*}

In the top panel of Figure~\ref{fig:stacks}, we show the median spectrum for the bulk of the sample, which comprises 395 visually identified QSOs at $z<0.95$. This stacked spectrum resembles a typical broad-line (Type~1) QSO spectrum with a blue continuum attributed to the accretion disk and several broad permitted lines such as \mgii, \hgam, \hb\ and \ha\ \citep[e.g.,][]{VandenBerk2001}. In the bottom panel, the stacked spectrum was built by using the 90 QSOs at $z\geq0.95$, for which we obtain a spectral range probing further into the rest-frame UV. The resulting stack is again typical and strongly resembles the lower redshift counterpart except with a slightly bluer continuum. The bluer wavelength coverage allows 
us to probe additional broad lines such as \lya, \civ\ and \ciii. Together, these two stacked spectra represent 93\% of the BGS-AGN sample. We examine spectra for other classes such as narrow-line (Type~2) AGN, galaxies and stars separately in Section~\ref{sec:bpt} and in Appendix~\ref{app:contamination}.

\begin{figure*}
\begin{centering}
\includegraphics[width=0.95\textwidth]{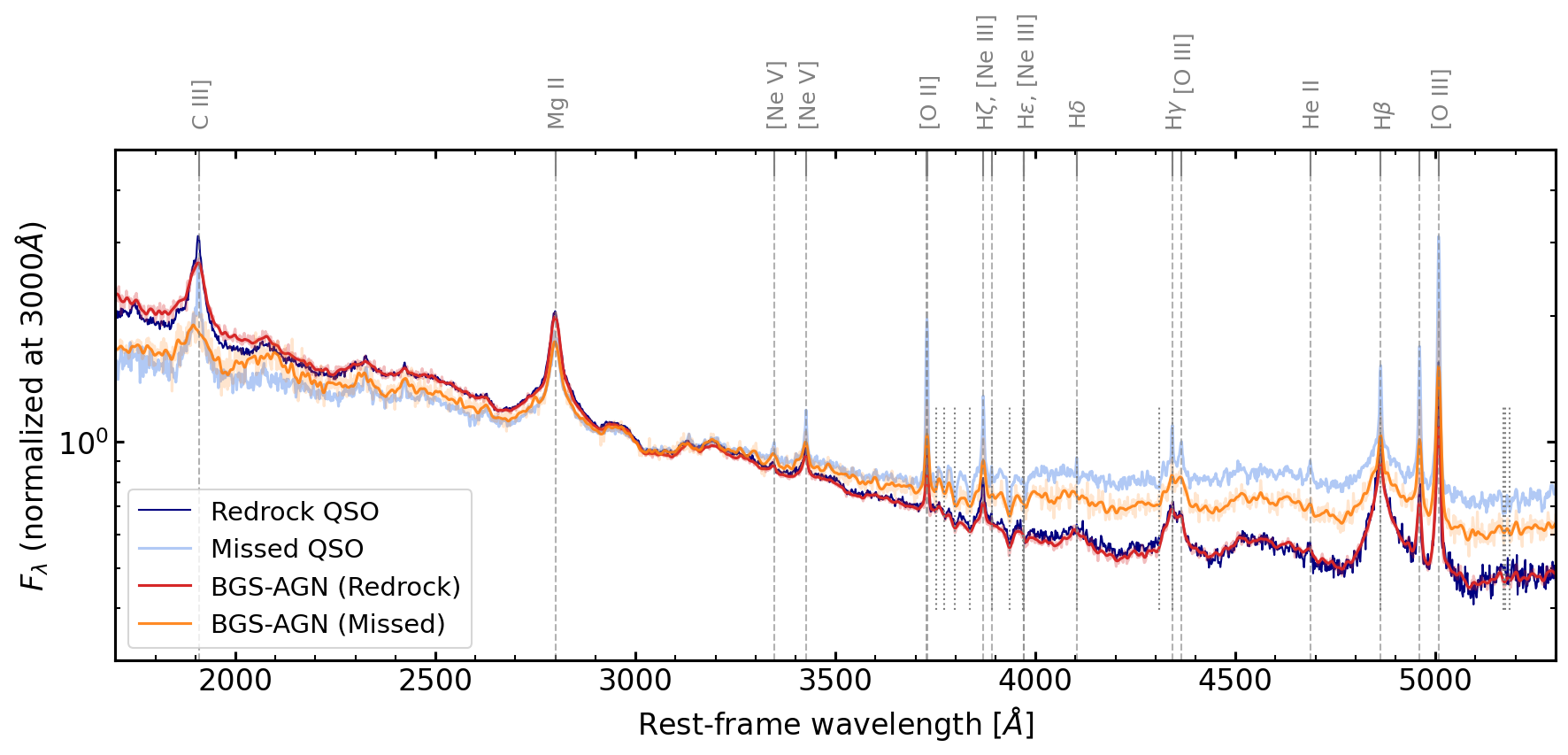}
\caption{Stacked spectra of BGS-AGN QSOs identified by Redrock (red) or missed by Redrock but identified by VI (orange) compared to the average spectra of DESI QSO targets identified by Redrock (dark blue) and those missed by Redrock but identified from VI (light blue). In each case, the light line shows the stacked spectrum, and the darker line is a smoothed version with a Gaussian kernel ($\sigma$ = 2~pixels). Vertical dashed lines mark the expected location of emission lines, as labeled at the top. Short vertical dotted lines mark the location of absorption lines.
\label{fig:stacks_rr}}
\end{centering}
\end{figure*}

Next, we compare directly with the stacked spectra from the visually inspected DESI QSO sample to further confirm that the BGS AGN sample indeed consists of \emph{bona fide} quasars. Similarly to \citet{Alexander+2023}, we created new stacks to separate QSOs identified with Redrock from those missed by the default Redrock pipeline that were however recovered visually. As Figure~\ref{fig:stacks_rr} demonstrates, the QSO spectra identified by Redrock are very similar between the BGS AGN sample presented here (red) and the main QSO survey of DESI (dark blue). In both cases, the stacked spectrum of the missed QSOs are redder in terms of spectral slope and with slightly more pronounced stellar absorption features (e.g., between 3500-4000~\AA). This difference is larger for the QSO sample (light blue) and more moderate for the BGS AGN missed sample (orange). Altogether, this confirms that, on average, the BGS AGN sample largely comprises typical QSOs. However, individual objects can still significantly deviate from these median spectra. 

\subsection{Emission-line classification}
\label{sec:bpt}

Optical emission line ratios are commonly used to identify the primary source of ionization in galaxies. The well-known BPT diagnostic diagram \citep{Baldwin+1981} combines the \oiii/\hb\ and \nii/\ha\ ratios, which are sensitive to a combination of ionization parameter and gas-phase metallicity and can be enhanced when gas is ionized by non-stellar sources such as AGN and/or shocks \citep[e.g., see review by][]{Kewley+2019}. Those line ratios are further selected to be closely spaced in wavelength and therefore have little to no sensitivity to dust obscuration. When using optical spectra, the BPT is only applicable at low redshifts given that \ha\ and \nii\ shift to the near infrared starting at redshift around $z>0.45$. 

A few alternative diagnostic diagrams have been suggested by keeping the \oiii/\hb\ ratio which is accessible in the optical range up to redshift $z\sim1$ but replacing the redder \ha\ and \nii\ lines. These include substituting \nii/\ha\ with absolute $H$-band magnitude \citep{Weiner+2007}, the rest-frame $U-B$ color as part of the Color-Excitation diagram \citep[CEx;][]{Yan+2011}, the stellar mass as part of the Mass-Excitation diagram \citep[MEx;][]{Juneau+2011,Juneau+2014} or the \oiii\ emission line width as part of the Kinematic-Excitation diagram \citep[KEx;][]{Zhang+2018}. 

In this work, we use both the original KEx diagram \citep{Zhang+2018} and a modified KEx diagram that we adapt to depend on the maximum line width (\sigmax) of \hb, \ha\ and \mgii\ instead of the originally proposed \oiii\ line width. The motivation for this change is to visually and quantitatively distinguish narrow-line (Type 2) and broad-line (Type 1) AGN as part of a single diagram.\footnote{also see \citet{Mullaney+2013}, where the authors used the \ha\ width} In all cases, the \oiii/\hb\ refers to only the narrow component of the \hb\ line, while \oiii\ is assumed to be a single component in the \texttt{FastSpecFit} fitting algorithm (Section~\ref{sec:svsample}).
The maximum line width corresponds to:
\begin{equation}\label{eq:sigmax}
\sigmax\ = \textrm{max}\{\sigma_{\hb}(\rm n), \sigma_{\hb}(\rm b), \sigma_{\ha}(\rm b), \sigma_{\mgii}\},
\end{equation}
where $\sigma_{\hb}(\rm n)$ and $\sigma_{\hb}(\rm b)$ are respectively the narrow and broad \hb\ components, $\sigma_{\ha}(\rm b)$ is the broad \ha\ component, and $\sigma_{\mgii}$ is the \mgii\ line width. In all cases, a value of line width is only considered if the line flux reaches S/N$>3$. The minimum requirement for inclusion on the KEx diagram is the availability of narrow \hb\ and \oiii\ emission lines, for which we also apply a S/N$>3$ detection threshold. Among the 519 unique BGS-AGN targets, there are 416 that cover the necessary spectral range and 390/416 (94\%) with a line detection for both \oiii\ and \hb.

\begin{figure}
\begin{centering}
\includegraphics[width=0.47\textwidth]{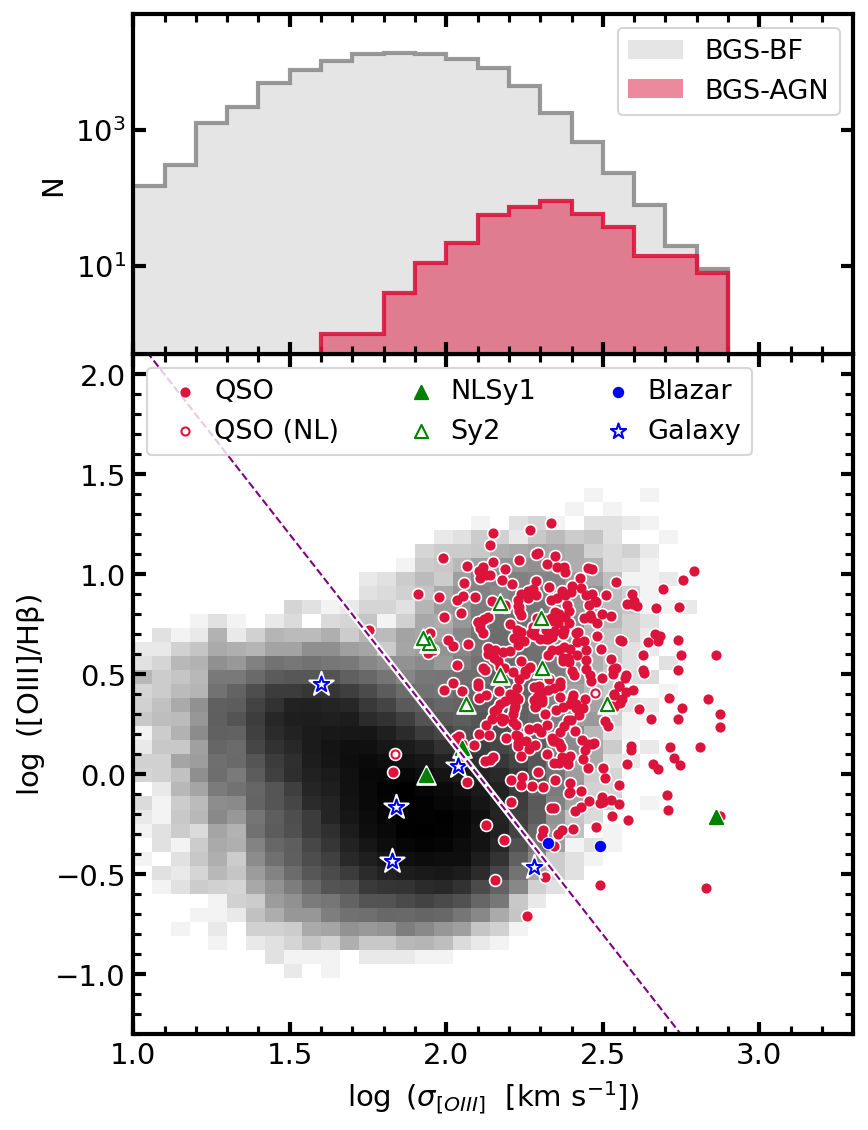}
\caption{(Top) Distribution of \oiii\ line widths for the BGS-BF (grey) and BGS-AGN (red) subsamples with \hb\ and \oiii\ detections. Overall the BGS-AGN sample is characterized by broader lines compared to the BGS-BF sample.  (Bottom) Original KEx AGN diagnostic diagram consisting in the \oiii/\hb\ line ratio versus the line width for the \oiii\ line, for the BGS-BF sample (shaded gray) and the BGS-AGN sample shown with symbols coded to their final spectral classification as labeled.
We note two QSOs for which the visually identified broad lines were not detected by the automated spectral fitting due to their low signal-to-noise spectra (open red circles). The diagonal dashed line is the demarcation of the original KEx diagram.
 \label{fig:kex_orig}}
\end{centering}
\end{figure}

\begin{figure}
\begin{centering}
\includegraphics[width=0.47\textwidth]{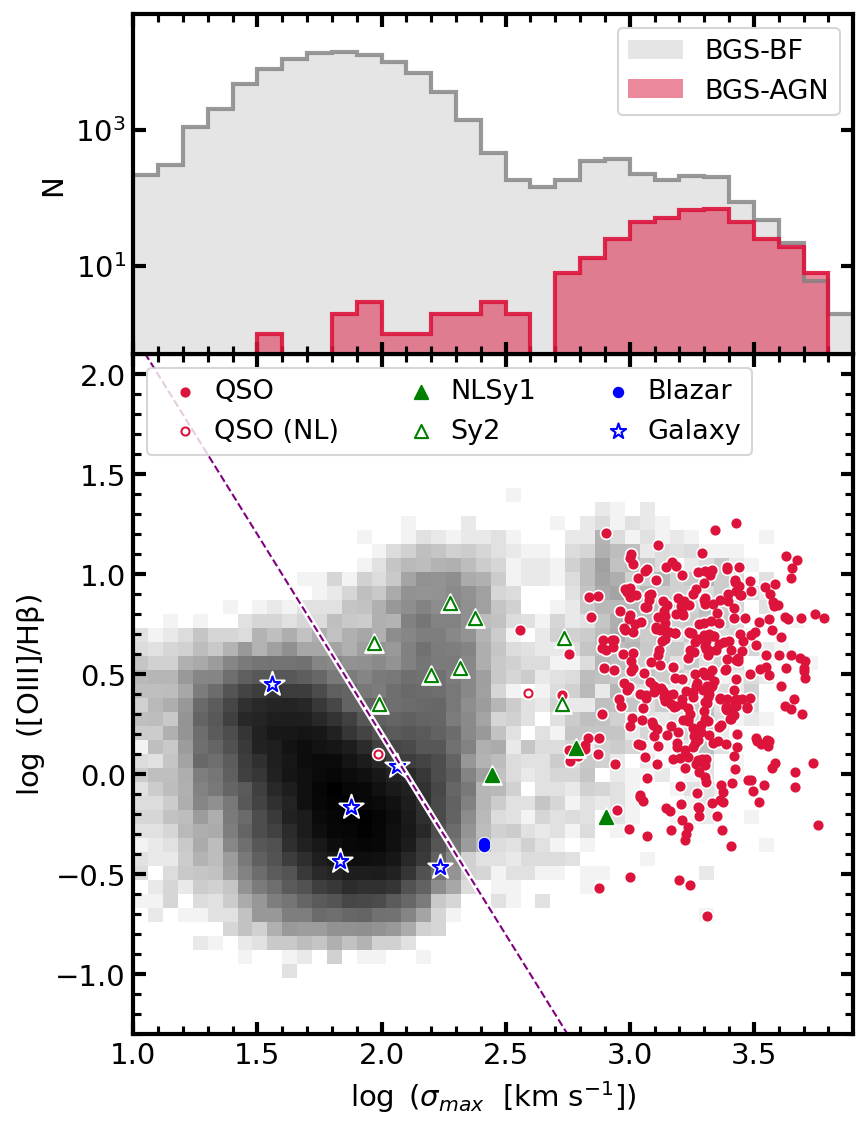}
\caption{(Top) Distribution of the maximum line widths (\sigmax; Equation~\ref{eq:sigmax}) for the BGS-BF (grey) and BGS-AGN (red) subsamples with \hb\ and \oiii\ detections. Overall the BGS-AGN sample is characterized by broader lines compared to the BGS-BF sample.  (Bottom) Modified KEx AGN diagnostic diagram consisting in the \oiii/\hb\ line ratio versus \sigmax, shown for the BGS-BF sample (shaded gray) and the BGS-AGN sample split per spectral type classification as labeled and as described in Figure~\ref{fig:kex_orig}. The diagonal dashed line is the demarcation of the original KEx diagram.
 \label{fig:kex}}
\end{centering}
\end{figure}

First, we show the original KEx diagnostic diagram in Figure~\ref{fig:kex_orig}. The top panel displays a comparison of the \oiii\ line width distribution between the BGS-BF sample (grey) and the BGS-AGN sample (red). One can see that the BGS-AGN sample is overall biased toward broader lines ($\sigma>$100~\kms) but there is no clear dichotomy in either line width distribution. The bottom panel shows the KEx diagram consisting of the \oiii/\hb\ line flux ratio (based on the narrow \hb\ component only) as a function of $\sigma_{\oiii}$. The dashed line is the empirical division proposed by \citet{Zhang+2018} to distinguish between star-forming galaxies to the lower left and AGN host galaxies to the upper right. The underlying grey bivariate distribution corresponds to the BGS-BF sample and we find that only 13.8\% is located above and to the right of the dividing line in the AGN side. The color symbols show the BGS-AGN sample and in contrast to the BGS-SF sample, we find that the majority (96.2\%) are on the AGN side.

The color and shape of the symbols used for the BGS-AGN sample are based on a revised spectral classification that involved examining all individual spectra and images as well as the emission line fitting results for all non-QSO spectral types as well as for all cases with conflicting results from various classification methods (see Appendix~\ref{app:finalclass} and \ref{app:contamination} for details and notes on individual objects). The final categories are as follows:
\begin{itemize}
\item QSO: visually identified QSOs from the presence of broad lines in the spectrum including both normal blue quasars and red quasars with reddened continua (red circles). For plotting, we use open circles to mark two cases for which the visually identified broad lines are not detected by the automated fitting routines;
\item NLSy1 (narrow-line Seyfert 1; filled triangles): visually flagged as Type~2 candidates but found to have line widths in the NLSy1 range ($400<$FWHM$<2000$~\kms, which is $170<\sigma<850$~\kms) with \oiii/\hb$<3$;
\item Sy2: Type~2 from the visual inspection campaign except for two cases relabeled based on their \oiii/\hb\ ratio and \sigmax\ (one Sy2 relabeled as galaxy, and one galaxy relabeled as Sy2);
\item Blazar: visually classified as galaxy candidates but with an unusual combination of weak to no emission lines, blue continuum and prominent radio (VLASS 3~GHz) detection;
\item Galaxy: visually classified as galaxy (except for one relabeled case from Type~2) with either no emission lines or narrow emission lines consistent with stellar photoionization according to the KEx diagrams. 
\end{itemize}
The KEx diagram classification is mostly consistent with the spectral types listed above\footnote{This is by definition for two objects out of 390 for which we used the line width and \oiii/\hb\ values to relabel} with a few discrepancies at the level of 2\% (8/375) QSO or NLSy1 in the star-forming galaxy side (filled red circles and green triangles). We tabulate the total number of objects in each spectral category in Table~\ref{tab:spectype} noting that the subsets with \oiii\ and \hb\ emission lines include: 370/485 QSO, 2/2 QSO (NL), 3/3 NLSy1, 8/8 Sy2, 2/3 blazars and 5/8 galaxies.

By design due to being based on a forbidden line width, the original KEx diagram does not distinguish between the narrow line (Type~2) and broad line QSO objects. For this purpose, we next investigate using the maximum line width introduced above (Equation~\ref{eq:sigmax}) as a substitute for $\sigma_{\oiii}$. We compare the \sigmax\ distributions for the BGS-BF sample (in gray) and the BGS-AGN sample (in red) in the top panel of Figure~\ref{fig:kex}. We find that the BGS-AGN sample is strongly biased toward broad lines. Even among just the Type 1 AGN from both samples ($\log(\sigmax)>2.6$), there is still a tendency for BGS-AGN targets to have broader lines relative to the BGS-BF sample, which may indicate more massive black holes.

In the bottom panel of Figure~\ref{fig:kex}, we show the modified KEx diagram with the maximum line width versus the \oiii/\hb\ line flux ratio from the narrow lines. The underlying distribution of BGS-BF galaxies (shaded grey) is characterized by a dominant population of star-forming galaxies with a faint plume of Type~2 AGN (with \sigmax$<$500~\kms; 2.7~dex) and a yet fainter plume of Type~1 AGN with  \sigmax$>1000$~\kms (3~dex; gray shaded distribution under the colored symbols). The BGS-AGN sample is explicitly divided according to the spectral classification listed above and with identical plotting symbols and demarcation line as in Figure~\ref{fig:kex_orig}. The modified KEx diagram allows us to compare the relative importance of star forming galaxies, narrow-line AGN and broad-line AGN at a glance, showing a more striking difference between the BGS-AGN sample and the BGS-BF samples compared to the original KEx diagram. That said, the original still had an excellent success rate at identifying AGN signatures.

\begin{deluxetable*}{lrlcl} 
\tablecaption{Spectral type classification}\label{tab:spectype}
\tablehead{
\colhead{Spectral type} & \colhead{N} & \colhead{Definition}  & \colhead{AGN?} & \colhead{Figure}}
\startdata
QSO & 485 & Broad lines among \lya, \civ, \mgii, \hb, \ha, with either a typical blue or an & Yes & Fig.~\ref{fig:stacks} \\
    &     & attenuated red continuum &  &  \\
QSO (NL) & 2 & Visually a QSO but the broad lines are not detected with \texttt{FastSpecFit} & Yes & Fig.~\ref{fig:balQsoMakerMissed}, \ref{fig:missBL} \\
NLSy1 & 3 & Visually flagged as Type~2 but with moderate broad lines (\sigmax$<$850~\kms) & Yes & Fig.~\ref{fig:nls1} \\
      &   & and \oiii/\hb$<3$  &  &  \\
Sy2 & 8 & AGN-like values or \oiii/\hb\ and \sigmax (possibly visually flagged as Type~2)  & Yes & Fig.~\ref{fig:sy2qm}, \ref{fig:sy2} \\
Blazar & 3 & Blue continuum with weak or no emission lines and a radio counterpart in VLASS & Yes & Fig.~\ref{fig:blazar} \\
Galaxy & 9 & Stellar continuum without or with emission lines consistent with stellar photoionization & No & Fig.~\ref{fig:gal_blue}-\ref{fig:hiz_gal} \\
Star & 9 & Best-fit by a stellar template with Redrock and either confirmed visually (N=6) or & No & Fig.~\ref{fig:starspsf}-\ref{fig:stars_lowq} \\
     &   & assigned a low confidence (N=3) &  &  \\
\hline
Total & 519 & & \\
\enddata
\end{deluxetable*}

\section{Properties of the BGS AGN sample}
\label{sec:properties}

Now that we have checked that the majority of the BGS AGN sample consists of quasars, we study a few properties such as their redshift distributions, infrared colors, and AGN luminosities. To put these properties in a broader context, we consider the BGS AGN sample relative to what we expected from the precursor SDSS-LS sample, and in relation to the BGS-BF and to QSO samples from DESI SV3.

\subsection{Redshift distribution N(z)}
\label{sec:zdistr} 

\begin{figure*}
\begin{centering}
\includegraphics[width=0.45\textwidth]{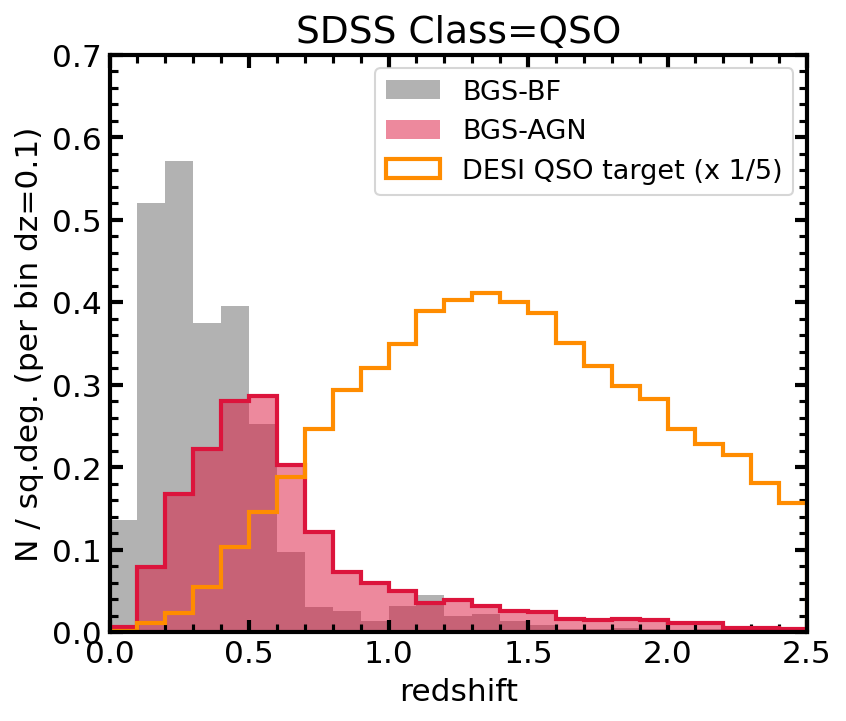}
\includegraphics[width=0.45\textwidth]{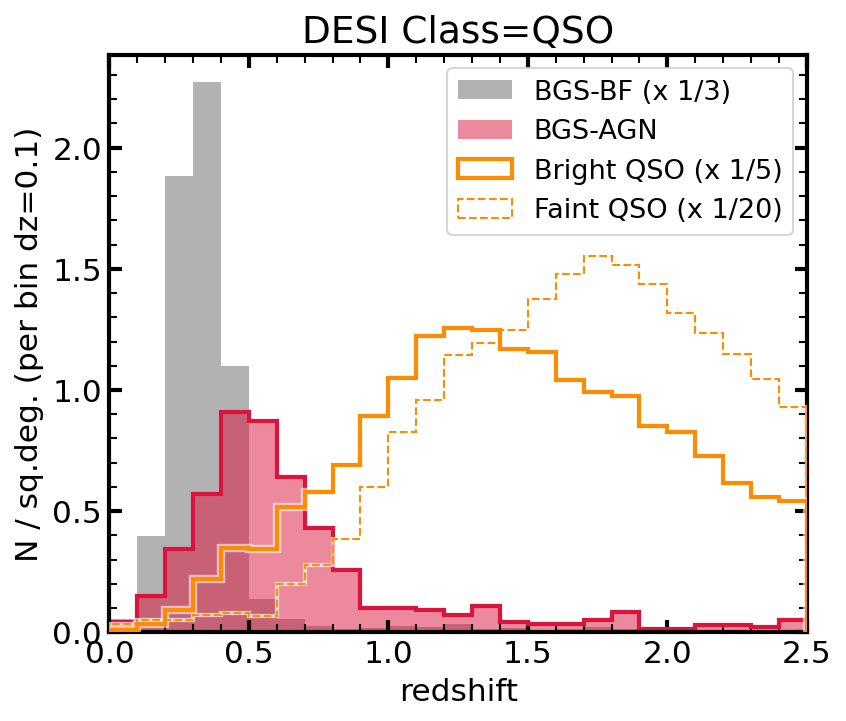}
\caption{Redshift distributions as predicted from the SDSS QSOs  (left-hand panel), and as observed with DESI SV3 (right-hand panel) for the QSO spectral type. In both panels, the population that fulfills the selection criteria for the BGS-BF sample is shown in gray while the BGS-AGN sample is shown in red. On the right-hand side panel, the distribution of QSOs from the BGS-BF sample was scaled down by a factor of three. We attribute the much larger number density to the difference in magnitude limits for SDSS and DESI. This figure truncates the tail of the distribution at $z=2.5$ to focus on the bulk of the sample. Nine additional BGS-AGN QSOs at $2.5<z<3.3$ are included in Figure~\ref{fig:z_w1w2}, which shows the complete sample of 519 unique BGS AGN targets in SV3. To compare with higher redshift QSOs, we add the redshift distribution for objects selected as DESI QSO targets in orange histograms. The solid orange histograms are QSO targets that meet the magnitude limits of the BGS sample. On the right-hand side, we further show QSO targets fainter than BGS with a dashed orange histogram scaled by a factor of 20 due to the high target density.}
\label{fig:zdists}
\end{centering}
\end{figure*}

We compute the redshift distributions per unit area expressed as $N(z)$ in redshift slices of 0.1 as shown in Figure~\ref{fig:zdists}. The left-hand panel shows the expected distributions for the SDSS-LS sample that is spectroscopically classified as QSO and which meets the BGS-BF selection (filled grey), BGS-AGN selection (filled red) in comparison with 
the QSOs that were removed from BGS due the \Grr$<0.6$ cut but could be potential DESI QSO targets (orange histogram). The latter is scaled down by a factor of 5 to more easily compare the shape of each distribution. This panel suggests that QSOs from the BGS-BF sample are mostly at low redshifts ($z<0.5$), the DESI QSO targets are mostly at high redshifts ($z>1$) while the BGS-AGN sample somewhat bridges the gap between the two with a peak around $z\sim0.5$.

The right-hand panel shows a version of the same figure but with the DESI SV3 data for the corresponding samples (same color scheme). While the trends noted above remain applicable, we note that the BGS-BF redshift distribution is even more strongly limited to $z<0.5$ and reaches a much higher number density peaking around 7 per square degree (shown scaled down by a factor 3). 
For the DESI QSO targets, we split them between the subset that would meet the BGS magnitude limits (solid orange histogram) to be directly comparable with the subsample shown for the SDSS-LS sample. Those bright QSO targets show a very similar distribution shape as the SDSS-LS sample, with the solid line orange histograms peaking around $z\sim1.3$ in both panels. As a further comparison, we show DESI QSO targets that are fainter than the BGS limits with a dashed line histogram. As expected, the redshift distribution of the faint QSO targets peak at a higher redshift (around $z\sim1.8$), and reaches a much higher target density (the histogram was scaled down by a factor of 20). 
Overall, modulo some quantitative normalization differences, the SDSS predictions were accurate in the sense that the BGS-AGN quasars peak around $z\sim0.5$ making them intermediary between the lower redshift BGS-BF quasars ($z<0.5$) and the higher redshift DESI QSO targets ($z>0.5$).

\subsection{AGN luminosity vs. redshift}
\label{sec:luminosity}

\begin{figure*}
\begin{centering}
\includegraphics[width=0.9\textwidth]{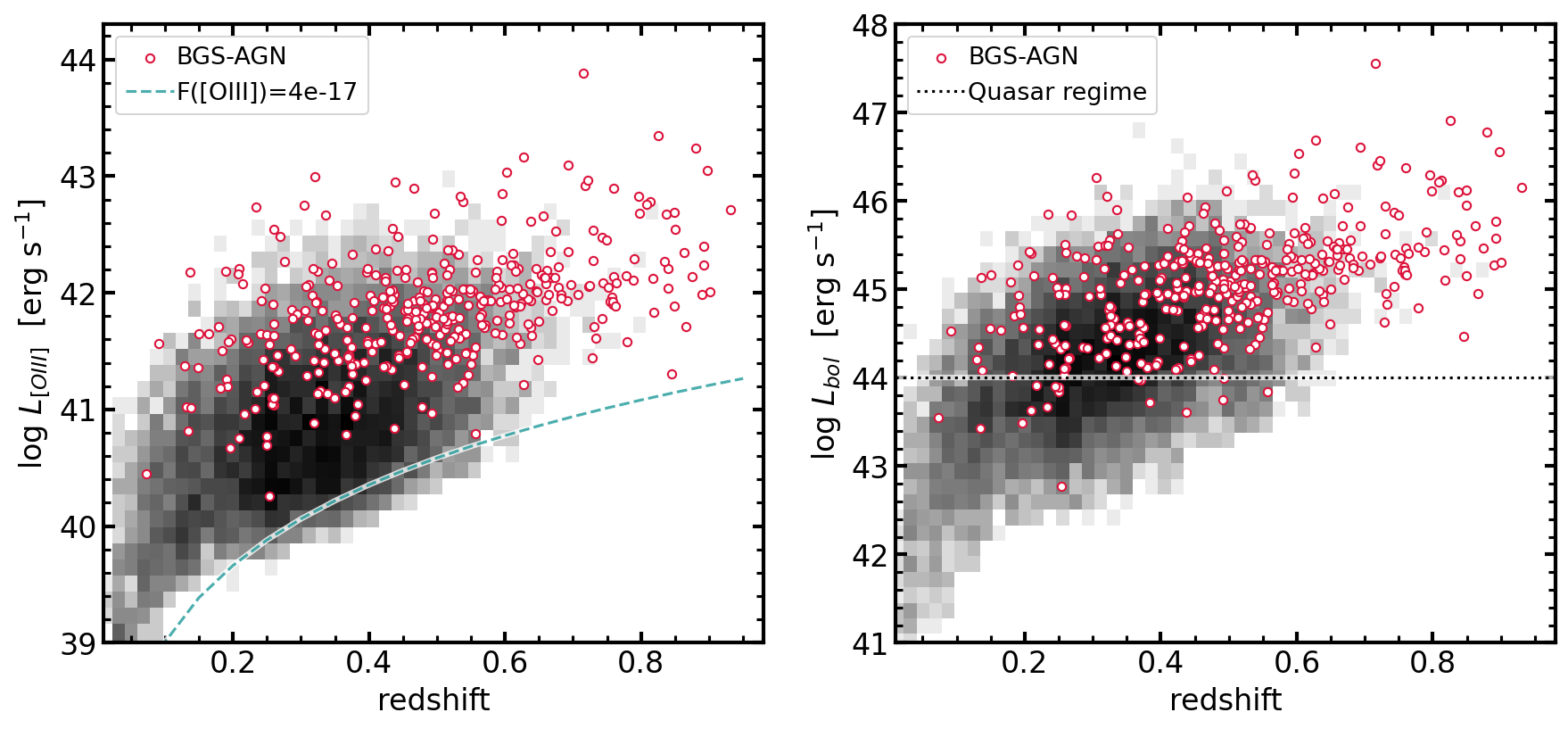}
\caption{(Left) Uncorrected \oiii\ luminosity for the BGS-BF (gray) and BGS-AGN (red symbols) subsamples that have \oiii\ and \hb\ measurements and were classified as AGN from the modified KEx diagnostic diagram (Figure~\ref{fig:kex}). The dashed line corresponds to the \oiii\ luminosity for a fixed line flux of $4\times10^{-17}$~\ergscmsq. (Right) Similar to the left panel but showing the dust-corrected bolometric luminosities of the AGN as a function of redshift. The horizontal dotted line marks a common threshold used to define the quasar regime with $L_{bol}>10^{44}$~\ergs. 
 \label{fig:Lbol}}
\end{centering}
\end{figure*}

In this section, we estimate the AGN bolometric luminosity based on the \oiii\ luminosity, which is appropriate for low-redshift massive galaxies because their \oiii\ flux is dominated by the narrow line region gas when an AGN is present \citep{Kauffmann+2003}. We follow the approach of \citet{Lamastra+2009} and first apply a dust obscuration correction based on the \ha/\hb\ Balmer Decrement assuming an intrinsic dust-free ratio of 3.0 \citep{Osterbrock+2006} and the dust attenuation prescription from \citet{Bassani+1999}. When the Balmer Decrement is not available, we use the typical value of 4.0 found for the BGS sample. In the small number of cases with a measured ratio below 3, we assume that the dust obscuration is negligible and do not attempt to apply a correction.

We restrict this part of the analysis to the subsample of galaxies that pass the AGN criterion from the modified KEx diagram (above the dashed line from Figure~\ref{fig:kex}). This choice is motivated to further increase the likelihood that the \oiii\ luminosity is indeed dominated by the NLR gas rather than by ionized gas from star-forming regions within the 1.5-arcsecond fiber aperture. We show both the uncorrected \oiii\ luminosities (left panel) and the dust-corrected bolometric luminosities (right panel) as a function of redshift in Figure~\ref{fig:Lbol}. There is a general increase in luminosity with redshift due in part to the sensitivity limit for emission line flux measurements (with S/N$>$3), which modulates the lower envelope of the distribution. For reference, the dashed line shows the luminosity corresponding to a constant line flux of $4\times10^{-17}$~\ergscmsq, which highlights that the BGS-AGN sample mostly comprises well detected emission lines. Interestingly, we note that the BGS-AGN lower envelope is higher indicating instead a trend toward higher \oiii\ luminosities compared to BGS-BF galaxies at the same redshifts. However, the BGS-AGN sample has a slightly lower average dust obscuration and this reduces the apparent offset between the BGS-BF and BGS-AGN samples in their AGN bolometric luminosities (right-hand panel). The BGS-AGN bolometric luminosity values seem to be nearly consistent with those for the BGS-BF quasars at a given redshift. But because they are located preferentially at higher redshifts, they also exhibit higher luminosities on average, reaching well into the quasar regime at $L_{\rm bol}>10^{44}~\ergs$ (horizontal dotted line).

\subsection{Infrared $W1-W2$ color trend}
\label{sec:IRcol}

The WISE $W1-W2$ color has been suggested as a criterion to identify AGN by, e.g., \citet{Stern+2012} and \citet{Wu+2012}. The former study selects objects redder than $(W1-W2)_{\rm Vega}\geq0.8$, which corresponds to $(W1-W2)_{AB}\geq0.16$. The latter employs a slightly lower cut of $(W1-W2)_{\rm Vega}\geq0.57$, which corresponds to $(W1-W2)_{AB}\geq-0.07$. In this section, we compare the trend of the WISE $W1-W2$ color as a function of redshift for the BGS AGN sample with that of the SV3 sample divided into galaxies and QSOs based on the Redrock pipeline spectral type. The comparison samples are required to have $S/N>10$ in both WISE bands, good quality Tractor photometry (MASKBITS=0) and reliable spectra (COADD\_FIBERSTATUS=0), yielding 336,067 galaxies at $0.025<z<1.6$ and 27,999 QSOs at $0.025<z<3.5$.

\begin{figure*}
\begin{centering}
\includegraphics[width=0.8\textwidth]{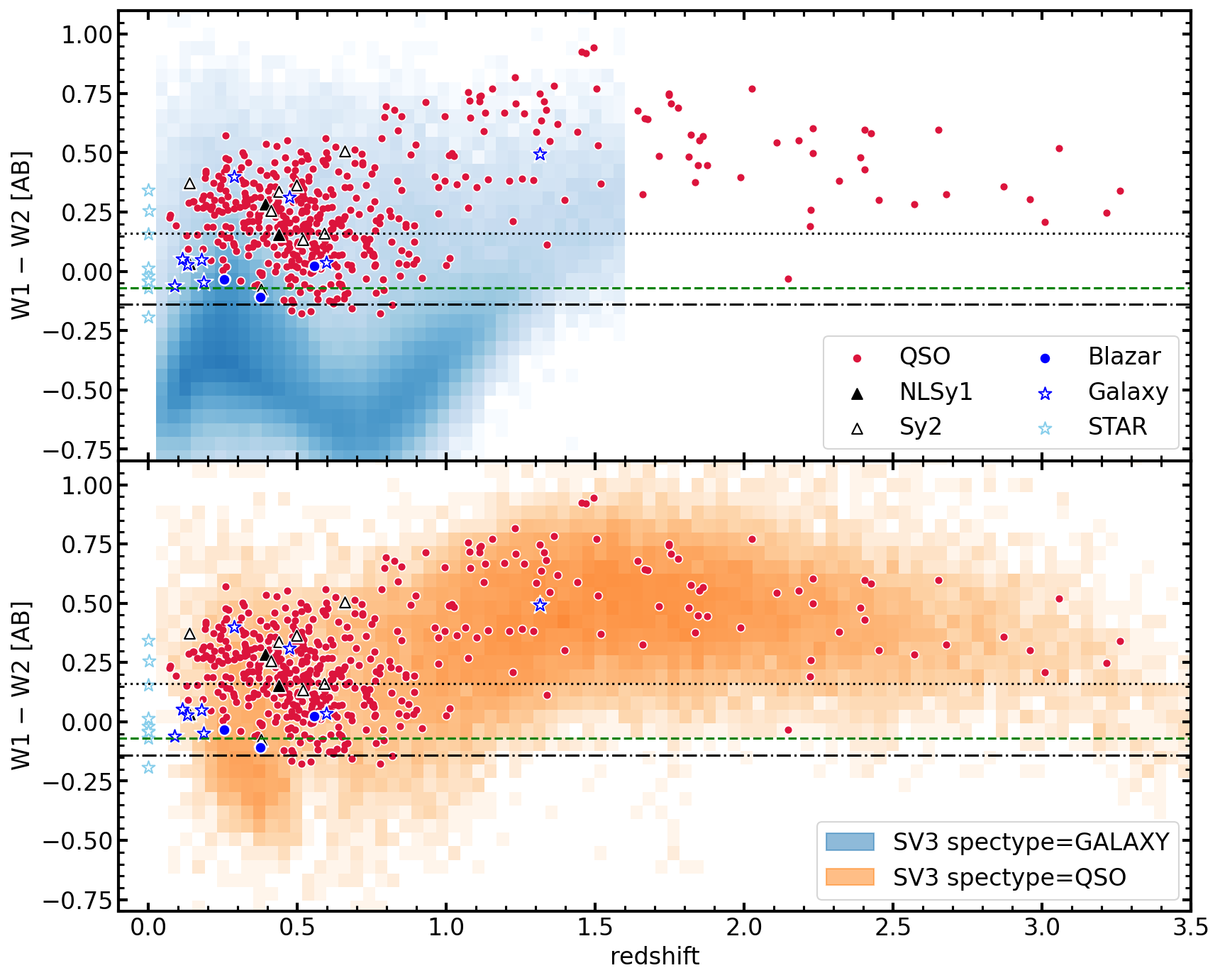}
\caption{Observed $W1-W2$ color as a function of redshift for the BGS AGN sample (colored symbols) compared to SV3 galaxies (top panel) and SV3 QSOs (bottom panel). The BGS AGN symbols are placed at the location of the VI redshift and are coded to indicate the final spectral classification: stars in light blue star symbols, galaxies with no signatures of AGN in dark blue star symbols, blazars in filled blue circles, galaxies with Type~2 AGN signatures in black open triangles, NLSy1 in filled black triangles, and QSOs in red filled circles. The majority of the BGS AGN sample comprises QSOs at $0.1<z<1$ and have WISE colors consistent with QSOs from the overall SV3 sample. Horizontal lines show color cuts from the literature: \citet[][dotted black line]{Stern+2012}, \citet[][green dashed line]{Wu+2012} and \citet[][black dash-dotted line]{Wang+2016}.
 \label{fig:z_w1w2}}
\end{centering}
\end{figure*}

As Figure~\ref{fig:z_w1w2} illustrates, there is a different trend and locus for the bulk of galaxies (top panel; in blue) and QSOs (bottom panel; in orange). The BGS AGN sample (colored and black symbols) tends to follow the rest of the QSOs. However, there are galaxies scattered over the full range of WISE colors (blue shaded bivariate distribution in the top panel) overlapping with the bulk of the QSOs. This overlap implies that a small fraction of inactive galaxies exhibit AGN-like infrared colors, which could naturally explain the small galaxy contamination rate found in the BGS-AGN sample ($\sim2$\% after correcting for narrow-line AGN and blazars as described in Section~\ref{sec:bpt} and Appendix~\ref{app:finalclass}).

An alternative explanation could be that some AGN host galaxies possess IR signatures of AGN heated dust (red $W1-W2$ colors) but lack optical QSO signatures due to, e.g., dust obscuration. This is consistent with at least the Type~2 AGN (black triangles), which are characterized by spectra with only narrow lines and lacking a blue power-law continuum, overlapping with the QSO in terms of their infrared WISE colors but usually classified as a galaxy spectral type from Redrock (although five of the eight were classified as QSO by QN or MgII; Appendix~\ref{app:sy2}). Yet another explanation could be that the WISE photometry of some objects is affected by nearby blended sources. Among the small number of spectra visually identified as galaxies with no obvious AGN contribution (dark blue star symbols), we note that 5/9 are located on the edge of the distribution toward low redshift and blue WISE colors, which leaves only a small number of four galaxies (at $z>0.25$) with AGN-like infrared colors. We show the spectra and color images of all galaxy type in Appendix~\ref{app:contamination} and two out of the four galaxies at $z>0.25$ appear to have an additional red component on their images (TARGETID=39632961917682386 and 39632951826190082 in Figure~\ref{fig:gal_rest}).

By selection, BGS AGN targets are redder than $W1-W2 > -0.2$ (Section~\ref{sec:selection}). For reference, we draw the \citet{Stern+2012} criterion with a dotted black line. At redshift $z<1$, we find that nearly half of BGS AGN with visually identified QSO or Type~2 AGN signatures have colors bluer than the Stern criterion. In contrast, at $z>1$ the bulk of QSOs and of BGS AGN targets have redder colors. We also compare with the $W1-W2$ criteria from \citet{Wu+2012} (green dashed line) and from \citet{Wang+2016} (black dot-dashed line; $W1-W2 > -0.14$~AB). The latter was used together with other cuts to search for high-redshift ($z>5$) QSOs \citep[e.g.,][]{Yang+2016,Yang+2023}. They are similar to the threshold of $W1-W2 > -0.2$ used in this work. In all cases, we can see that a single $W1-W2$ criterion would not be sufficient to disentangle QSOs from galaxies given their overlap at $z<0.5$ and possibly also at $z>1$ but at such redshifts, DESI primarily selects ELGs which tend to have faint continuum and therefore usually not strongly detected in the WISE bands \citep{Raichoor+2023}. 

A commonly used strategy is to augment the WISE photometry with the longer wavelength channel $W3$ (12~$\mu$m) and use the $W1-W2$ color together with the $W2-W3$ color \citep[e.g.,][]{Jarrett+2011,Mateos+2012,Hviding+2022}. However, most DESI targets are not detected in the comparatively much less sensitive $W3$ map so it was not used for target selection. Nonetheless, future work could investigate the infrared properties for the subset of objects that do have a $W3$ detection, which we could expect to be relevant at the low and moderate redshifts of the BGS-AGN targets.

\subsection{Quasars at $z>2$}
\label{sec:lya}

High redshift quasars can be used as probes of the intervening structures along the line of sight to these bright background sources. 
In particular, the DESI QSO target class includes QSOs at $z > 2.1$ that will be observed at higher signal-to-noise in order to study the \lya\ Forest \citep{Schlafly+2023,Ramirez-Perez+2023}. 
While the redshift distribution of the BGS-AGN sample peaks at $z\sim0.5$, there is a small tail reaching $z>2$, which we examine in this section. The 23 quasars can be divided into two strongly distinct categories. On the one hand, 11 of them are bright ($r<17.5$) and were assigned a PSF morphological type. The spectra of those targets (Figure~\ref{fig:lya_qso}) are all blue with prominent broad emission lines, and further display \lya\ Forest absorption lines as well as metal absorbers such as \mgii\ absorbers \citep[e.g.,][]{Napolitano+2023}. Two of them have obvious broad absorption lines \citep[BALs;][]{Filbert+2023}. 

\begin{figure*}
\begin{centering}
\includegraphics[width=0.88\textwidth]{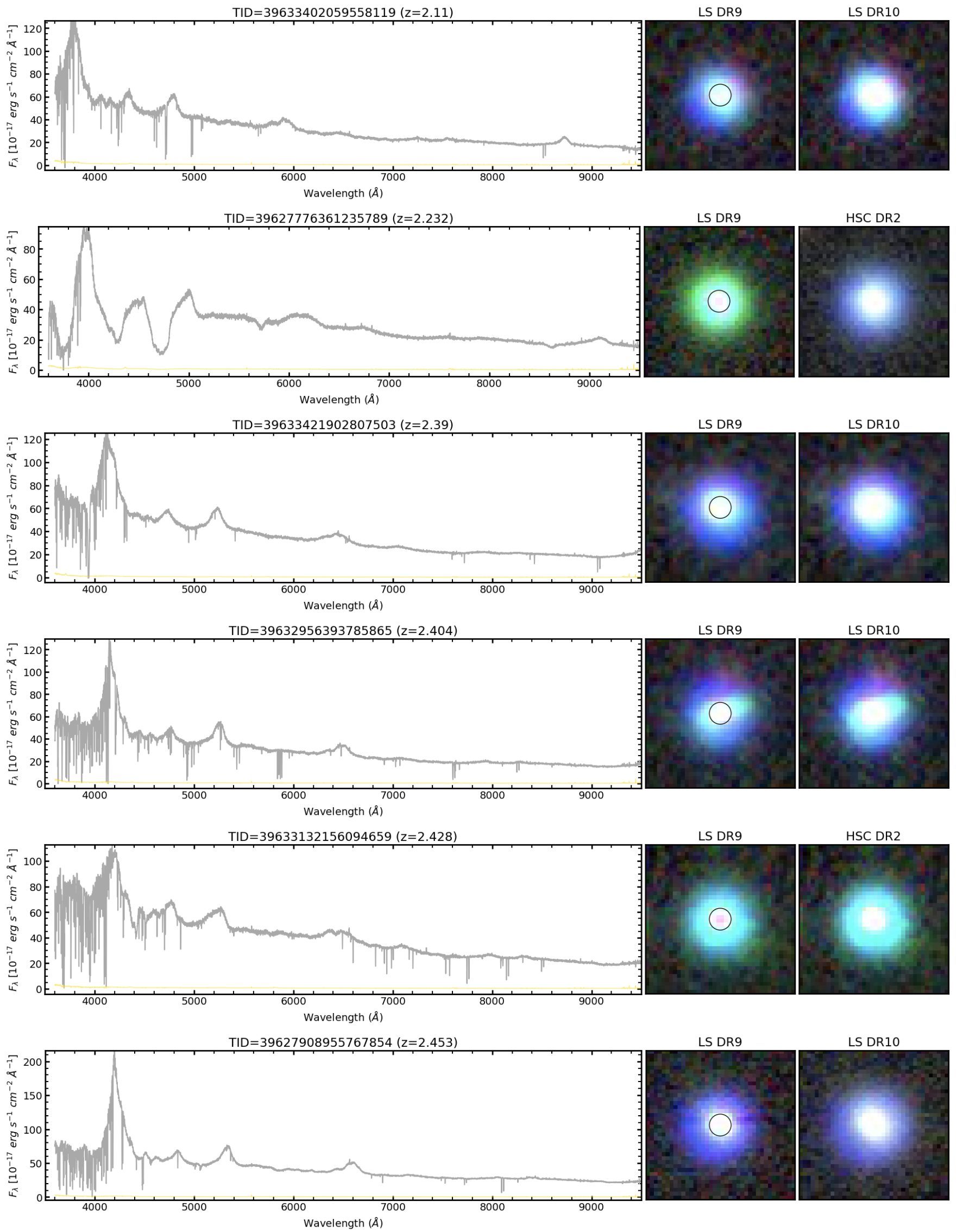}
\caption{Observed frame DESI spectra of $z>2$ QSOs with a Tractor PSF morphology (left-hand panels). The observed spectrum is in dark gray; the error spectrum is shown in yellow, and is comparatively negligible in the case of these bright QSOs. The high average signal-to-noise ratio allows us to display the original spectra without any smoothing thus maintaining narrow absorption features. Color images come from LS DR9 (middle panels) and either HSC DR2 if available or otherwise LS DR10 (right-hand panels). The 1.5~arcsec DESI fiber aperture is marked with a circle on the LS DR9 images. QSOs are displayed in order of increasing redshifts from top to bottom.
 \label{fig:lya_qso}}
\end{centering}
\end{figure*}

\addtocounter{figure}{-1}
\begin{figure*}
\begin{centering}
\includegraphics[width=0.9\textwidth]{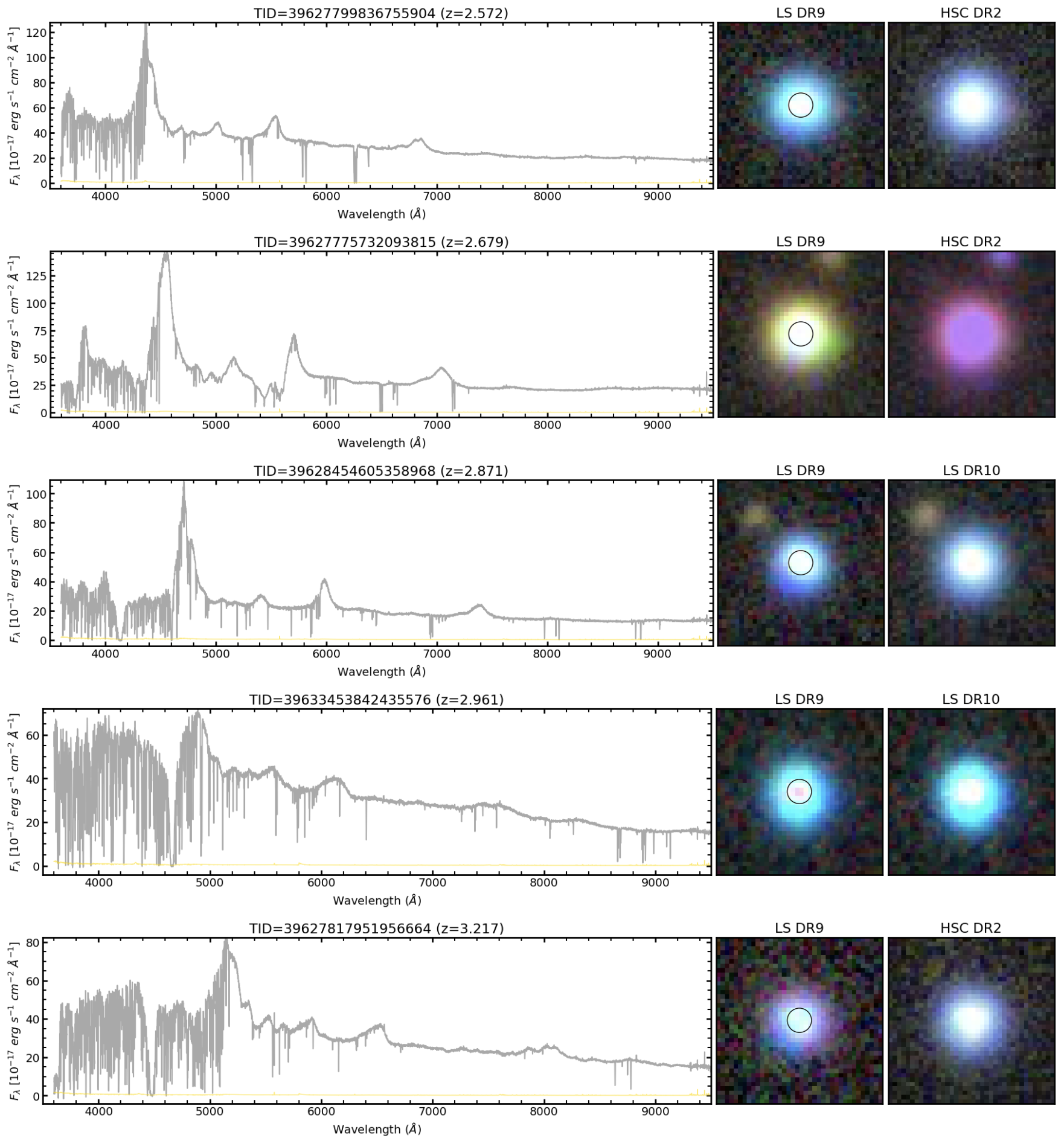}
\caption{Continued.}
\end{centering}
\end{figure*}

On the other hand, the remaining 12 quasars at $z>2$ are faint and were not assigned a PSF morphological type. Their spectra and images are displayed in Figure~\ref{fig:lya_qso_nopsf} in order of increasing redshift. Strikingly, nearly all the images indicate close angular proximity or even a possible overlap with other sources or components.  We attribute the departure from PSF morphology to this apparent presence of companions at small angular separation (most likely in projection). Nine of the 12 are best-fit with a Sersic profile while two are fit with a DeVaucouleurs profile and one with a round exponential profile.\footnote{The LS DR9 morphological types are defined here: {\url https://www.legacysurvey.org/dr9/description/\#morphological-classification}}

Despite this trend, the spectra are all confidently classified as QSO and display the usual broad emission lines (e.g., \lya, \ciii, \civ, \mgii). Some of them further show a \lya\ forest, metal absorbers and/or BALs. We conclude that they are genuine quasars and that the potential chance alignment with other sources mean that they risk being excluded from traditional QSO target selection algorithms.

\begin{figure*}
\begin{centering}
\includegraphics[width=0.9\textwidth]{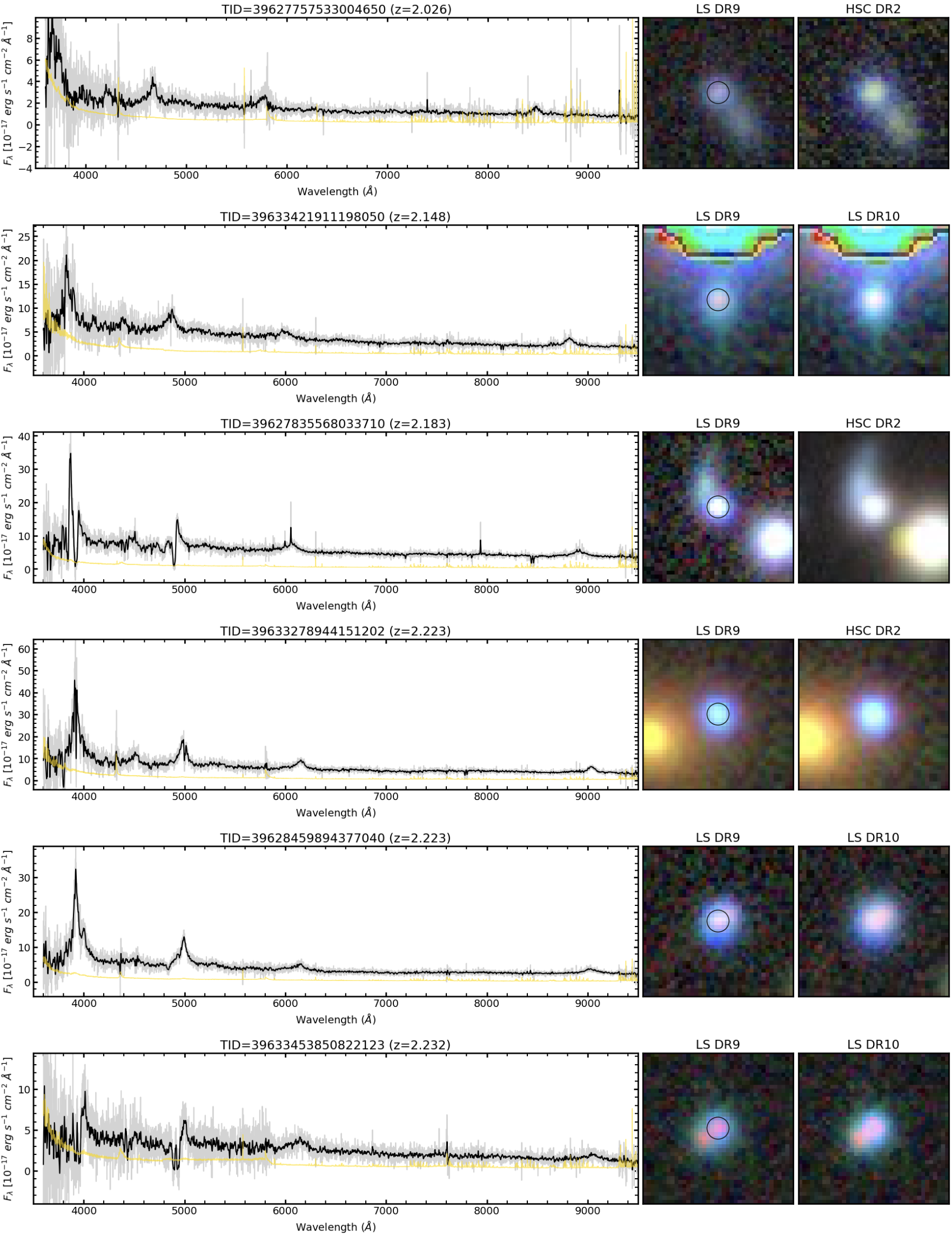}
\caption{Observed frame DESI spectra of $z>2$ QSOs with a Tractor non-PSF morphology (left-hand panels). The observed spectrum is in gray, a Gaussian-smoothed spectrum in black and the error spectrum is in yellow. Color images come from LS DR9 (middle panels) and either HSC DR2 if available or otherwise LS DR10 (right-hand panels). The 1.5~arcsec DESI fiber aperture is marked with a circle on the LS DR9 images. QSOs are displayed in order of increasing redshifts from top to bottom.
 \label{fig:lya_qso_nopsf}}
\end{centering}
\end{figure*}

\addtocounter{figure}{-1}
\begin{figure*}
\begin{centering}
\includegraphics[width=0.95\textwidth]{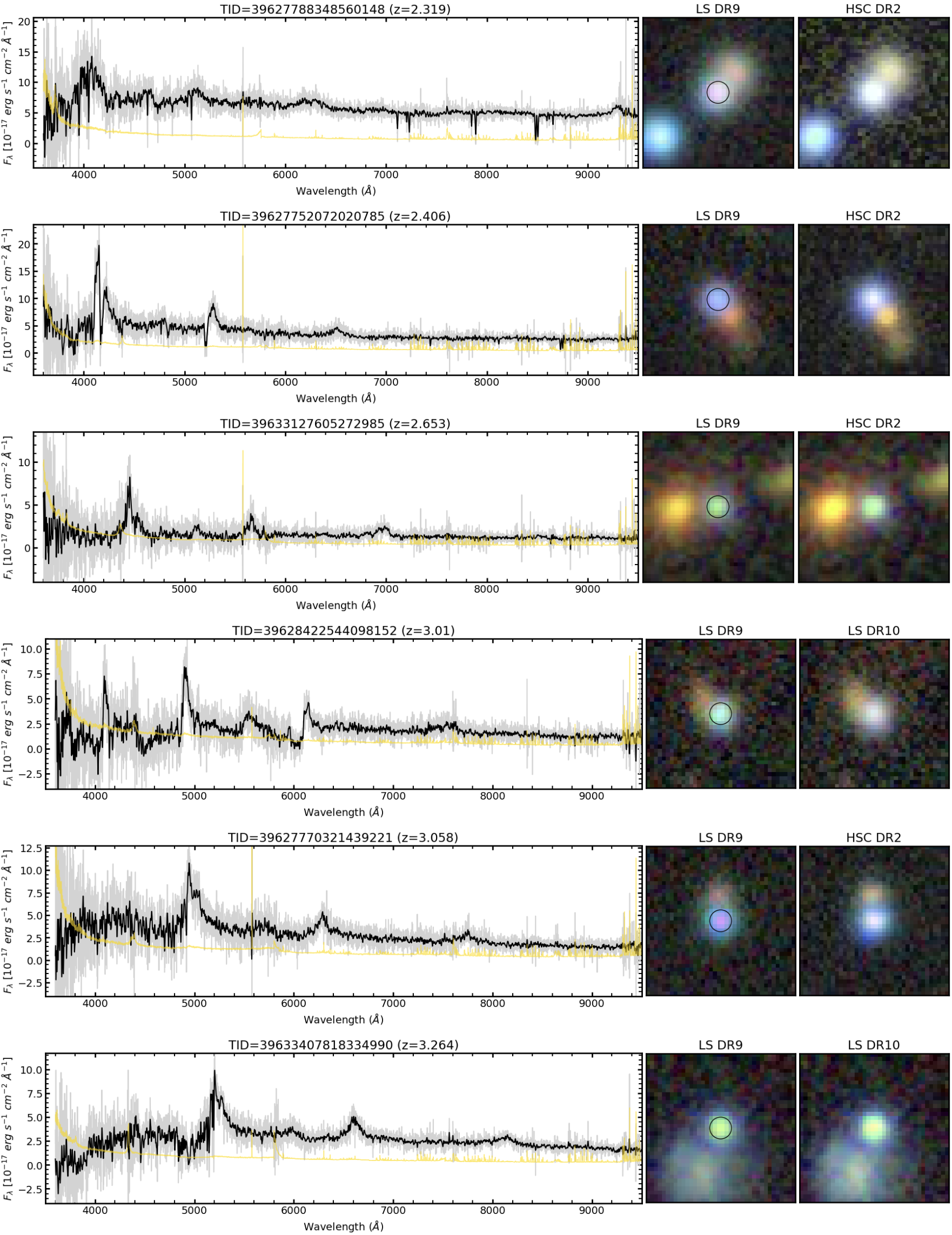}
\caption{Continued.}
\end{centering}
\end{figure*}

\section{Conclusions}
\label{sec:conclusions}

This manuscript presents the motivation and selection criteria for the BGS AGN target class of the DESI survey. This target class was introduced to supplement the BGS bright and faint surveys, which were missing AGN hosts that appear point-like according to the Gaia and Tractor photometry comparison. The main conclusions are as follows:

\begin{itemize}
    \item To select the BGS AGN while keeping the stellar contamination low, we employ optical and infrared color cuts, namely infrared colors typically expected from AGN-heated dust. The DESI SV3 sample comprises 782 spectra of 519 unique BGS AGN target;

    \item The resulting target number density of $3-4$ per square degree is modest but uniform across the full DESI footprint, as expected;

    \item Visual inspection of the BGS AGN spectra revealed a high success rate in terms of quasar classification (94\%) with small percentages of narrow-line AGN or Blazars (3\%). galaxy contaminants (2\%) and stellar contaminants (2\%);

    \item Using the default Redrock pipeline along only identified 76\% of the targets as QSO, whereas adding the QuasarNet and MgII afterburner classifiers raises this fraction to 94\%. In all cases, the stellar contamination remains around 2\%, indicating that the missed quasars were instead assigned to a galaxy spectral type;

    \item Stacking the spectra of BGS AGN results in a typical QSO spectrum. Additionally, splitting the subset identified (missed) by Redrock result in stacked spectra that are typical of (redder than) regular blue quasars;

    \item Applying an emission-line diagnostic diagram modified from the KEx diagram shows that BGS AGN targets lacking broad lines (Type 1 AGN) were typically found to exhibit emission line properties consistent with narrow-line (Type 2) AGN, confirming that the majority host an active black hole. Furthermore, this new version of the KEx diagram (Section~\ref{sec:bpt}) could become a useful tool for other AGN and galaxy studies;

    \item We computed the redshift distribution of the BGS AGN targets classified as QSOs and found it to be peak around $z\sim0.5$, which is intermediate between BGS QSOs from the bright and faint survey ($z<0.5$), and that of the QSO targets $z>0.5$. The redshift ranges are similar to our expectations from the SDSS test sample (Fig.~\ref{fig:zdists});

    \item There is an interesting high-redshift tail of $z>2$ quasars (\lya\ QSOs), which are included either because they are among the brightest (a PSF morphology is allowed only at $r<17.5$ for BGS-AGN targets) or because the had a non-PSF morphology due to close projected neighbors (e.g., foreground star, galaxy, or possible background lensed galaxies). We feature other rare AGN subcategories and conflicting cases in Appendix~\ref{app:finalclass}.

\end{itemize}

Looking forward, the first year data of the main DESI survey (Year 1 observations obtained until June 2022) already yield 18,302 good quality spectra of 15,523 unique BGS-AGN targets. This corresponds to a sample size increase by a factor 30 relative to the 519 unique BGS-AGN targets from DESI SV3 presented in this work. They have a very similar redshift distribution as shown in Figure~\ref{fig:zdists} with a marked peak at $z\sim0.5$ and a tail reaching $z>2-2.5$ so we expect much better statistics of the same types of interesting objects that bridge the gap between the lowest redshift quasars and the bulk of the DESI (or even SDSS) QSO targets. The DESI Year 1 data will be published as part of Data Release 1 similarly to the currently public DESI EDR.\footnote{\url{https://data.desi.lbl.gov/doc/releases/}}

\section*{Data Availability}
 
The Data Release 9 of the DESI Legacy Imaging Surveys is available at \url{https://www.legacysurvey.org/dr9/}. 

Documentation of DESI data access is maintained at
\url{https://data.desi.lbl.gov/doc/access/}. DESI EDR data can also be 
accessed via searchable catalog databases at the Astro Data
Lab \citep{Fitzpatrick+2014,Nikutta+2020} at \url{https://datalab.noirlab.edu} and via a spectral database SPARCL \citep{Juneau+2024} at \url{https://astrosparcl.datalab.noirlab.edu/}, both of which were used in this work.

Besides the images and spectra available above, the data shown in the
figures are available at
\url{https://doi.org/10.xxxxx/zenodo.xxxxxxxx} (URL will be updated upon manuscript acceptance).

\section*{Acknowledgements}

SJ's research is supported by the U.S. NSF NOIRLab, which is operated by the Association of Universities for Research in Astronomy (AURA) under a cooperative agreement with the National Science Foundation. DMA thanks the Science Technology Facilities Council (STFC) for support from the Durham consolidated grant (ST/T000244/1). RP acknowledges support from the University of Arizona and Astro Data Lab (NSF NOIRLab, AURA). VAF acknowledges funding from an United Kingdom Research and Innovation grant (code: MR/V022830/1). ADM was supported by the U.S.\ Department of Energy, Office of Science, Office of High Energy Physics, under Award Number DE-SC0019022. JM gratefully acknowledges funding support for this work from the U.S. Department of Energy, Office of Science, Office of High Energy Physics under Award Number DE-SC0020086. SMC acknowledges the support of STFC grant ST/X001075/1.

This material is based upon work supported by the U.S. Department of Energy (DOE), Office of Science, Office of High-Energy Physics, under Contract No. DE–AC02–05CH11231, and by the National Energy Research Scientific Computing Center, a DOE Office of Science User Facility under the same contract. Additional support for DESI was provided by the U.S. National Science Foundation (NSF), Division of Astronomical Sciences under Contract No. AST-0950945 to the NSF’s National Optical-Infrared Astronomy Research Laboratory; the Science and Technology Facilities Council of the United Kingdom; the Gordon and Betty Moore Foundation; the Heising-Simons Foundation; the French Alternative Energies and Atomic Energy Commission (CEA); the National Council of Science and Technology of Mexico (CONACYT); the Ministry of Science and Innovation of Spain (MICINN), and by the DESI Member Institutions: \url{https://www.desi.lbl.gov/collaborating-institutions}. Any opinions, findings, and conclusions or recommendations expressed in this material are those of the author(s) and do not necessarily reflect the views of the U. S. National Science Foundation, the U. S. Department of Energy, or any of the listed funding agencies.

The authors are honored to be permitted to conduct scientific research on Iolkam Du’ag (Kitt Peak), a mountain with particular significance to the Tohono O’odham Nation.

The DESI Legacy Imaging Surveys consist of three individual and complementary projects: the Dark Energy Camera Legacy Survey (DECaLS), the Beijing-Arizona Sky Survey (BASS), and the Mayall z-band Legacy Survey (MzLS). DECaLS, BASS and MzLS together include data obtained, respectively, at the Blanco telescope, Cerro Tololo Inter-American Observatory, NSF’s NOIRLab; the Bok telescope, Steward Observatory, University of Arizona; and the Mayall telescope, Kitt Peak National Observatory, NOIRLab. NOIRLab is operated by the Association of Universities for Research in Astronomy (AURA) under a cooperative agreement with the National Science Foundation. Pipeline processing and analyses of the data were supported by NOIRLab and the Lawrence Berkeley National Laboratory (LBNL). Legacy Surveys also uses data products from the Near-Earth Object Wide-field Infrared Survey Explorer (NEOWISE), a project of the Jet Propulsion Laboratory/California Institute of Technology, funded by the National Aeronautics and Space Administration. Legacy Surveys was supported by: the Director, Office of Science, Office of High Energy Physics of the U.S. Department of Energy; the National Energy Research Scientific Computing Center, a DOE Office of Science User Facility; the U.S. National Science Foundation, Division of Astronomical Sciences; the National Astronomical Observatories of China, the Chinese Academy of Sciences and the Chinese National Natural Science Foundation. LBNL is managed by the Regents of the University of California under contract to the U.S. Department of Energy. The complete acknowledgments can be found at \url{https://www.legacysurvey.org/acknowledgment/}

This research uses services and data provided by the SPectra Analysis and Retrievable Catalog Lab (SPARCL) and the Astro Data Lab, which are both part of the Community Science and Data Center (CSDC) program at NSF NOIRLab.


%

\vspace{5mm}
\facilities{Mayall (DESI), Mayall (Mosaic-3), Blanco (DECam), Bok (90Prime), Astro Data Lab, 
Sloan, WISE, NEOWISE, Gaia}



\software{astropy \citep{astropy1, astropy2, astropy3}, 
          desispec \citep{Guy+2023}, desitarget \citep{Myers+2023}
          \texttt{FastSpecFit} \citep{Moustakas2023},
          Redrock \citep{Bailey+2023}, SPARCL \citep{Juneau+2024}}




\appendix


\section{Refined classification from emission lines}
\label{app:finalclass}

As described in Section~\ref{sec:class}, the spectra were classified based on different methods including a campaign of visual inspection (Section~\ref{sec:vi}) and applying emission line diagnostic diagrams (Section~\ref{sec:bpt}). In this Appendix, we present additional details that led to the adopted spectral classification and display individual spectra and images to illustrate each category.
In particular, we examine the spectra 
of all cases that were subject to a conflicting classification between various methods, and we highlight rare AGN subpopulations such as red quasars (Section~\ref{app:redqso}), NLSy1 (Section~\ref{app:nls1}) and blazars (Section~\ref{app:blazar}). 

\subsection{QSO missed by the modified QSO classifier pipeline}

\begin{figure*}
\begin{centering}
\includegraphics[width=0.98\textwidth]{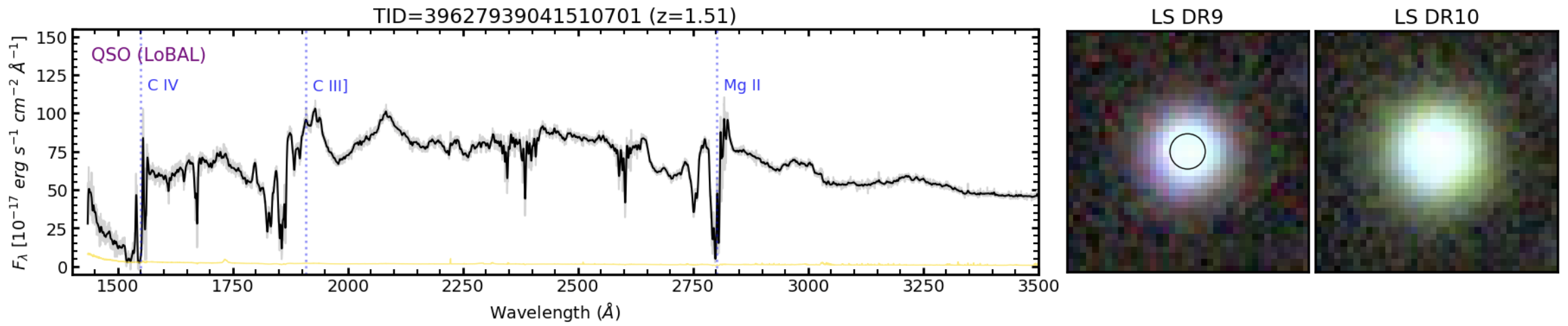}
\caption{Rest-frame DESI spectrum of a LoBAL quasar.
The unsmoothed spectrum is in gray with a smoothed spectrum overplotted in black. The error spectrum is shown in yellow. Color images come from LS DR9 (middle panels) and either HSC DR2 or LS DR10 (right-hand panels). The 1.5~arcsec DESI fiber aperture is marked with a circle on the LS DR9 images.
 \label{fig:balQsoMakerMissed}}
\end{centering}
\end{figure*}

In Section~\ref{sec:afterburn}, we describe the classification results from the modified QSO pipeline that combines Redrock with the afterburner QN and MgII classifiers. There were five conflicting cases not classified as QSO by the pipeline but classified as such visually. One striking case is a $z\sim1.51$ QSO with prominent low-ionization BALs in the spectrum as shown in Figure~\ref{fig:balQsoMakerMissed}. The original redshift as determined by Redrock was inaccurate likely due to the BAL features. Even though it is challenging to determine a precise redshift for such a spectrum, the visual inspection classification is secure in terms of assigning a QSO spectral type. 

\begin{figure*}
\begin{centering}
\includegraphics[width=0.98\textwidth]{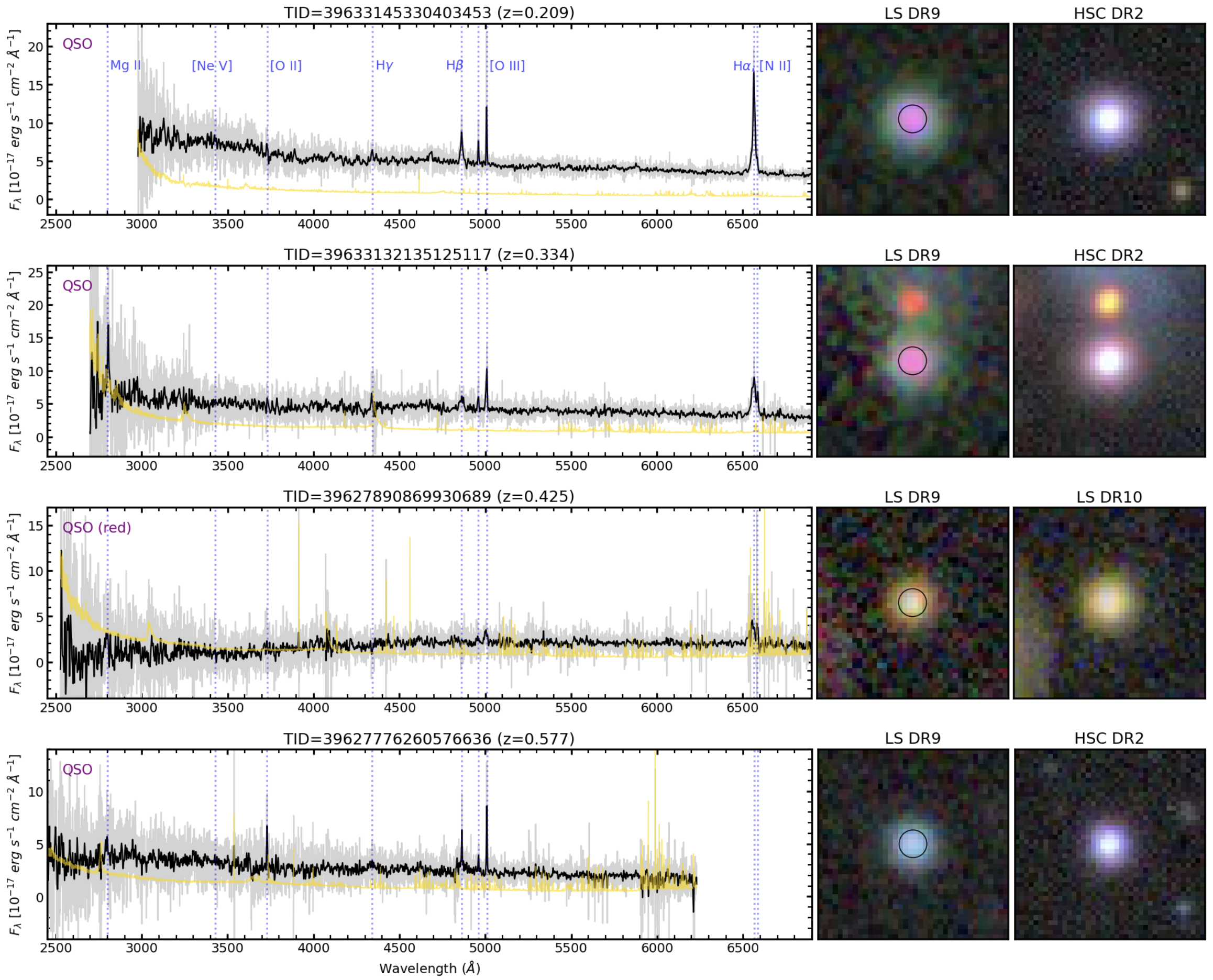}
\caption{Rest-frame DESI spectra of visually identified QSOs that were missed by the afterburner pipeline. The top two examples may be NLSy1 given that their broad lines appear fairly narrow. The spectrum in the third row is reddened in addition to having relatively faint (low signal-to-noise) features. The spectrum in the bottom row has a blue continuum with faint broad \mgii\ and \hb\ lines which are apparent when interactively examining the spectrum but have a low significance in the unsmoothed spectrum. In all rows, the unsmoothed spectrum is in gray with a smoothed spectrum overplotted in black. The error spectrum is shown in yellow. Color images come from LS DR9 (middle panels) and either HSC DR2 or LS DR10 (right-hand panels). The 1.5~arcsec DESI fiber aperture is marked with a circle on the LS DR9 images.
 \label{fig:QsoMakerMissed}}
\end{centering}
\end{figure*}

As shown in Figure~\ref{fig:QsoMakerMissed} the other four cases are a combination of faint spectra with low signal-to-noise, comparatively weak and/or narrow broad lines. The object in the top row (TARGETID=39633145330403453) has a maximum line width of \sigmax=362~\kms, which is in the NLSy1 regime ($\sigma=170-850$~\kms) but it has \oiii/\hb=5.3, which is higher than the formal definition for NLSy1 (\oiii/\hb$<3$). The second and third row show cases with weak broad line and we note that TARGETID=3962789086993068 may be a red QSO (shown in the third row;) also see Section~\ref{app:redqso}). 

Lastly, we note that TARGETID=39627776260576636 (bottom row) has tentative broad \mgii\ and \hb\ lines in its spectrum with otherwise a low signal to noise ratio. This target has a higher quality spectrum with more exposure time in DESI DR1 (in preparation), which shows the lines more clearly and that spectrum is assigned a QSO spectral type by the default Redrock pipeline. Therefore, this gives us confidence in the visual classification. Overall, we keep the QSO visual classification for all five targets shown in this Section but we understand that these cases are challenging to identify with an automated pipeline due to BALs or faint signal. Still, these missed QSOs are only at the 1\% level for the BGS-AGN sample (5/519) so we also conclude that the modified pipeline performs adequately.

\subsection{Broad line identification}
\label{app:linewidth}

A clear signature of typical QSOs is the presence of broad permitted lines in their spectra and those were the main features identified visually. In the rest-frame optical, the main broad features are the Balmer series of Hydrogen. In the rest-frame UV, the main features are \mgii, \ciii, \civ, and \lya. In this section, we discuss the quantitative identification of broad lines from automated spectral fitting, and our motivation for combining both UV and optical lines when adopting the maximum line width \sigmax\ from the \mgii, \hb, and \ha\ lines in Section~\ref{sec:bpt}. 

\begin{figure}
    \centering
    \includegraphics[width=0.47\textwidth]{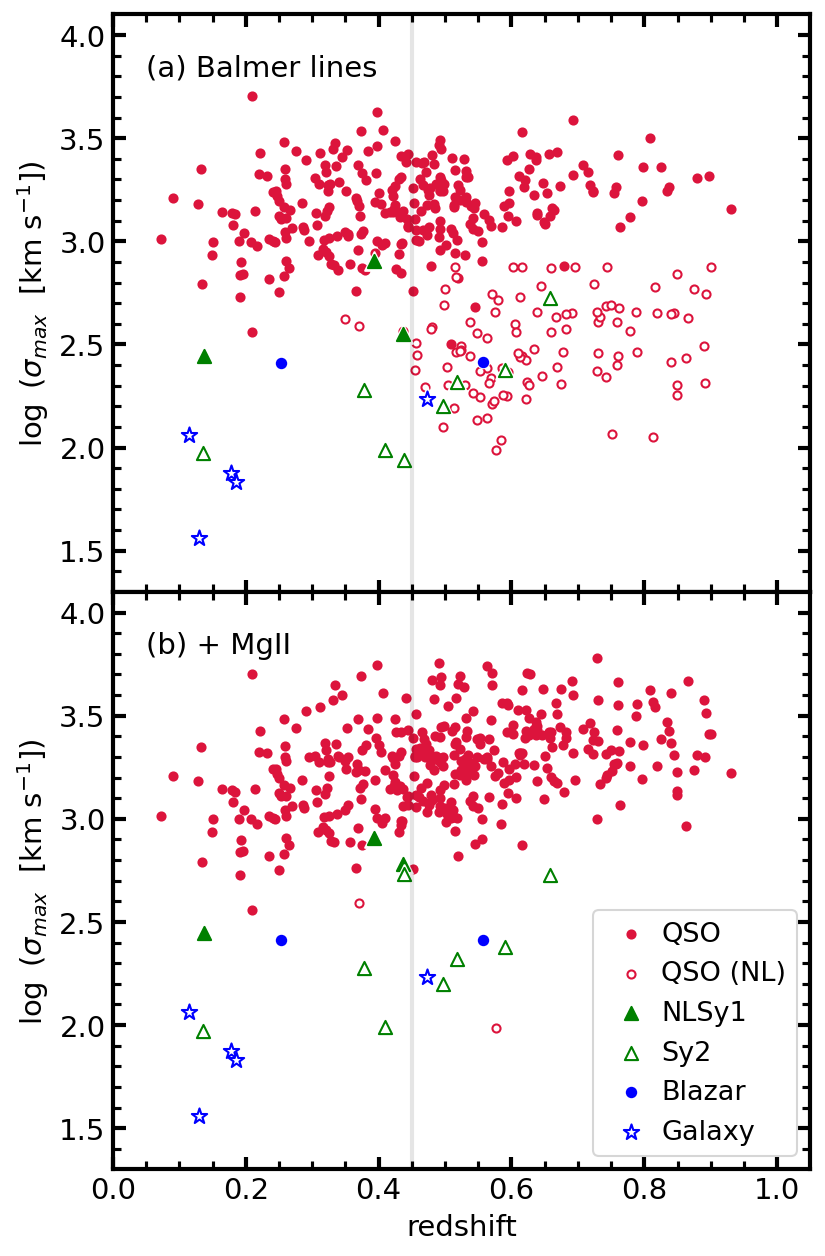}
    \caption{Maximum line width as a function of redshift. The top panel is the result when considering only the Balmer lines \ha\ and \hb, while the bottom panel shows the result after adding the MgII doublet. In both panels, the points are color-coded with the original VI classification as labeled. Furthermore, the plotting symbols are either filled or open depending on whether a broad line was detected (Type 1) or not (Type 2), respectively.}
    \label{fig:sig_z}
\end{figure}

In the \texttt{FastSpecFit} emission line fitting algorithm, the \hb\ and \ha\ Balmer line profiles are tied together in terms of their width and position. However, the presence of a broad component is comparatively more challenging to detect for \hb\ due to being $>3$ times fainter line than \ha. Consequently, when relying on the Balmer lines alone, there is a marked decrease of cases with detected broad lines starting at $z>0.45$ where \ha\ is redshifted outside of the spectral range (Figure~\ref{fig:sig_z}a). At $z<0.45$ where both \ha\ and \hb\ are available, most visually identified QSOs are also assigned a broad line (solid red circles). However, at $z>0.45$, there are many visually classified QSOs that lack a significant broad Balmer component (open red circles). 

Luckily, the \mgii\ doublet becomes available within the DESI spectral range at $z>0.3$ and is characterized by an easily detectable broad profile in Type~1 AGN. Therefore, combining the line width information from \ha, \hb, and \mgii\ allow us to classify most of the BGS-AGN samples with emission lines as broad-line QSOs (solid red circles in Figure~\ref{fig:sig_z}b). 

\begin{figure*}
\begin{centering}
\includegraphics[width=0.98\textwidth]{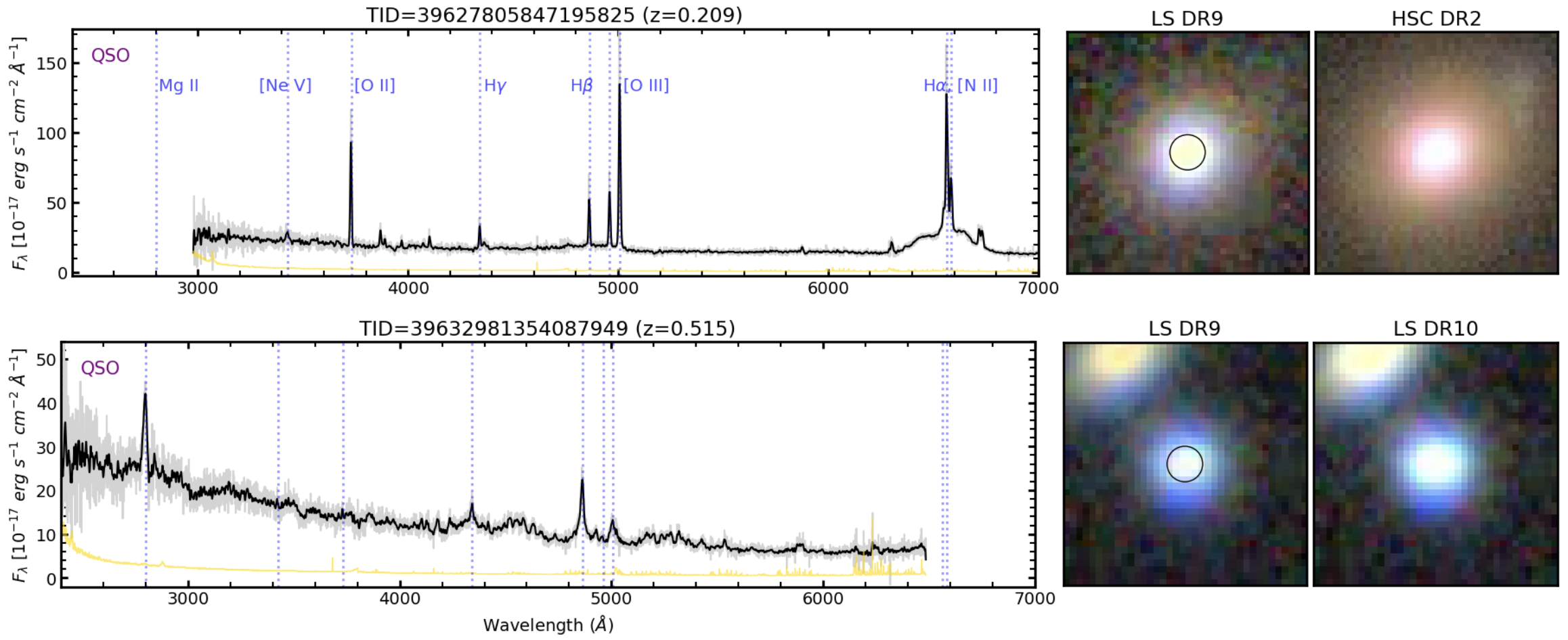}
\caption{Rest-frame DESI spectra for which we obtain a new fit with \texttt{FastSpecFit} for the following reasons: (top) very broad \ha\ line which required different priors. Noticeably, the \hb\ broad line is much weaker and harder to visually disentangle from the continuum; (bottom) originally had an erroneous redshift but here we can see obvious \mgii\ emission as well as a broad \hb.  
The unsmoothed spectrum is in gray with a smoothed spectrum overplotted in black. The error spectrum is shown in yellow. Color images come from LS DR9 (middle panels) and either HSC DR2 or LS DR10 (right-hand panels). The 1.5~arcsec DESI fiber aperture is marked with a circle on the LS DR9 images.
 \label{fig:badfit}}
\end{centering}
\end{figure*}

\begin{figure*}
\begin{centering}
\includegraphics[width=0.98\textwidth]{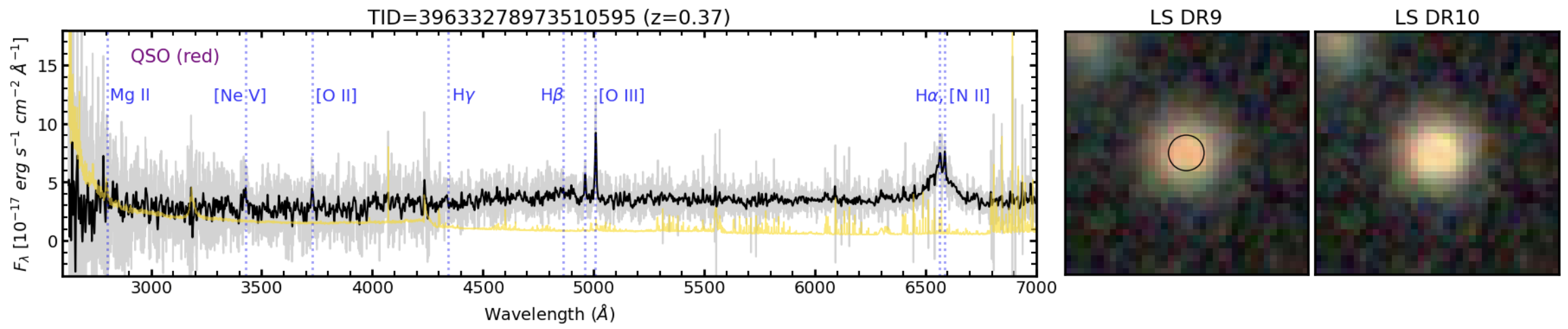}
\caption{Rest-frame DESI spectrum for a low signal-to-noise case which fails our S/N$>$3 criterion due to \texttt{FastSpecFit} finding S/N=2.8 for the broad \ha\ component. 
The unsmoothed spectrum is in gray with a smoothed spectrum overplotted in black. The error spectrum is shown in yellow. Color images come from LS DR9 (middle panel) and LS DR10 (right-hand panel). The 1.5~arcsec DESI fiber aperture is marked with a circle on the LS DR9 image.
 \label{fig:missBL}}
\end{centering}
\end{figure*}

Among spectra visually classified as QSO and with \hb\ and \oiii\ lines, there were originally 14 for which the emission line measurements from \texttt{FastSpecFit} failed to detect a broad line with S/N$>$3 in any of \hb, \ha, \mgii. Inspecting the emission-line fitting results of those 14 spectra revealed the following:
\begin{itemize}
\item One case failed due to the extreme broad line being outside the normal prior range (top panel of Figure~\ref{fig:badfit}). The solution was to refit the spectrum with a new prior, which converged with \sigmax=5051.5~\kms\ (and $\sigma_{\oiii}$=188.5~\kms). In detail, the broad line profile deviates from a single Gaussian but a more complex characterization is beyond the scope of this work.
\item Two cases had an inaccurate redshift from the Redrock pipeline. The solution was to refit them with the corrected redshifts, and we show the case within our redshift range of interest in the bottom panel of Figure~\ref{fig:badfit}. 
\item Nine cases have broad lines with low signal-to-noise and thus are sensitive to small variations of the smooth continuum component of the \texttt{FastSpecFit} model. The solution was to refit these nine without a smooth continuum correction, which successfully recovered at least one of broad lines for those objects.
\item The last two were not recovered when fitting without a smooth continuum contribution. One has a broad \ha\ line that could be measured at a significance of S/N=2.8 (\sigmax=2555.6~\kms), which fails the formal requirement S/N$>$3. The spectrum is shown in Figure~\ref{fig:missBL}. The last case also has a low signal-to-noise per pixel and due to having a redshift at $z>0.45$, it lacks coverage of \ha\ and fails to be detected in other broad lines (bottom panel of Figure~\ref{fig:QsoMakerMissed}). Visually, one can spot tentative broad \mgii\ and \hb\ lines and the blue continuum is typical of quasars and Type 1 AGN. Furthermore, a deeper spectrum to be published with DESI DR1 confirms its QSO spectral type (not shown). Thus, we preserve the QSO spectral type for these two low-S/N spectra but they are plotted with open red circles in Figures~\ref{fig:kex_orig}, \ref{fig:kex} and \ref{fig:sig_z}.
\end{itemize}

\subsection{Red QSOs}\label{app:redqso}

The majority of normal quasars are characterized by blue continuum and colors, which is attributed to emission from the accretion disk. We indeed found a typical blue quasar spectrum when averaging the spectra in Section~\ref{sec:bgsstacks}. However, dust attenuation can impact these colors and lead to red quasars \citep[e.g.,][]{Richards+2003,Glikman+2004} or even extremely red quasars \citep[e.g.,][]{Ross+2015,Hamann2017+}. These may in turn correspond to special phases of galaxy evolution such as the blow out phase following a gas-rich major merger \citep{Sanders+1988}, according to which dusty gas-rich material is entrained thereby obscuring the central AGN with a foreground dust screen relative to the line of sight.

\begin{figure*}
\begin{centering}
\includegraphics[width=0.98\textwidth]{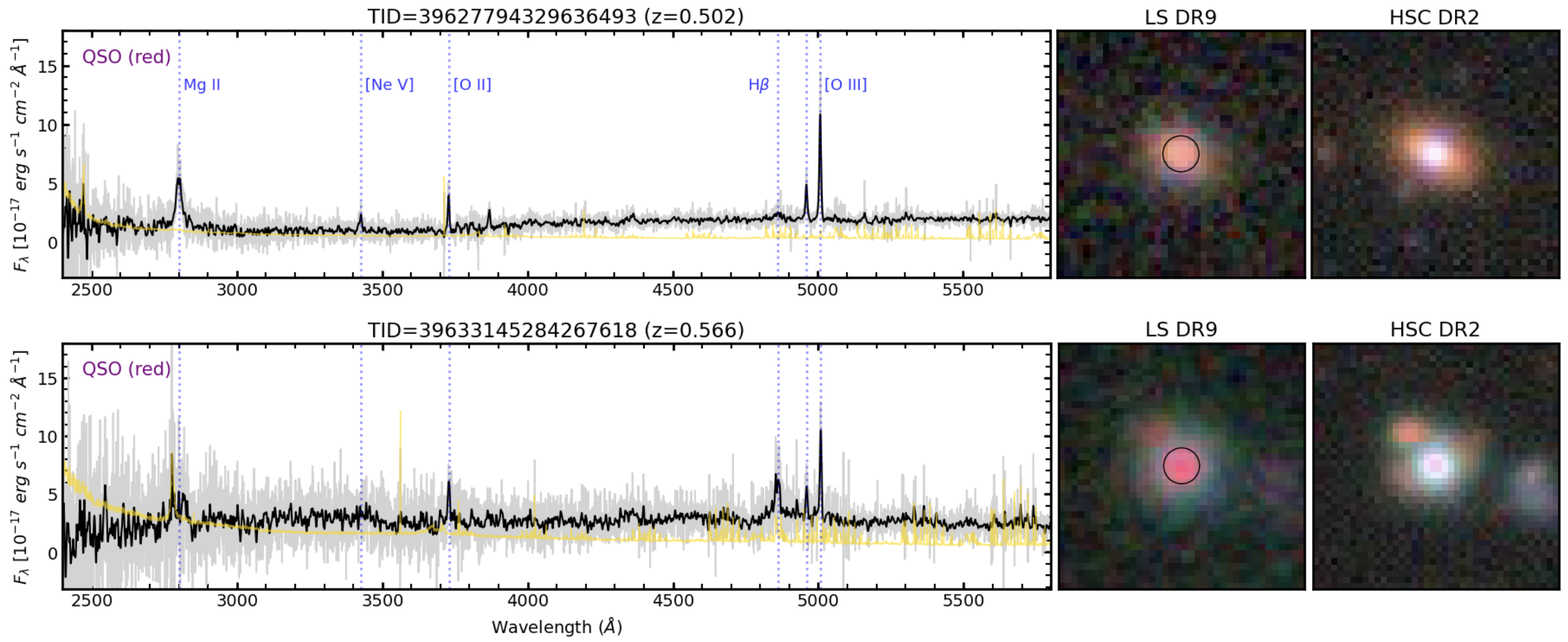}
\caption{Rest-frame DESI spectra of two red QSOs. The unsmoothed spectrum is in gray with a smoothed spectrum overplotted in black. The error spectrum is shown in yellow. Both spectra present emission lines such as \mgii, \hb, and \oiii\ as well as stellar absorption features along a reddened continuum. Color images come from LS DR9 (middle panels) and HSC DR2 (right-hand panels). The 1.5~arcsec DESI fiber aperture is marked with a circle on the LS DR9 images.
 \label{fig:redqso}}
\end{centering}
\end{figure*}

We show two example spectra of red quasars in Figure~\ref{fig:redqso}. The top panel shows a spectrum which was classified by Redrock as a galaxy but had a Type~2 label from visual inspection. However, the \mgii\ line is clearly a broad line so we relabel it as a QSO despite the red continuum. The bottom panel shows TARGETID=39633145284267618, which was also classified by Redrock as a galaxy but had a ``Type 1?'' flag from visual inspection. A closer look reveal the presence of broad but faint lines like \mgii\ and a broad component to \hb. The DESI survey now has a better quality spectrum for this object for DR1 (not shown) that the Redrock pipeline identifies as a QSO, and the presence of these broad lines is confirmed with a high confidence. 

There are a few other candidate red QSOs in other figures (e.g., Figures~\ref{fig:QsoMakerMissed}, \ref{fig:missBL}). However, we do not attempt to systematically differentiate between blue and red quasars in this work and we consider them all as QSOs when computing statistics. DESI includes a secondary target program dedicated to observing red quasars \citep{Fawcett+2023}, which is finding both red and extremely red cases. \citet{Fawcett+2023} created high quality stacked spectra showing similar colors as the examples reported here (their figure 12).


\subsection{Narrow-line Seyfert 1}\label{app:nls1}

\begin{figure*}
\begin{centering}
\includegraphics[width=0.98\textwidth]{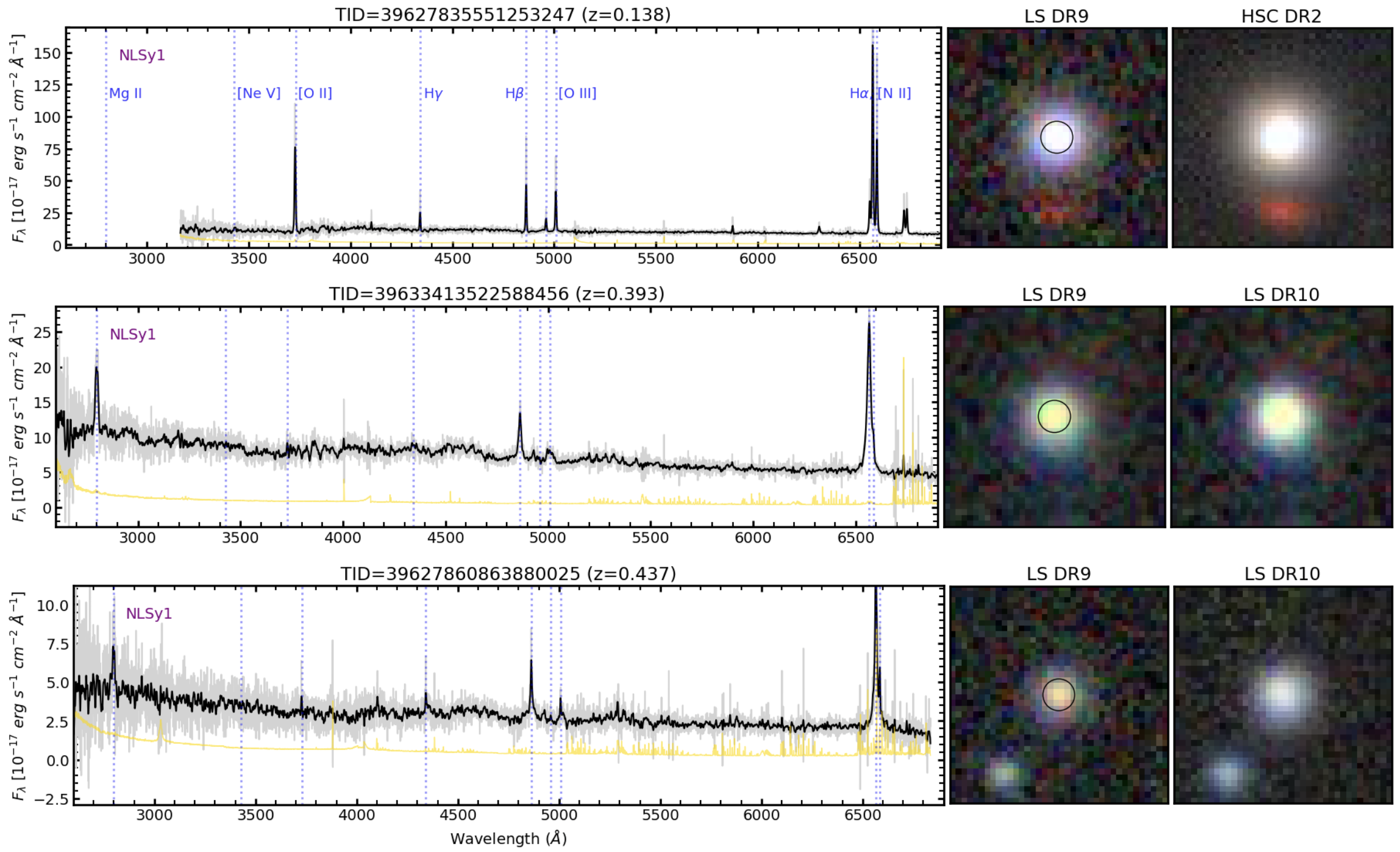}
\caption{Rest-frame DESI spectra of three candidate NLSy1 galaxies. They were visually identified as possibly being Type~2 AGN but the line width fitting found a line to the NLSy1 range ($170<\sigma<850$~kms). The unsmoothed spectrum is in gray with a smoothed spectrum overplotted in black. The error spectrum is shown in yellow. All spectra present clear emission lines as well as stellar continuum, and noticeably the \ha\ line at 6563~\AA\ is partly blended with the \niilam\ doublet. The bottom two spectra cover \mgii\ (2800~\AA), which also appear consistent with a NLSy1 classification. Color images come from LS DR9 (middle panels) and either HSC DR2 or LS DR10 (right-hand panels). The 1.5~arcsec DESI fiber aperture is marked with a circle on the LS DR9 images.
 \label{fig:nls1}}
\end{centering}
\end{figure*}

Three objects that were labeled as Type~2 AGN from VI were found to have a broad \hb\ or \ha\ component by \texttt{FastSpecFit} (filled green triangles in Figures~\ref{fig:kex_orig} and \ref{fig:sig_z}). Their spectra resemble that of Narrow Line Seyfert 1's (NLSy1) and are displayed in Figure~\ref{fig:nls1}. These spectra are distinct from both regular broad-line Seyfert 1's and Seyfert 2's but we had not defined a separate NLSy1 category during the visual inspection campaign. Early work on NLSy1's reported \hb\ line widths in the range 400-1000~\kms \citep{Osterbrock+1985}, and the commonly used definition is $400<$FWHM(\hb)$<2000$~\kms\ (which corresponds to a Gaussian line width $170<\sigma<850$~\kms) with \oiii/\hb$<3$ \citep{Osterbrock+1983,Goodrich1989}. More recent studies have also used \mgii\ with the same line width threshold to expand the range of redshift for NLSy1 search \citep[e.g.,][]{Rakshit+2021}. The three spectra below are consistent with these definitions, and we note that one of them was also identified as a QSO by the QN classifier (TARGETID=39633413522588456; middle row of Figure~\ref{fig:nls1}). 

In detail, recent work found that careful spectral fitting might require a third component resulting in a narrow, intermediate and broad component \citep{Mullaney+2008}. While there are only a few examples shown here from the BGS-AGN survey validation sample, the full DESI survey is likely well suited to study this subpopulation. NLSy1's may be at an interesting evolutionary phase such as young AGN that have recently switched on and are accreting with high Eddington rates \citep[see, e.g., review by][]{Komossa2008}.

\subsection{Seyfert 2}\label{app:sy2}

We use the label Sy2 to refer to Type~2 AGN and QSO candidates and meaning that only narrow lines are visible in their spectra. The VI campaign originally assigned a Type~2 label to 12 spectra. One of them was relabeled as galaxy while in the reverse direction one galaxy was relabeled as Type~2 due, in both cases, to their \oiii/\hb\ ratio and line width as measured by \texttt{FastSpecFit}. In addition, three were found to have a narrow broad line consistent with a NLSy1 category introduced above (Section~\ref{app:nls1}, bringing the number down to nine. Lastly, one of those was relabeled as a red QSO upon further examination showing a broad but faint line together with a reddened continuum (top row of Figure~\ref{fig:redqso}). So this results in a final number of eight Sy2.

\begin{figure*}
\begin{centering}
\includegraphics[width=0.98\textwidth]{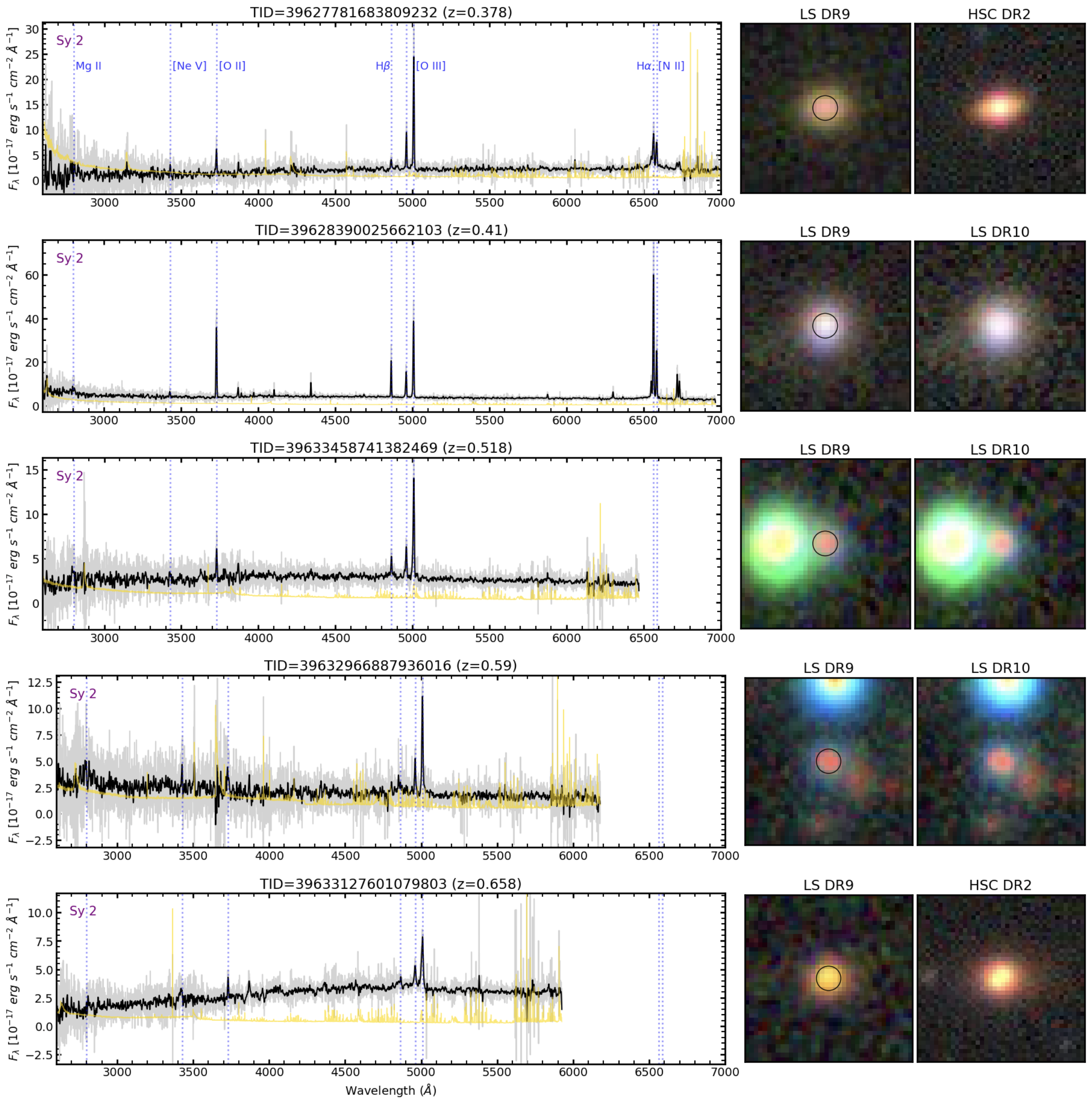}
\caption{Rest-frame DESI spectra visually classified as Seyfert~2 but which were classified as QSO by either the QN or MgII classifier. The unsmoothed spectrum is in gray with a smoothed spectrum overplotted in black. The error spectrum is shown in yellow. Color images come from LS DR9 (middle panels) and either HSC DR2 if available or otherwise LS DR10 (right-hand panels). The 1.5~arcsec DESI fiber aperture is marked with a circle on the LS DR9 images.
 \label{fig:sy2qm}}
\end{centering}
\end{figure*}

First, we show 5/8 that were classified as QSO from the QN or MgII classifiers in Figure~\ref{fig:sy2qm}. Those five spectra all have emission lines with either a red or flat continuum. Some of them display an \nevlam\ emission line, which is a clear evidence of AGN due to its high ionization potential of 97~eV. However, we cannot visually rule out whether some of these spectra are host-reddened Type 1 AGN (i.e., red QSOs) rather than regular Type 2 AGN attributed to small-scale obscuration of the BLR according to the AGN unification model \citep[e.g.,][but also see \citet{RamosAlmeida+2011}]{Antonucci1993,Urry+1995}. We count them as narrow line AGN in this work, but we remind that the modified QSO classification pipeline based on the QN and MgII classifiers assigns them a QSO spectral type. If we switched those five as well as the three NLSy1, our QSO success rate would reach (495/519) = 95.4\%.

\begin{figure*}
\begin{centering}
\includegraphics[width=0.98\textwidth]{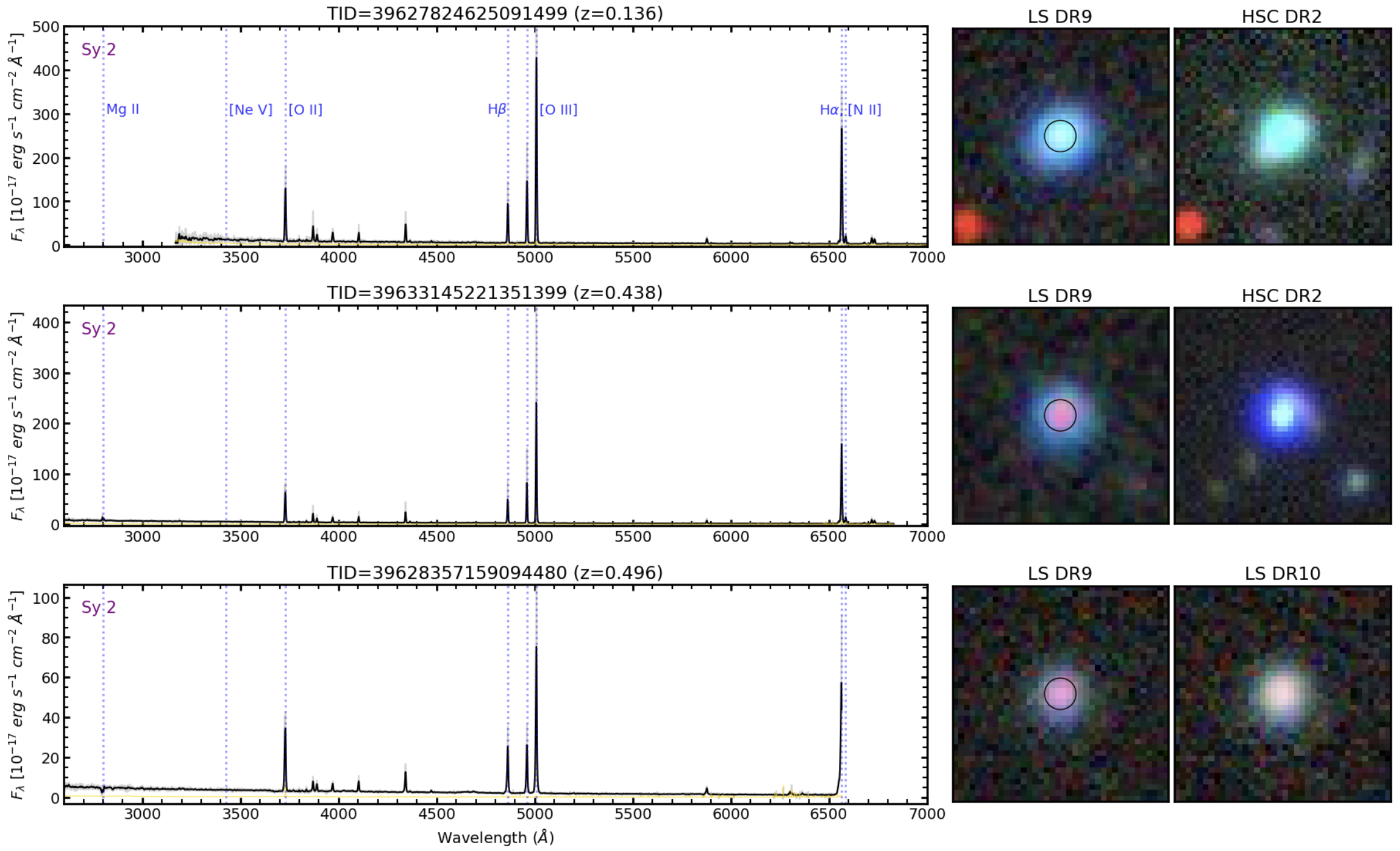}
\caption{Rest-frame DESI spectra of three candidate Seyfert~2, which were classified as a galaxy spectral type by Redrock and the QN and MgII classifiers. The unsmoothed spectrum is in gray with a smoothed spectrum overplotted in black. The error spectrum is shown in yellow. The expected locations of a subset of emission lines are marked with vertical dotted lines. All spectra present strong emission lines but no [Ne V] emission and the Mg II doublet is detected once in emission (middle row) and once in absorption (bottom row). Color images come from LS DR9 (middle panels) and either HSC DR2 if available or otherwise LS DR10 (right-hand panels). The 1.5~arcsec DESI fiber aperture is marked with a circle on the LS DR9 images.
 \label{fig:sy2}}
\end{centering}
\end{figure*}

Last, we show the remaining 3/8 that were not classified as QSO by the afterburners in Figure~\ref{fig:sy2}. While they all very strong narrow emission lines, none of them have a \nevlam\ detection. They could concurrently host a starburst and/or have low-metallicity gas similarly to Green Pea galaxies \citep{Cardamone+2009}, some of which have been found to host an AGN \citep{Harish+2023}.


\subsection{Blazars}\label{app:blazar}

\begin{figure*}
\begin{centering}
\includegraphics[width=0.98\textwidth]{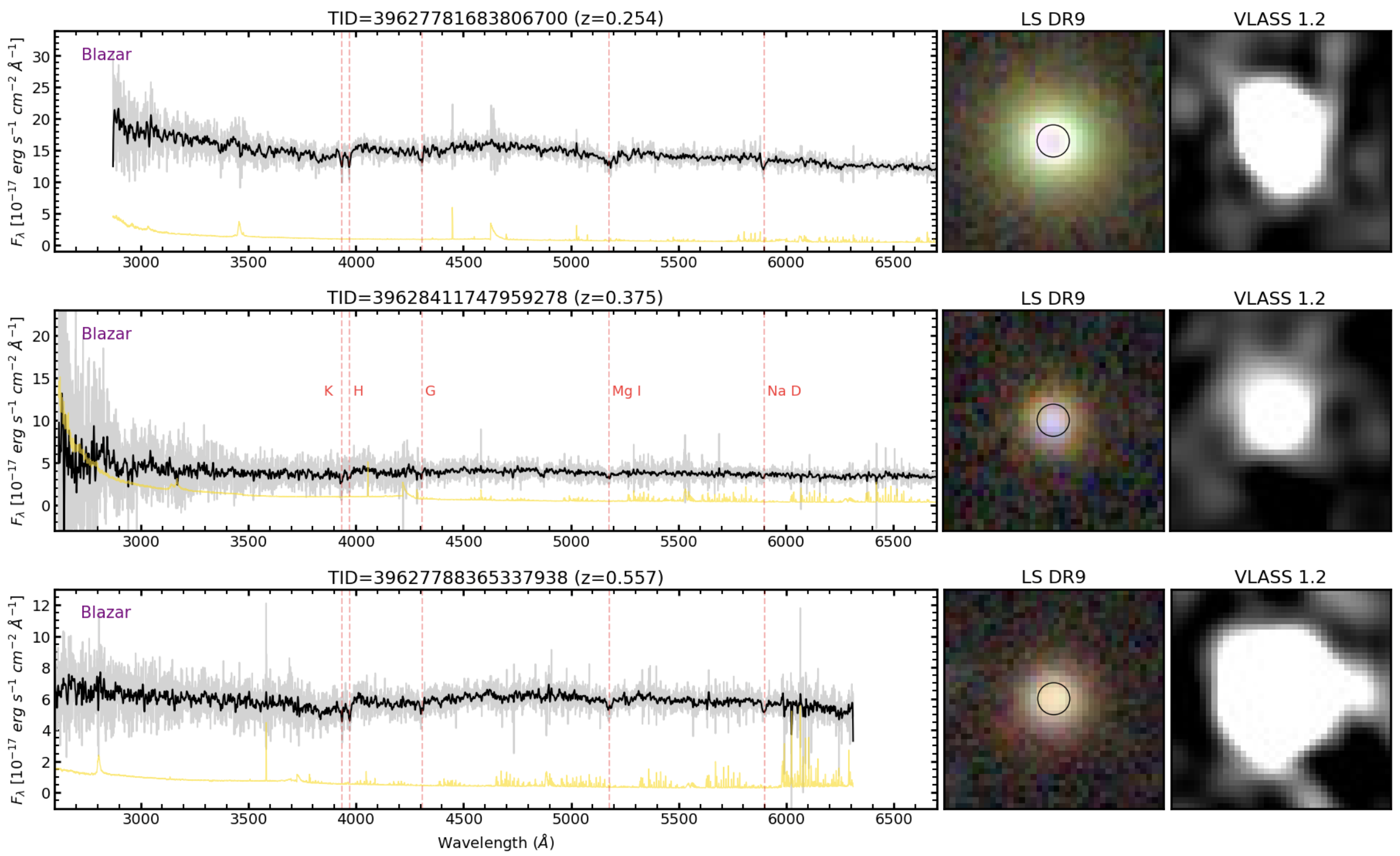}
\caption{Rest-frame DESI spectra of three candidate Blazars, which were identified by Redrock and visually as having a GALAXY spectral type. The unsmoothed spectrum is in gray with a smoothed spectrum overplotted in black. The error spectrum is shown in yellow. All spectra present weak to no emission lines and have noticeable absorption features as labeled by the vertical dashed lines. Color images come from LS DR9 (middle panels) and radio maps from VLASS 1.2 (right-hand panels). The 1.5~arcsec DESI fiber aperture is marked with a circle on the LS DR9 images.
 \label{fig:blazar}}
\end{centering}
\end{figure*}

Blazars are a class including BL Lac objects (with a nearly featureless spectra) and compact radio sources with a flat spectrum thought to arise when the observer is looking toward the radio jet \citep{Angel+1980,Antonucci+1985}. This chance alignment makes this a rare class of objects, and is usually associated with high variability. \citet{Plotkin+2008} classify optical properties based on SDSS spectra and found a range of spectral shapes and features. The cases shown in Figure~\ref{fig:blazar} resemble the BL Lac in being nearly featureless except that host galaxy stellar absorption lines are noticeable (vertical dashed red lines). Those three spectra were assigned a galaxy spectral type by Redrock and by the visual inspection campaign. A closer look showed an unexpected blue rise of the continuum given the otherwise near absence of emission lines so we searched for the presence of radio emission from the VLASS2.1 which covers 3~GHz and all three have obvious and fairly compact emission, consistent with a blazar classification. None of the remaining nine galaxy spectral types show such radio emission. We thus classify those three as the only three blazar candidates but did not systematically search for radio emission in the QSO spectra even though previous work has reported some blazars with broad optical lines \citep[e.g.,][]{Marcha+1996,Shaw+2012}.

%

\clearpage

\section{Star and galaxy contamination}
\label{app:contamination}

As mentioned in Section~\ref{sec:validation}, there were small percentages of stellar and galaxy contaminants in the BGS AGN sample. In this Appendix, we show example spectra and color images for those objects.

\subsection{Stars}

\begin{figure*}
\includegraphics[width=0.95\textwidth]{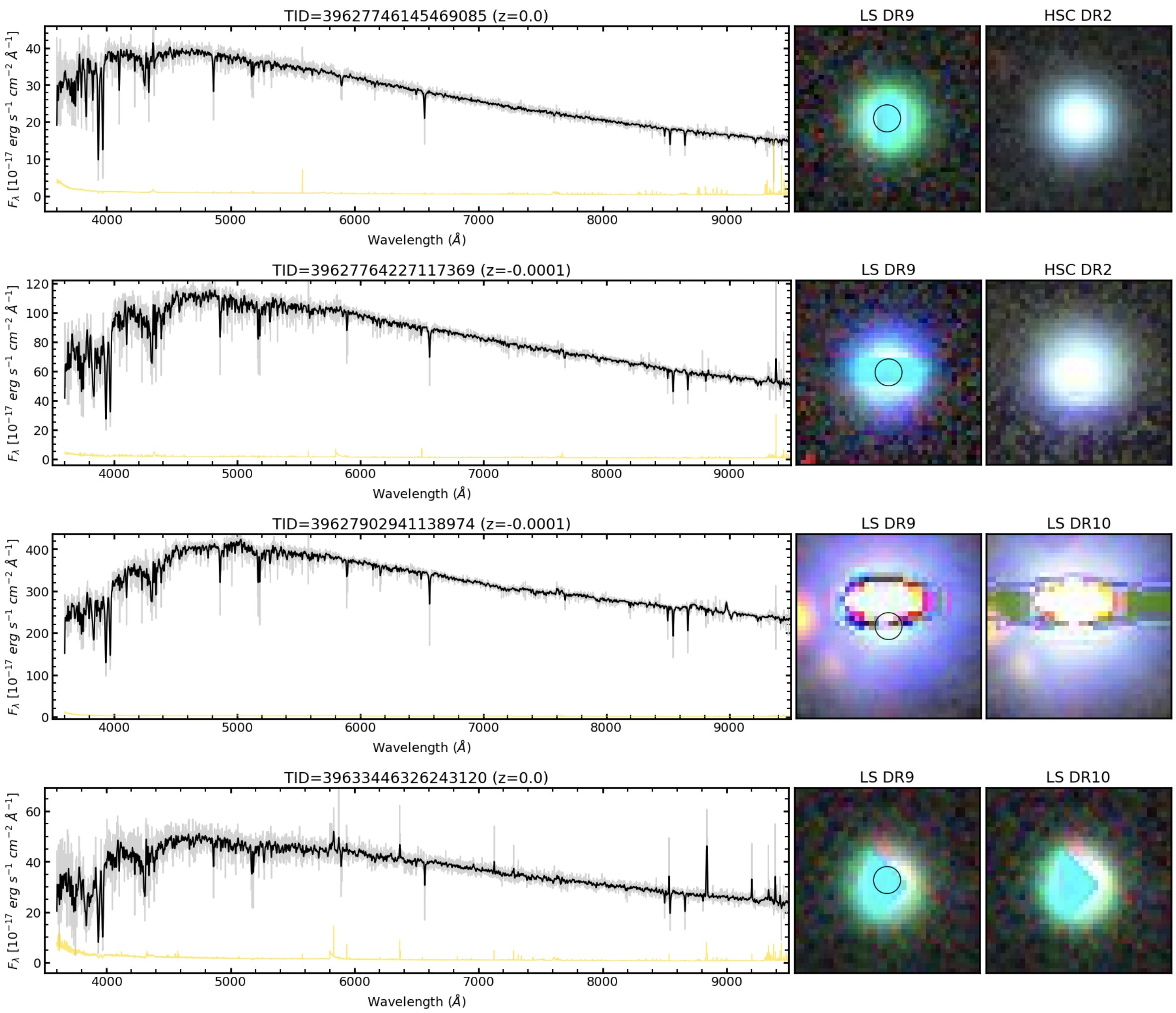}
\caption{Observed frame DESI spectra of the four stellar contaminants with a PSF morphology (left-hand panels). The observed spectrum is in gray with a smoothed spectrum overplotted in black. The error spectrum is shown in yellow, and is comparatively negligible in the case of the four bright targets shown here. Color images come from LS DR9 (middle panels) and either HSC DR2 if available or otherwise LS DR10 (right-hand panels). The 1.5~arcsec DESI fiber aperture is marked with a circle on the LS DR9 images.}\label{fig:starspsf}
\end{figure*}

We split the nine stars based on their photometric morphology from the Tractor fitting and according to the visual inspection confidence. Figure~\ref{fig:starspsf} shows the four stars with a point source (PSF) morphological type. They are all bright by definition (PSF was only allowed for $r<17.5$) and were all assigned a high confidence.

\begin{figure*}
\includegraphics[width=0.95\textwidth]{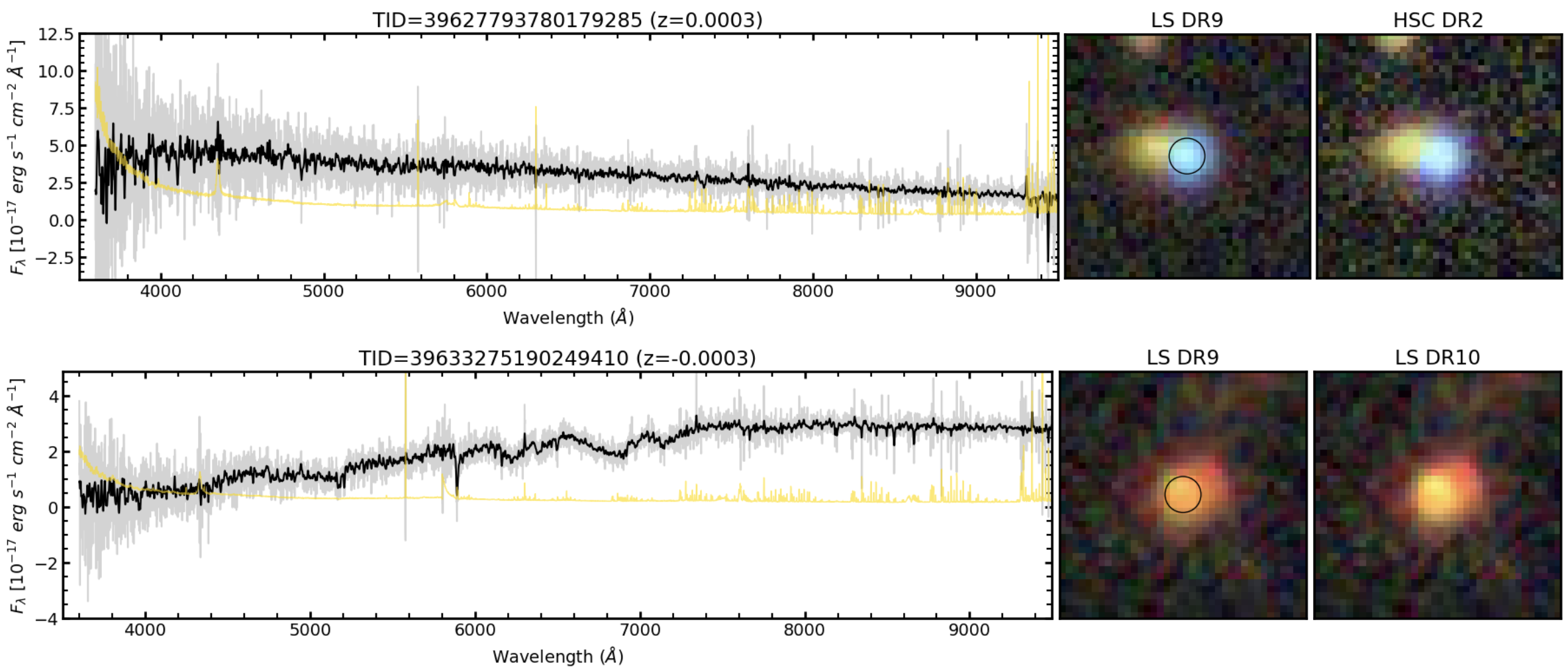}
\caption{Similar to Figure~\ref{fig:starspsf} but for the observed frame DESI spectra of the two stellar contaminants with a non-PSF morphology and with a high confidence redshift quality (left-hand panels). Color images are shown from LS DR9 (middle panels) and either HSC DR2 if available or otherwise LS DR10 (right-hand panels). The images suggest possible two or more blended objects with distinct colors.}\label{fig:stars}
\end{figure*}

Figure~\ref{fig:stars} shows the two stellar contaminants with a non-PSF morphological type and a high redshift confidence rating. In both cases, there are possibly two (or more) sources in the optical images with different colors. The top row shows a blue spectrum and target and while there is possibly a second object that contributes to the spectrum, no secondary redshift identification was made. Zooming in the spectrum interactively suggests subtle absorption lines at the expected location of the Balmer lines but there are no obvious strong features seen in the noisy spectrum. In contrast, the second row shows a red stellar spectrum with clear absorption lines and bands. The image shows a central orange object (star) but the nature of the redder extended feature is unknown. Potential contributions from more than one target to the resulting spectra have not been ruled out but no secondary redshifts were visually flagged and a more detailed analysis is beyond the scope of this work. The bottom line is that we suspect that these stellar contaminants were included in the BGS-AGN sample based on the unusual combination of colors and morphology affected by blending with overlapping neighbors in angular distance, which may be projection effects and not necessarily associated pairs or multiples.

\begin{figure*}
\includegraphics[width=0.95\textwidth]{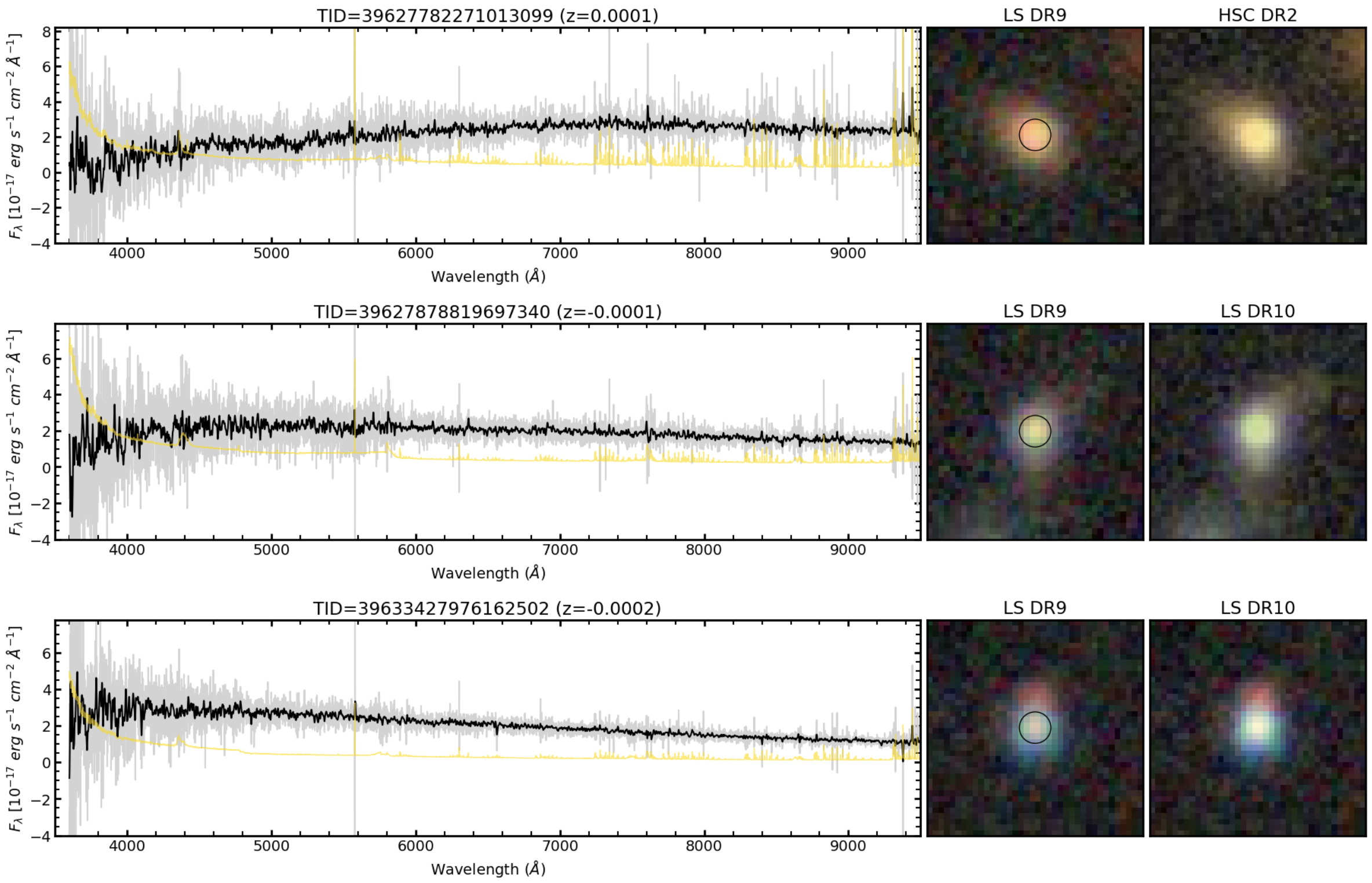}
\caption{Observed frame DESI spectra of the three candidate stellar contaminants with a non-PSF morphology and a low confidence in the redshift and/or classification (left-hand panels). Color images are shown from LS DR9 (middle panels) and either HSC DR2 if available or otherwise LS DR10 (right-hand panels). Some images suggest two or more blended objects while the top and third row appear to have a possible diffuse component but with faint spectra leading to a low confidence visual classification.}\label{fig:stars_lowq}
\end{figure*}

Lastly, we show the three cases with non-PSF morphological types and with a low confidence redshifts, which also tend to correspond to a low-confidence classification due to poorer spectral quality (Figure~\ref{fig:stars_lowq}). 
They tend to be faint and the images show either a possible extended component (in 2/3 cases) or a close projected neighbor with a different color (in 1/3 case). In the two cases with possible extended components, the spectra are noisy and highly uncertain and the objects appear unlikely to be stars due to their morphologies. Therefore, no confident redshift or classification could be obtained for them. The third case with a close projected neighbor may show \ha\ absorption line consistent with a near zero redshift but remains overall a low confidence case. Luckily, these uncertain cases represent only 0.6\% (3/519) of the sample.

\subsection{Galaxies}

\begin{figure*}
\begin{centering}
\includegraphics[width=0.95\textwidth]{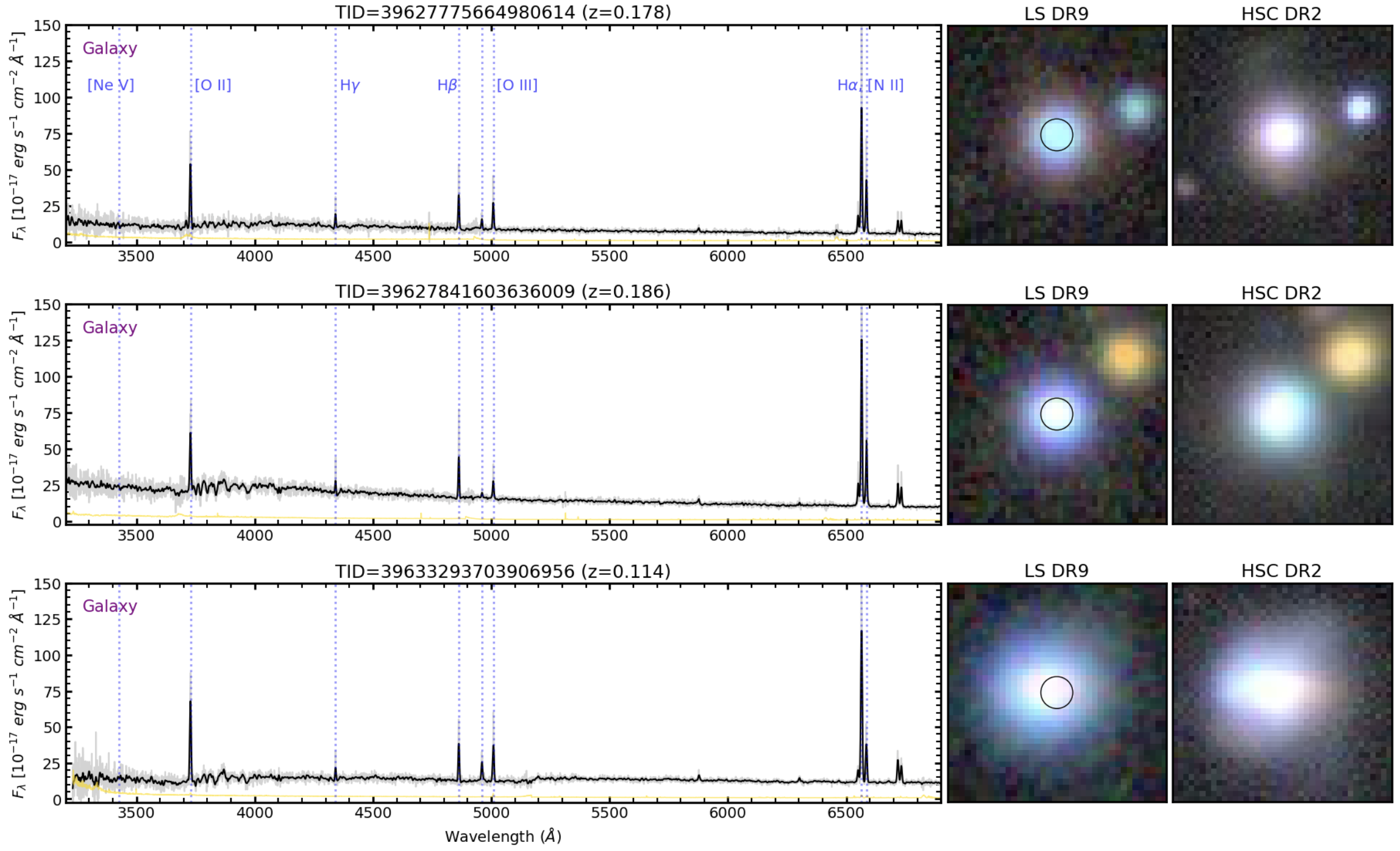}
\caption{Rest-frame DESI spectra visually classified as galaxies. The three cases shown here share common characteristics such as strong emission lines and overall blue continuum and colors. The observed spectrum is in gray, a Gaussian-smoothed spectrum in black and the error spectrum is in yellow. Color images come from LS DR9 (middle panels) and HSC DR2 (right-hand panels). The 1.5~arcsec DESI fiber aperture is marked with a circle on the LS DR9 images. 
 \label{fig:gal_blue}}
\end{centering}
\end{figure*}

In this section we display spectra and images of the nine BGS-AGN targets that were visually classified as galaxies and without any AGN subcategory such as Type~2 or blazar candidates. There remains a lot of variety among both the spectra and the color images, ranging from star-forming galaxies with blue images and strong narrow emission lines to redder passive galaxies with spectra dominated by stellar continuum and stellar absorption lines. First, we create a montage for the three bluest galaxies characterized by strong emission lines (Figure~\ref{fig:gal_blue}). In addition to the obvious emission lines, all three spectra exhibit stellar absorption features which are especially noticeable in the 3700-4000~\AA\ region (H\&K lines, Balmer absorption, etc.). These features suggest that the host galaxies are fairly massive and may include a range of stellar population ages in addition to ongoing star formation or starbursting activity. 

\begin{figure*}
\begin{centering}
\includegraphics[width=0.95\textwidth]{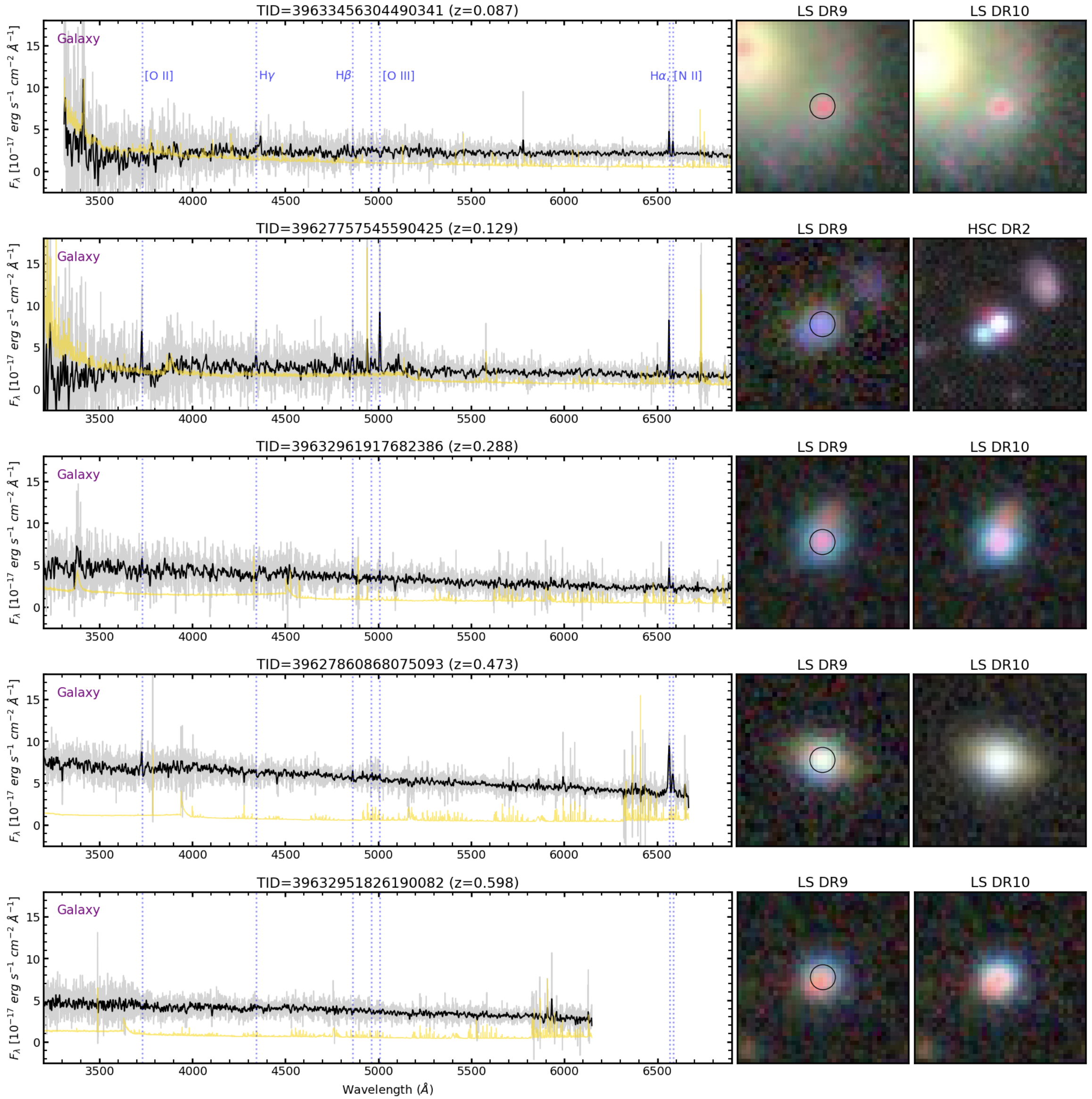}
\caption{Rest-frame DESI spectra visually classified as galaxies. The three cases shown here share common characteristics such as strong emission lines and overall blue continuum and colors. The observed spectrum is in gray, a Gaussian-smoothed spectrum in black and the error spectrum is in yellow. Color images come from LS DR9 (middle panels) and either HSC DR2 if available or otherwise LS DR10 (right-hand panels). The 1.5~arcsec DESI fiber aperture is marked with a circle on the LS DR9 images. Galaxies are displayed in order of increasing redshifts from top to bottom.
 \label{fig:gal_rest}}
\end{centering}
\end{figure*}

Next, we display five galaxies with weak or no detectable emission lines in order of increasing redshift (Figure~\ref{fig:gal_rest}). The color images of the first case indicates a clear overlap with a background galaxy, and it is not clear whether the low redshift of $z=0.087$ applies to the intended target. The remaining four galaxies in Figure~\ref{fig:gal_rest} tend to show variations in their color images suggesting possible multiple components (associated or only in projection). The spectra still tend to show a blue continuum but the weakness of the emission lines are generally consistent with either no or very weak AGN. Additionally, we verified that there is no VLASS detection as this could indicate radio AGN like the Blazar candidates (Figure~\ref{fig:blazar}).

\begin{figure*}
\begin{centering}
\includegraphics[width=0.95\textwidth]{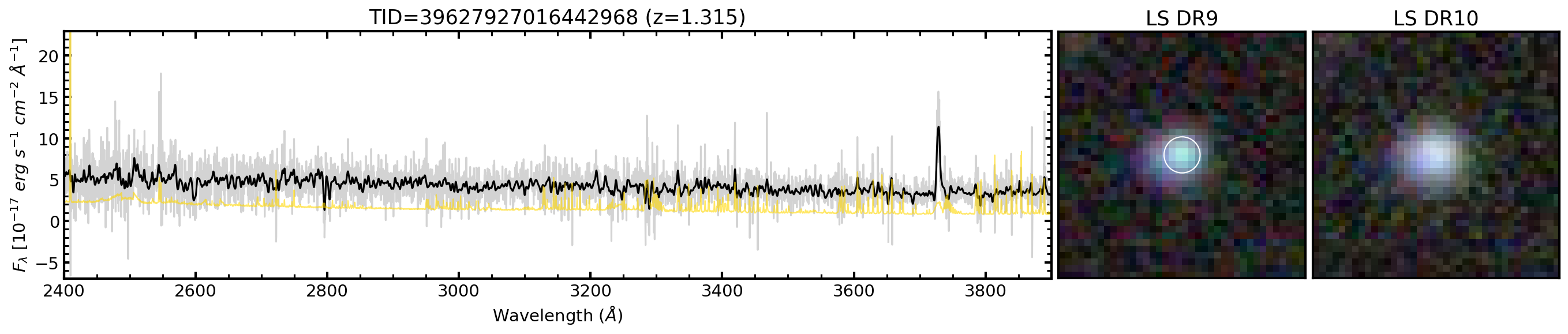}
\caption{Rest-frame spectrum of the only $z>1$ target visually classified as a galaxy (TARGETID=39627927016442968). The spectrum is shown both unsmoothed (grey) and smoothed with a Gaussian kernel ($\sigma=3$; black line). It is characterized by a faint blue continuum, a MgII doublet absorption feature (2800\,\AA), and the \oii\ doublet emission line (3727\,\AA). Color images are shown for the LS DR9 and LS DR10 as labeled and measure 10 arcseconds on a side. The DESI 1.5 arcsecond fiber aperture is marked with a circle.
 \label{fig:hiz_gal}}
\end{centering}
\end{figure*}

Lastly, we show the rest-frame spectrum and color images of the single VI-identified galaxy with a redshift $z>1$ in Figure~\ref{fig:hiz_gal}. The spectrum is characterized by a faint blue continuum with a clear \mgii\ doublet absorption feature ($\sim2800$\,\AA) and an obvious \oiilam\ emission feature though the doublet components are not well resolved. The galaxy seems marginally resolved in the images with the LS DR9 image showing bright and blue central emission. Given the lack of AGN signatures in the spectrum, we assume that this emission is due to star forming or starbursting regions in the host galaxy.

\clearpage

\bibliography{biblio}{}
\bibliographystyle{aasjournal}

\end{document}